\newenvironment{eq}[1]
{\[\begin{array}{#1}}{\end{array}\]}

\let\rvec=\vec        

\def\intq3{\int{d^3q\over q_0(2\pi)^3}} 
\def\intp3{\int_{\R^3}{d^3p\over p_0(2\pi)^3}}

\def\plintaq3{{}^\pl\hskip-2.5mm\int{d^3q\over |\rvec q|(2\pi)^3}} 
\def\plintaqq3{{}^\pl\hskip-2.5mm\int{d^3q\over 2|\rvec q|(2\pi)^3}} 
\def\plintq3{{}^\pl\hskip-2.5mm\int{d^3q\over q_0(2\pi)^3}} 
\def\plintp3{{}^\pl\hskip-2.5mm\int_{\R^3}{d^3p\over 2p_0(2\pi)^3}}

 \def\20strich{ \rule[.5mm]{20mm}{.5mm}  }
 \def\40strich{\rule[.5mm]{40mm}{.5mm}}
 \def\12strich{\rule[.5mm]{12mm}{.5mm}}

\def\plint {\hskip-2mm{^\pl\hskip-3mm\int}}

 \def\({\Bigl(}
\def\){\Bigr)}   
 \def\|{\Big|}
\def\then{\Rightarrow}  
 \def\o{\circ}
\def\m{\bullet}    
\def\x{\times}
   
\def\ox{\otimes}

\def\pl{{~\oplus~}}

\def\SUM{\displaystyle \sum}

\def\mid{\big\bracevert}

\def\sub{\subseteq}
\def\subnoteq{\subset}
\def\sup{\supseteq}
\def\supnoteq{\supset}
\def\and{\wedge}

\def\rin{{\,\in\kern-.42em\in}}

\def\rank{\,{\rm rank}}

\def\res{{\rm res}}

\def\det{\,{\rm det }\,}

\def\centr{\,{\rm centr}\,}

\def\sx{~\rvec\x~\!}

\def\irrep{{{\bf irrep\,}}}

\def\A{{\,{\rm A\kern-.55emA}}}
\def\B{{\,{\rm I\kern-.2emB}}}
\def\C{{\,{\rm I\kern-.55emC}}}
\def\E{{\,{\rm I\kern-.2emE}}}
\def\G{{\,{\rm I\kern-.55emG}}}
\def\H{{{\rm I\kern-.2emH}}}
\def\I{{\,{\rm I\kern-.2emI}}}
\def\K{{\,{\rm I\kern-.2emK}}}
\def\L{{\,{\rm I\kern-.2emL}}}
\def\M{{\,{\rm I\kern-.16emM}}}
\def\N{{\,{\rm I\kern-.16emN}}}
\def\Q{{\,{\rm I\kern-.5emQ}}}
\def\R{{{\rm I\kern-.2emR}}}
\def\S{{\,{\rm I\kern-.42emS}}}
\def\T{{\,{\rm I\kern-.37emT}}}
\def\UU{{\,{\rm I\kern-.51emU}}}
\def\Z{{\,{\rm Z\kern-.32emZ}}}

\def\p{\partial}

\def\al{\alpha}  \def\be{\beta} \def\ga{\gamma}
\def\de{\delta}  \def\ep{\epsilon}  \def\ze{\zeta}
\def\th{\theta}   \def\vth{\vartheta} 
\def\ka{\kappa}   \def\la{\lambda}   \def\si{\sigma}
\def\De{\Delta}   \def\om{\omega} \def\Om{\Omega}
\def\phi{\varphi} 
 \def\Ga{\Gamma}  
\def\Si{\Sigma}

\def\vec#1{\underline{\bf vec}_{#1}}

\def\GL{{\bf GL}}  
\def\SL{{\bf SL}}
\def\U{{\bf U}} 
 
\def\O{{\bf O}}   
\def\SU{{\bf SU}} 
 
\def\SO{{\bf SO}}

 \def\D{{\bl D}}

\def\d#1{{\check{#1}}}
\def\angle#1{\langle#1\rangle}

\def\rstate#1{|#1\rangle}
\def\lstate#1{\langle#1|}

\def\brack#1{\lbrack#1\rbrack}

\def\ty#1{{\tt #1}}
\def\ro#1{{\rm #1}}
\def\bl#1{{\bf {#1}}}
\def\cl#1{{\cal #1}}

\def\ol#1{\overline{#1}}

\def\dprod#1#2{\langle#1,#2\rangle}
\def\sprod#1#2{\langle#1|#2\rangle}
\def\rsprod#1#2{\lbrace#1|#2\rbrace}

\def\com#1#2{\lbrack#1,#2\rbrack}

\def\acom#1#2{\{#1,#2\}}

\def\expv#1#2#3{\langle#1|#2|#3\rangle}
 
\def\map{\longrightarrow}
\def\inmap{\hookrightarrow}
\def\lrmap{\leftrightarrow}

\def\lmap{\longleftarrow}

\def\mape{\longmapsto}

\def\ccon{~\stackrel{2R}*\hskip-2mm_\ka~}

\documentclass[12pt]{article}

\advance\topmargin by -1.6cm

\begin{document}

\begin{titlepage} 
$~$
\vskip5mm
\hfill MPP-2005-39 
\vskip25mm

\centerline{\bf GAUGE COUPLING CONSTANTS}
\vskip5mm
\centerline{\bf  AS RESIDUES OF SPACETIME REPRESENTATIONS}
\vskip15mm
\centerline{
Heinrich Saller\footnote{\scriptsize hns@mppmu.mpg.de} }
\centerline{Max-Planck-Institut f\"ur Physik}
\centerline{Werner-Heisenberg-Institut}
\centerline{M\"unchen, Germany}

\vskip25mm

\centerline{\bf Abstract}
\vskip5mm

The gauge coupling constants in the electroweak standard model can be written as mass
ratios, e.g. the  coupling constant for isospin
interactions  $g_2^2=2{m_W^2\over m^2}\sim
2({80\over169})^2\sim{1\over 2.3}$ 
with the mass of the  charged weak boson  and the 
mass parameter characterizing  the ground state degeneracy. A theory is given which
relates the two masses in such a  ratio to invariants 
which characterize the re\-pre\-sen\-ta\-tions 
of a noncompact nonabelian group with real rank 2.
The two noncompact abelian subgroups are operations
for time and  for a hyperbolic
position space in a model for spacetime, homogeneous under 
dilation and  Lorentz group action. 
 The re\-pre\-sen\-ta\-tions of the spacetime model
embed the bound state re\-pre\-sen\-ta\-tions of hyperbolic position space
as seen in the nonrelativistic hydrogen atom.
Interactions 
like Coulomb or Yukawa interactions are described by 
Lie algebra re\-pre\-sen\-ta\-tion coefficients.
A quantitative determination
of the  ratio 
of the invariants for position and time related operations,
determined  by the spacetime re\-pre\-sen\-ta\-tion,  gives the right order of magnitude for the gauge
coupling constants.

\end{titlepage}

\vskip15mm
{\small\tableofcontents}
\newpage
\setcounter{page}{5}

\section{Introductory remarks}

In the following, the radical point of view is taken that
all basically relevant physical properties, e.g. 
energies, momenta,  masses, spin, helicity, charges  and also 
coupling constants, can be understood as invariants 
and eigenvalues connected with
the action of  operations from real finite dimensional  Lie groups.
It helps on this way and will be shown in this paper, that 
the basically relevant  wave functions, like
the hydrogen bound states or the on-shell part of the particle Feynman propagators,
 are re\-pre\-sen\-ta\-tion matrix elements 
of  operation groups and that the 
basic interactions, like Coulomb or Yukawa potentials
or the off-shell part of Feynman propagators,
are re\-pre\-sen\-ta\-tion coefficients of the corresponding  Lie algebras as 
tangent operations. Sometimes, there will arise quite a new and, perhaps,
unfamiliar  
group and operation oriented language for well known physical concepts.

Spacetime is an operational concept, physically interpretable by its
re\-pre\-sen\-ta\-tions. It will be modelled by a homogeneous space 
of a  group. Therefore, in the end, an understanding
 of spacetime 
is reduced to an understanding of the
re\-pre\-sen\-ta\-tions of the underlying group
and - for its interactions
and its particles - of the tangent translations, i.e. of the  corresponding Lie
algebra re\-pre\-sen\-ta\-tions.

Masses of particles and coupling constants, especially for
mass zero particles, will be related to invariants of
 translation re\-pre\-sen\-ta\-tions\cite{WIG}
and the normalization of such  re\-pre\-sen\-ta\-tions as arising from  product
re\-pre\-sen\-ta\-tions of an underlying nonabelian group. The formulation in terms of
`residual re\-pre\-sen\-ta\-tions'\cite{S002} leads to an
interpretation of invariants and normalizations as complex poles and their 
residues.

The numerical values of gauge coupling constants
seem not to be rational numbers as shortly sketched below
(section 3) in the context of the standard model for the electroweak and strong
interactions\cite{WEIN}.
If the values are from a continuous spectrum,
the related operations
have to include a noncompact group since all properties from
compact operations are given
by rational numbers\cite{FULHAR}, e.g. (hyper)charge numbers,
angular momenta (spin), color dimensions etc.
The nontrivial re\-pre\-sen\-ta\-tion structure of a Lie group, 
which is both noncompact and  nonabelian,  
e.g. of the Lorentz group,
is - for re\-pre\-sen\-ta\-tions with a probability inducing Hilbert product -
infinite dimensional.

In a certain sense the sections 4, 5 and 6 serve  as an introduction
to the theory of noncompact group re\-pre\-sen\-ta\-tions 
as applied for an understanding of the gauge coupling constants
as residues of spacetime re\-pre\-sen\-ta\-tions.
They contain a formulation
 of  quantum mechanical free scattering
and bound states  and of free particles in quantum field
theory to exemplify - without interaction -
the use of translation re\-pre\-sen\-ta\-tions
and - for interaction -
the use of nonabelian hyperbolic re\-pre\-sen\-ta\-tion structures.
 
After this preparation the spacetime re\-pre\-sen\-ta\-tion 
 theory for relativistic  particles and interactions can be seen as a
 generalization and an embedding of these structures. 

\section{Basic re\-pre\-sen\-ta\-tion theory and notations}

The level of the mathematical tools to treat
noncompact nonabelian Lie groups  is not undergraduate:
It is difficult from the conceptual point of view\cite{KNAPP} 
and looks complicated in the
explicit  formulation. In this section 
some basic re\-pre\-sen\-ta\-tion theory\cite{FOL,KNAPP,KIR} is shortly summarized which will be used 
below  in many physical examples.

There will be considered re\-pre\-sen\-ta\-tions of  
real Lie groups $G$ on complex vector
spaces with action $\rstate a\mape g\m\rstate a$ for $g\in G$.
If not stated otherwise, all those re\-pre\-sen\-ta\-tions are meant as
Hilbert re\-pre\-sen\-ta\-tions, i.e. 
acting upon a   Hilbert space
with probability inducing invariant scalar product $\sprod ab$.
Faithful re\-pre\-sen\-ta\-tions of noncompact Lie groups are infinite
dimensional. 

For a locally compact group $G$ all re\-pre\-sen\-ta\-tions are
direct sums of cyclic ones
 and direct integrals of irreducible ones.
A cyclic re\-pre\-sen\-ta\-tion space is the closed complex
span of the group orbit of a vector $G\m\rstate c$ - called a cyclic vector.
Irreducible re\-pre\-sen\-ta\-tions are cyclic -  not necessarily vice versa.

For a  compact group $U$ all re\-pre\-sen\-ta\-tions are
direct sums  of irreducible finite dimensional ones 
- there, the additional concept `cyclic' is not important.
With the Plancherel measure a counting measure the direct integrals are direct
sums.

All  re\-pre\-sen\-ta\-tions of a locally compact group
involve complex group functions acted upon with the both sided regular group re\-pre\-sen\-ta\-tion.
Their values are re\-pre\-sen\-ta\-tion matrix elements (coefficients)
$G\ni g\mape d(g)=\expv a{g\m}b$.
The dual of the algebra with the  continuous 
compactly supported
functions $\cl C_c (G)$ is the convolution algebra 
with the Radon measures 
$\cl M(G)$  which are  generalized functions
(distributions)  with Haar measure basis. 
The Dirac measures embed the group.
The Lebesgue spaces $L^p(G)$, $1\le p\le\infty$, 
contain the selfdual Hilbert space $L^2(G)$ with the square integrable,
the convolution algebra $L^1(G)$ with the absolute integrable 
 and its dual $L^\infty(G)$ with 
the essentially bounded function classes.
$L^1(G)$ can be considered 
to constitute a both sided
ideal in the 
Radon distributions.
For a compact group the algebra $L^1(U)$ contains all Lebesgue spaces,
$L^p(U)\sup L^q(U)$ for $p\le q$.
The  functions, basically relevant for  free and interacting
physical structures in spacetime,   will be given explicitly below. 

The Hilbert  spaces with re\-pre\-sen\-ta\-tions of a locally compact group
can be constructed with equivalence classes of functions from $L^1(G)$.
They are characterized by their scalar product:
There is a bijection between 
functions which induce a scalar product on $L^1(G)$
\begin{eq}{l}
\sprod{f}{f'}_d=\int _{G\x G}dg_1dg_2\ol{f(g_1)}d(g^{-1}_1g_2)f'(g_2)
\end{eq}the so called positive type functions\cite{GELRAI}  
$d\in L^\infty(G)_+$ 
from the dual of the Lebesgue convolution group algebra,
and equivalence classes of cyclic $G$-re\-pre\-sen\-ta\-tions.
Not all Hilbert spaces with the action of a
noncompact locally compact group have to be constituted
by square integrable function classes.

The positive type functions constitute a  cone.
Their conjugation goes with the group inversion
${d(g)}=\ol{d(g^{-1})}$ and they are bounded by the value at the neutral element
$|d(g)|\le d(e)$.
The conjugated partner of a positive type functions
for a cyclic re\-pre\-sen\-ta\-tion 
characterizes the dual re\-pre\-sen\-ta\-tion\cite{LIE13} $d\lrmap \ol d$
and a real positive type function $d=\ol d$  
a selfdual  cyclic re\-pre\-sen\-ta\-tion. 
Many explicit examples will be given below.

A positive type function gives the  group expectation values 
of a cyclic vector $g\mape d(g)=\expv c {g\m}c$.
If normalized at the group unit,
it is called  a state. 
With a normalized cyclic vector,   group re\-pre\-sen\-ta\-tion and  
quantum probability normalization are related to each other.

An extremal element in the convex set of states, i.e.
not  combinable  with strictly positive numbers from other states,
is called a pure state. A  cyclic vector for a pure state
is called  a  pure cyclic vector.
There is a bijection 
between pure states and  equivalence classes of 
irreducible  $G$-re\-pre\-sen\-ta\-tions.

Summarizing these concepts and their notations

{\scriptsize
\begin{eq}{c}
\begin{array}{|c |}\hline
\hbox{positive type function (cyclic) }d\in L^\infty(G)_+\cr
\hbox{with cyclic vector }\cr
\hfill G\ni g\mape d(g)=\expv c {g\m}c\hfill\cr\hline
\hbox{pure state (irreducible) }\cr
\hbox{with normalized pure cyclic vector}\cr
\hfill G\ni e\mape d(e)=\sprod \Om\Om=1\hfill\cr\hline

\end{array}\cr\cr
G\sub\cl C_c (G)'=\cl M(G)\sup L^1(G),~~L^1(G)'=L^\infty(G)

\end{eq}}

Group functions 
contain the functions of  group classes 
$G/H$  with a closed subgroup, i.e. functions 
on symmetric (homogeneous) $G$-spaces.
The $H$-intertwiners on $G$, valued in
a $H$-re\-pre\-sen\-ta\-tion space, are acted upon with
induced $G$-re\-pre\-sen\-ta\-tions\cite{MACK1}, in general not irreducible.

Some notation for symmetric spaces, relevant in the following:
The spheres and hyperboloids
parametrize the orientation manifolds of compact rotations $\SO(s)$ 
either  in compact rotations $\SO(1+s)$ or in noncompact Lorentz transformations  
$\SO_0(1,s)$. They get  symbols 
(instead of the elsewhere  also used $S^s$ and $H^s$)
which look a bit similar to the 1-dimensional circle 
$\Om^1$ and one branch hyperbola $\cl Y^1$
\begin{eq}{rll} 
\hbox{spheres:}&\Om^s\cong\SO(1+s)/\SO(s),& s=1,2,\dots,~~\Om^0=\{\pm 1\}\cr  
&\hfill\Om^1\cong\SO(2)\cong\U(1),&\Om^3\cong\SU(2)\cr
\hbox{hyperboloids:}&\cl Y^s\cong\SO_0(1,s)/\SO(s),& s=1,2,\dots\cr  
&\hfill\cl Y^1\cong\SO_0(1,1)\cong\D(1),&\cl Y^3\cong\SL(\C^2)/\SU(2)\cr
\end{eq}The 3-dimensional hyperboloid $\cl Y^3$ is the
orientation manifold of position spaces in Minkowski spacetime, it characterizes
special relativity.
In contrast to the hyperboloids, the spheres have a finite measure
(area, volume)
$|\Om^{s}|={2\pi^{1+s\over 2}\over \Ga({1+s\over 2})}$.
A semidirect affine group $G\sx\R^n$ has cosets $G\sx\R^n/G\cong\R^n$ with
respect to the  homogeneous group $G\sub\GL(\R^n)$. They can be parametrized by
$\R^n$ as symmetric space, in general not as additive group.

The two abelian Lie groups, the compact 1-torus $\U(1)$ 
and its noncompact cover,
the additive group $\R$ with
 multiplicative notation $\D(1)=\exp \R$ (dilations),
 are the Cartan subgroup types. The abelian subgroups 
  in semisimple groups come in their
 selfdual re\-pre\-sen\-ta\-tions, i.e.  as
abelian axial rotations   $\SO(2)$ and as abelian Lorentz transformations 
$\SO_0(1,1)$ respectively.

There is a big difference between orthogonal groups $\SO_0(t,s)$
for  even dimensions $t+s=2R$ and odd dimensions $t+s=1+2R$
with rank $R=1,2,\dots$ as seen, e.g., 
in the centers of their
covering groups and
the number  of spinor re\-pre\-sen\-ta\-tions. 
This difference can be illustrated also 
with the Lorentz group in four dimensions\cite{GELNAI,NAIM} 
 $\SO_0(1,3)\sim\SL(\C^2)$ with one type of Cartan subgroups 
 $\SO(\C^2)=\SO(2)\x\SO_0(1,1)$
and the Lorentz group in three dimensions\cite{BARG} 
$\SO_0(1,2)\sim\SL(\R^2)$ with two Cartan subgroup types
$\SO(2)$ and $\SO_0(1,1)$.
The theory below is built
for odd dimensional position  rotations $\SO(2R-1)$, nontrivial for $R\ge2$,
as subgroups of orthochronous Lorentz groups $\SO_0(1,2R-1)$
acting on  even dimensional spacetimes  with 1-dimensional time.   
For an odd dimensional spacetime with even dimensional
position, e.g.  Bargman spacetime
 with $\SO_0(1,2)\sx \R^3$\cite{BARG,S041} replacing
Minkowski spacetime with $\SO_0(1,3)\sx\R^4$, 
the theory would look very different.

The cyclic groups $\Z_R\cong\Z/R\Z$ 
(rest classes $k{\rm mod} R$ with additive notation)
will  be used also in a multiplicative notation $\I(R)$ (complex unit roots
$z^R=1$). The sign and step functions
are re\-pre\-sen\-ta\-tions of
the reals in $\I(2)\cong\Z_2\cong\Om^0$ and will be  denoted as follows
\begin{eq}{lcccl}
\R\ni x\mape& \ep(x)&=&{x\over |x|}&\in\{\pm 1\}\cong\I(2)\cong\Om^0\cr
\R\ni x\mape& \vth(\pm x)&=&{1\pm\ep(x)\over2}&\in\{0,1\}\cong\Z_2
\end{eq}The step functions give the 
characteristic functions for 
future $\R_+$ and past $\R_-$.

With a group $G$-re\-pre\-sen\-ta\-tion and its complex functions there is also
the re\-pre\-sen\-ta\-tion of its Lie algebra, denoted  as its logarithm $\log G$.
The tangent structure of  a noncompact nonabelian group
and its homogeneous spaces $G/H$  
will be interpreted 
as interactions for the spaces $G/H$. The more abstract formulation of 
Lie algebra re\-pre\-sen\-ta\-tion coefficients and the
physical realization as interactions for position space and spacetime
is taken up again in
section 10.

\section{Standard model of gauge interactions} 

For an experimental orientation,
the  order of magnitude of  gauge coupling constants  is given 
as used in the standard model of 
 particle interactions.
 
Electrodynamics connects the translations
$\R^4$ of Minkowski spacetime
with internal $\U(1)$-trans\-formations.
The  charge $iQ$ implements the   action of the Lie algebra $\log\U(1)$.
Its position density leads
to currents $\bl J_k$, e.g. for a quantum Dirac 
field $\acom{\ol{\bl\Psi}}{\bl\Psi}(\rvec x)=\ga^0\de(\rvec x)$
with integer $\U(1)$-winding number 
(charge number)  $z$ 
\begin{eq}{rl}
\U(1):&\bl\Psi\mape e^{i z\al}\bl\Psi,~~
\ol{\bl\Psi}\mape  e^{-i z\al}\ol{\bl\Psi},~~z\in\Z\cr
\hbox{for }\log\U(1):&
\bl J_k= z{[\bl\Psi\ga_k, \ol{\bl\Psi}]\over2},
~Q=\int d^3 x\bl J_0(x),~~\left\{\begin{array}{ll}
[Q,\bl\Psi]&=z \bl\Psi\cr
[Q,\ol{\bl\Psi}]&=-z\ol{\bl\Psi}\end{array}\right.
\cr
\end{eq}

The $\U(1)$-gauge dynamics is characterized by the classical Lagrangian 
with  the individual kinetic  Lagrangians
and the electromagnetic interaction
\begin{eq}{l}
\bl L(\bl A)+\bl L(\bl\Psi)-\bl A^k\bl J_k
,~~\left\{\begin{array}{rl}

\bl L(\bl A)&=\bl F_{kj}{\p^k \bl A^j-\p^j \bl A^k\over2}
+ g ^2{\bl F_{kj} \bl F^{kj}\over4}\cr
\bl L(\bl\Psi)&=i\ol{\bl\Psi}\p^k\ga_k\bl\Psi+m\ol{\bl\Psi}\bl\Psi
\end{array}\right.
\end{eq}The constant $ g ^2$ is the 
{electromagnetic coupling  constant}, 
related to Sommerfeld's fine structure constant $\al_S$,
with the experimental value 
\begin{eq}{l}
\al_S={g^2\over 4\pi}\sim{1\over 137.036}
,~~  g^2 \sim {1\over 10.9}
\end{eq}The coupling  constant is
the normalization of the gauge field $\bl A$ and related to
the residue at the mass zero pole in the Feynman propagator 
\begin{eq}{rl}
\angle{\acom {\bl A^k}{\bl A^j}(x)-\ep(x_0)\com {\bl A^k}{\bl A^j}(x)}
&={i\over\pi}\int{d^4q\over 2\pi}
{-\eta^{kj}g^2\over q^2+io} e^{iqx}\cr
\angle{ \com{\ol{\bl\Psi}}{\bl\Psi}(x)
-\ep(x_0)\acom{\ol{\bl\Psi}}{\bl\Psi}(x)}
&={i\over\pi}\int {d^4q\over 2\pi}{\ga^kq_k+m\over
q^2+io-m^2} e^{iqx}\cr
\de(q^2-m^2)~~+\hskip5mm{i\over\pi}{1\over
q^2_\ro P-m^2}\hskip10mm&={i\over\pi}{1\over
q^2+io-m^2},~~
\ro P\hbox{ principal value}\cr
\hskip6mm\hbox{on-shell}\hfill\hbox{off-shell}\hskip9mm&\cr

\end{eq}All electromagnetic interactions are quantitatively determined 
with the
value of the gauge coupling constant $g^2$ and the
re\-pre\-sen\-ta\-tion characteristic
integer $\U(1)$-winding numbers $z\in\Z$.

The minimal {standard model} 
of the elementary interactions in Minkowski spacetime $\R^4$
is a theory of compatibly represented external 
Poincar\'e group and internal `chargelike' operations. It
 embeds the 
 electromagnetic interaction for an  electron Dirac field 
 (quantum electrodynamics)
 into
the electroweak and strong 
  interactions 
 of  lepton and quark Weyl fields. 
The fields involved are acted upon with
irreducible re\-pre\-sen\-ta\-tions $[2L|2R]$ of the  Lorentz (cover) group
$\SL(\C^2 )$ 
and irreducible re\-pre\-sen\-ta\-tions
$[y]$, $[2T]$ and $[2C_1,2C_2]$ of the hyperchar\-ge group $\U(1)$
(rational hyperchar\-ge number $y$), isospin group $\SU(2)$ 
(integer or  halfinteger isospin  $T$) and  color  group
$\SU(3)$ (characterized by two integers $2C_{1,2}$)
as given in the following table

{\scriptsize
\begin{eq}{c}
\begin{array}{|c||c|c|c|c|c|c|}\hline
\hbox{\bf field}&\hbox{\bf symbol}&\SL(\C^2 )&
\U(1)&\SU(2)&\SU(3)\cr
              & &[2L|2R]& [y] &[2T]& [2C_1,2C_2]\cr\hline\hline
\hbox{left lepton}&\bl l&[1|0]&-{1\over2}&[1]&[0,0]\cr\hline
\hbox{right lepton}&\bl e&[0|1]&-1&[0]&[0,0]\cr\hline
\hbox{left quark}&\bl q&[1|0]&{1\over6}&[1]&[1,0]\cr\hline
\hbox{right up quark}&\bl u&[0|1]&{2\over3}&[0]&[1,0]\cr\hline
\hbox{right down quark}&\bl d&[0|1]&-{1\over 3}&[0]&[1,0]\cr\hline\hline
\hbox{hyperchar\-ge gauge}&\bl A_0&[1|1]&0&[0]&[0,0]\cr\hline
\hbox{isospin gauge}&\rvec{\bl A}&[1|1]&0&[2]&[0,0]\cr\hline
\hbox{color gauge}&\bl G&[1|1]&0&[0]&[1,1]\cr\hline
\end{array}\cr\cr

\hbox{\bf fermion and gauge fields of the standard model}\cr
\end{eq} 
}
\noindent with the  dimensionalities of the re\-pre\-sen\-ta\-tions spaces 
\begin{eq}{l}
d_{\SL(\C^2)}=(1+2L)(1+2R),~\left\{ \begin{array}{rl}
d_{\SU(2)}&=1+2T\cr
d_{\SU(3)}&=(1+2C_1)(1+2C_2)(1+C_1+C_2)\end{array}\right.
\end{eq}

The  gauge interactions  
\begin{eq}{l}
\bl L(\bl A_0,\rvec{\bl A},\bl G)+
\bl L(\bl l,\bl e,\bl q,\bl u,\bl d)
- (\bl A_0^k\bl J_k+\bl A_a^k\bl J^a_k+\bl G_A^k\bl J^A_k)
\end{eq}involve the gauge  fields with their nonabelian self-interactions
\begin{eq}{rrlll}
\hbox{for }\U(1):&
\bl L(\bl A_0)
&=\bl F_{kj}{\p^k \bl A_0^j-\p^j \bl A_0^k\over2}
&+ g _1^2{\bl F_{kj} \bl F^{kj}\over4} 
\cr
\hbox{for }\SU(2):&
\bl L(\rvec{\bl A})
&=\bl F^c_{kj}{\p^k \bl A^j_\ka -\p^j \bl A^k_\ka 
-\ep^{ab}_\ka \bl A_a^k\bl A_b^j\over2}

&+ g _2^2{\bl F_{kj}^b \bl F^{kj}_b\over4} \cr
\hbox{for }\SU(3):&
\bl L(\bl G)
&=\bl F^C_{kj}{\p^k \bl G^j_\ka -\p^j \bl G^k_\ka -\ep^{AB}_\ka \bl G_A^k\bl G_B^j\over2}
&+ g _3^2{\bl F_{kj}^B \bl F^{kj}_B\over4}
\end{eq}and  the currents of the fermion fields  
as `densities' of the Lie algebra
 where a chiral basis 
 $\ga_k={\scriptsize\pmatrix{0&\si_k\cr\d \si_k&0\cr}}$
 uses Weyl matrices $\si_k=(\bl1_2,\rvec\si)$ and
 $\d\si_k=(\bl1_2,-\rvec\si)$.
$\bl 1_R$ denotes the $R$-dimensional unit (matrix)   
\begin{eq}{rll}  
\hbox{for $\log\U(1)$:}
&\bl J_k=&
-{1\over2}\bl l\d\si_k \bl l^\star
-\bl e\si_k \bl e^\star
+{1\over6}\bl q\d\si_k \bl q^\star
+{2\over3}\bl u\si_k\bl u^\star
-{1\over3}\bl d\si_k\bl d^\star\cr
\hbox{for $\log\SU(2)$:}
&\bl J^a_k=&
\bl l\d\si_k{\tau^a\over2} \bl l^\star
+\bl q\d\si_k{\tau^a\over2} \bl q^\star\cr
\hbox{for $\log\SU(3)$:}
&\bl J^A_k=&
\bl q\d\si_k{\la^A\over2}\bl q^\star
+\bl u\si_k{\la^A\over2}\bl u^\star
+\bl d\si_k{\la^A\over2}\bl d^\star
\end{eq}

The electromagnetic group $\U(1)$ is embedded into  
the product of the Abelian hyperchar\-ge group $\U(1)$
with the nonabelian isospin-color group $\SU(2)\x\SU(3)$
\begin{eq}{rl}
\U(1)\inmap
 \U(1)\o[\SU(2)\x\SU(3)]&\cong {\U(1)\x\SU(2)\x\SU(3)\over \I(2)\x\I(3)}\cr
 \U(R)=\U(\bl1_R)\o\SU(R)&\cong{\U(1)\x\SU(R)\over \I(R)},~~
 \U(\bl 1_R)\cap\SU(R)\cong \I(R)
\end{eq}In the standard model, the  re\-pre\-sen\-ta\-tions  of both factors are
centrally cor\-re\-la\-ted\cite{HUCKS,S921,S981}
via  the
$\SU(2)\x\SU(3)$-centrum, the cyclotomic group  $\I(2)\x\I(3)=\I(6)$
(hexality = two-triality, `David star').
E.g., the hypercharge invariant of the 
left handed quarks $\bl q$ with isospin-color multiplicity 6 is the inverse,
i.e. $y={1\over 6}$.
The central correlation of the internal symmetries
is expressed by the modulo-relations
for the rational 
invariants $[y||2T;2C_1,2C_2]$ 
 of $\U(1)\o[\SU(2)\x\SU(3)]$ 
carried by the standard
model fermion fields 
\begin{eq}{l}
\begin{array}{rl}
6y{\rm mod}2&=2T{\rm mod}2\in\Z_2\cr
6y{\rm mod}3&=2(C_1-C_2){\rm mod}3\in\Z_3\cr\end{array},~~
y\cdot  d_{\SU(2)}\cdot d_{\SU(3)}\in\Z
\end{eq}The hypercharge as  invariant of the abelian  group is connected
with the isospin-color re\-pre\-sen\-ta\-tion dimension as invariant
of the nonabelian group.

For the transition from interaction fields to particles\cite{S003} 
the centrally correlated hypercharge $\U(1)$  and isospin $\SU(2)$-transformations 
have to  be disentangled
\begin{eq}{l}
\U(2)=\U(\bl 1_2)\o\SU(2),~~
\U(\bl 1_2)\cap\SU(2)=\centr\SU(2)=\{\pm\bl1_2\}\cr
\end{eq}The transition from the
two  parameters $(\al_0,\al_3)\in[0,2\pi]^2$, correlated
at $(\pi,0)\cong(0,\pi)$, to the 
two uncorrelated ones $(\al_+,\al_-)\in[0,2\pi]^2$
for a maximal abelian subgroup (Cartan torus)
\begin{eq}{l}
\U(\bl1_2)\o\SO(2)=\U(1)_+\x\U(1)_-
,~~
e^{i(\bl 1_2\al_0+\tau^3\al_3)}=e^{i{\bl 1_2+\tau^3\over2}\al_+}
e^{i{\bl 1_2-\tau^3\over2}\al_-}\cr
 \end{eq}is performed 
 via the  Weinberg $\SO(2)$-rotation 
(`center of charge transformation' - in analogy to the center of mass
transformation in mechanics) which defines 
coupling constants for transformations from a direct product  Cartan subgroup 
\begin{eq}{l}
 g _1^2{\bl F_0^2\over4}+ g _2^2{\bl F_3^2\over4}
= g ^2{\bl F_+^2\over4}+\ga ^2{\bl F_-^2\over4}
 \end{eq}One  direct factor $\U(1)_+$ is gives the 
 electromagnetic $\U(1)$-action on particles.
Electromagnetic relativity is described by the Goldstone 
manifold $\U(2)/\U(1)_+$ and experimentally visible in
the  three weak interactions.

The central $\I(2)$-correlation in the internal hypercharge-isospin group
gives the necessarily integer winding numbers $z=y+T_3$ for 
the electromagnetic
group $\U(1)_+$ action on the color trivial lepton particles. 
The remaining non-integerness
of the $\U(1)_+$-numbers $y+T_3\in\{\pm {1\over3} ,\pm{2\over 3}\}$ for 
isospin trivial quark 
color triplet fields 
can be removed  in color trivial  product re\-pre\-sen\-ta\-tions as required  
for hadronic particles.

The  Weinberg angle $\th$ - for the form -
and the fine structure constant - for the area - determine the
electroweak orthogonal triangle for the coupling constants 
(notation 
$(a^2,b^2|h^2,c^2)$ with squared lengths of the sides and the height).  
The orthogonal sides 
are the hypercharge $\U(1)$ and 
isospin $\SU(2)$ coupling constants
and the height  the electromagnetic $\U(1)_+$ coupling constant   
\begin{eq}{l}
\hbox{experiment:}\left\{
\begin{array}{rl}
{4\pi\over g^2}&\sim  137\cr
{g_2^2\over g_1^2}
=\cot^2\th&\sim3.35\cr
\end{array}\right\}\then\left\{
\begin{array}{l}
\hbox{orthogonal triangle }\cr

({ g^2_1},{g^2_2}|{g^2},{\ga^2 })=g_1g_2
({ g_1\over g_2},{g_2\over g_1}|{g\over\ga},{\ga\over g} )\cr
\hbox{with } g_1g_2=g\ga,~g^2_1+ g^2_2=\ga^2\cr
\end{array}\right.
\end{eq}With the ground state degeneracy, 
implemented in the minimal standard model by a Higgs field 
$\angle{\bl\Phi^\star\bl\Phi(x)}=m^2$
and  experimentally given by the
weak interaction mass $m\sim 169 {{\rm GeV}\over c^2}$,
the electroweak triangle for the coupling 
constants can be written as a triangle for mass ratios,
involving the masses $m_W\sim 80 {{\rm GeV}\over c^2}$  and 
$m_Z\sim 91 {{\rm GeV}\over c^2}$ for the weak charged and neutral boson
\begin{eq}{rl}
({ g^2_1},{g^2_2}|{g^2},{\ga^2 })&={2\over m^2}({ m_1^2},{ m_W^2}|{m_h^2},{ m_Z^2})
\sim ({1\over 8.4},{1\over 2.5}|{1\over 10.9},{1\over 1.9})\cr
g_1g_2= g\ga&={2m_1m_W\over m^2}={2m_hm_Z\over m^2}={2g^2\over\sin2\th}\sim{1\over4.6}
\end{eq}It is this order of magnitude
which should be looked for squared  gauge coupling constants.

\section{Free scattering states and free particles}

In quantum mechanics, e.g. for a nonrelativistic potential,
there are bound states and scattering states.
Free scattering states and  free particles as their relativistic extension 
involve cyclic re\-pre\-sen\-ta\-tions of the additive translation groups
 which will be considered 
in this section.  
They collect irreducible  re\-pre\-sen\-ta\-tions,
e.g. for time, position and spacetime
translations $x\in\R^n$, $n=1,3,4$. The  
irreducible Hilbert spaces are 1-dimensional, the re\-pre\-sen\-ta\-tions are not faithful 
\begin{eq}{l}
\R^n\ni x\mape e^{iqx}\in\U(1)
\end{eq}The  (energy-)momenta $q$ as elements of the dual group
characterize the irreducible  re\-pre\-sen\-ta\-tions
\begin{eq}{l} 
q\in\irrep\R^n\cong\R^n
\end{eq}

The selfdual translation re\-pre\-sen\-ta\-tions are the direct sum
of the dual irreducible ones, i.e. of a re\-pre\-sen\-ta\-tion pair with reflected 
(energy-)momenta $\pm q\in\R^n$
\begin{eq}{l}
\R^n\ni x\mape 
{\scriptsize\pmatrix{ e^{iqx}&0\cr 0& e^{-iqx}\cr}}=e^{i\si_3 qx}
\cong e^{i\si_1 qx}={\scriptsize\pmatrix{ \cos qx&i\sin qx\cr i\sin qx & \cos
qx\cr}}\in\SO(2)
\end{eq}Both  $x\mape e^{iqx}$ 
and $x\mape \cos qx$ are states,
the exponentials are  pure states with a normalized vector 
$\sprod qq=1$ from the
1-dimensional Hilbert space  $\C\rstate q$ a  pure cyclic vector. 
The cosine is decomposable into  dual exponentials.
It is the basic  selfdual  spherical state and characterizes 
an $\R^n$-re\-pre\-sen\-ta\-tion on a  2-dimensional Hilbert space
$\C\rstate q\pl\C\lstate q\cong\C^2$
with dual basis vectors. $\rstate q\pm\lstate q\in\C^2$
are cyclic vectors - not pure.

The simplest quantum mechanical example is given by
 the time translation re\-pre\-sen\-ta\-tions
of the harmonic oscillator with its frequency (energy) as invariant
-  selfdual  spherical for the 
position-momentum pair $(x,p)$ (cyclic vectors) and
 irreducible  for creation and annihilation operator
$(\ro u,\ro u^\star)={{1\over \ell}x\mp i\ell p\over \sqrt 2}$ 
 as translation eigenvectors (pure cyclic vectors) 
  
\begin{eq}{l}
H={p^2\over 2M}+k{x^2\over 2}=
\om[\ell^2{p^2\over 2}+{x^2\over 2\ell^2}]=\om{\acom{\ro u}{\ro
u^\star}\over2}\cr
\hbox{with frequency }\om^2={k\over M}
\hbox{ and length }\ell^4={1\over kM}\cr
\R\ni t\mape e^{\pm i\om t}\in\U(1)\then
{\scriptsize\pmatrix{\hfill{1\over \ell}x(t)\cr -i\ell p (t)\cr}}
={\scriptsize\pmatrix{
\cos \om t & i\sin \om t\cr
i\sin \om t&\cos \om t\cr}}
{\scriptsize\pmatrix{\hfill{1\over \ell}x(0)\cr - i\ell p (0)\cr}}\cr
\end{eq}The intrinsic  length $\ell$ will be related below
to an invariant for the position re\-pre\-sen\-ta\-tion.

In general, the  positive type functions 
$d\in L^\infty (\R^n)_+$ for cyclic translation re\-pre\-sen\-ta\-tions
are, with Bochner's theorem\cite{BOCH},
Fourier transformed  positive Radon measures  
of (energy-)momenta 
$d(x)=\int d^n q~\tilde d(q)e^{iqx}$, $\tilde d\in\cl M(\R^n)_+$,
e.g. for the  irreducible and basic 
 selfdual spherical cases above with a Dirac distribution 
\begin{eq}{l}
t\in\R:~~e^{i\om t}=\int d q~\de(q-\om)e^{iqt},~~
\cos\om t=\int d q~|\om| \de(q^2-\om^2)e^{iqt}
\end{eq}Selfdual positive type functions have 
Radon measures which are symmetric under reflection $q\lrmap -q$ ($q^2$-dependent).

\subsection{Euclidean groups for nonrelativistic scattering}

The group theoretical framework for 
nonrelativistic scattering is the re\-pre\-sen\-ta\-tion theory of
the semidirect Euclidean group $\SO(3)\sx\R^3$
with the rotation $\SO(3)\cong\SU(2)/\I(2)$ 
acting on the position translations $\R^3$.
Their irreducible faithful  re\-pre\-sen\-ta\-tions
are induced by irreducible $\SO(2)\x\R^3$
re\-pre\-sen\-ta\-tions and labeled by an integer  invariant  $n$ 
(helicity) for axial rotations $\SO(2)$ 
(`little group') around the momentum directions and 
the modulus of the momenta 
$\rvec q^2=P^2>0$ as continuous translation invariant
\begin{eq}{l}
(n,P^2)\in\irrep[\SU(2)\sx\R^3]\cong\N\x\R_+
\end{eq}

The re\-pre\-sen\-ta\-tions are direct integrals of 
translation re\-pre\-sen\-ta\-tions 
\begin{eq}{rl}
\SO(3)\sx\R^3/\SO(3)\cong\R^3\ni \rvec x\mape
d^3(\rvec x)&= \int {d^3q\over  2\pi P}\de(\rvec q^2-P^2)e^{-i\rvec q\rvec x}\cr
&=\int{d^2\om\over 4\pi}\cos P\rvec \om\rvec x
={\sin Pr\over Pr}\cr 
d^3(0)&=1\end{eq}The integration goes over the  momenta 
on a 2-sphere
$\Om^2\cong\SO(3)/\SO(2)$ with  radius $P>0$
and directions $\rvec \om$ 
\begin{eq}{l}
\rvec \om={\rvec q\over |\rvec q|}\in\Om^2,~~\left\{\begin{array}{rl}
\int d^2\om&=\int_{0}^\pi \sin\chi d\chi\int_{-\pi}^{\pi}d\phi=|\Om^2|=4\pi\cr
\de(\rvec\om)&={1\over\sin\chi}\de(\chi)\de(\phi)\cr
\end{array}\right.
\end{eq}The 
not square integrable spherical 
Bessel function $\rvec x\mape {\sin Pr\over Pr}$  is 
a  state for a cyclic translation re\-pre\-sen\-ta\-tion  
and a pure state for an irreducible re\-pre\-sen\-ta\-tion of the Euclidean group.

The Hilbert space\cite{S051} is given by the 2-sphere square integrable 
function (classes) $f\in L^2(\Om^2)$ with wave packets for momentum
directions
\begin{eq}{l}
\sprod {P^2;f}{P^2;f'}=\int {d^2\om\over4\pi}~\ol {f(\rvec\om)} f'(\rvec\om) 
\end{eq}The eigenvectors are no Hilbert vectors. They
  constitute a distributive basis with  scalar product distribution
for the Hilbert space,
e.g. for a re\-pre\-sen\-ta\-tion with trivial rotation invariant (helicity)
$n=0$
\begin{eq}{rl}
\hbox{distributive basis:}&\{\rstate{P^2;\rvec \om}\mid\rvec\om\in\Om^2\}\cr
\hbox{scalar product distribution:}&
\sprod{P^2;\rvec\om'}{P^2;\rvec\om}=4\pi\de(\rvec\om-\rvec\om')\cr
\hbox{translation action:}& \rstate{P^2;\rvec\om}\mape e ^{iP\rvec\om\rvec
x}\rstate{P^2;\rvec\om}\cr
\hbox{Hilbert vectors:}&\rstate {P^2;f}=\int{d^2\om\over 4\pi}
f(\rvec \om)\rstate{P^2;\rvec\om}\cr
\hbox{pure cyclic vector:}
&\rstate{P^2;1}=\int{d^2\om\over
4\pi}\rstate{P^2;\rvec\om}\cr
&\int{d^2\om d^2\om'\over
(4\pi)^2}\lstate{P^2;\rvec\om'}\cos \rvec q\rvec x \rstate{P^2;\rvec\om}
={\sin Pr\over Pr}\cr
\end{eq}

The corresponding pure states for the
re\-pre\-sen\-ta\-tions of  the Euclidean group 
$\SO(s)\sx\R^s$ with  $s=2,3,\dots$
position dimensions 
integrate translation re\-pre\-sen\-ta\-tions over the momentum sphere 
$\Om^{s-1}\cong\SO(s)/\SO(s-1)$. They 
involve Bessel functions for integer and halfinteger
index (more below). 
The translation invariant $P^2>0$ is used as intrinsic momentum unit
\begin{eq}{rl}
\R^s\ni\rvec x\mape 
d^s(\rvec x)&=\int {2d^{s}q\over |\Om^{s-1}| }\de(\rvec q^2-1)
e^{-i\rvec q\rvec x}\cr
&=\int{d^{s-1}\om\over|\Om^{s-1}| }\cos \rvec\om\rvec x
={2\over |\Om^{s-1}|}{\pi\cl J_{s-2\over 2}(r)\over({r\over 2\pi})^{s-2\over
2}}
\cr
\hbox{ with }d^s(0)&=1,~~\rvec \om={\rvec q\over |\rvec q|}\in\Om^{s-1},~~
\int d^{s-1}\om=|\Om^{s-1}|={2\pi^{s\over 2}\over \Ga({s\over 2})}
\end{eq}$L^2(\Om^{s-1})$ is the Hilbert space
with the Euclidean group action.

\subsection{Poincar\'e groups for free particles}

For free relativistic particles 
the nonrelativistic scattering group $\SO(3)\sx\R^3$ is 
embedded in the semidirect
Poincar\'e group $\SO_0(1,3)\sx\R^4$ 
with the Lorentz group $\SO_0(1,3)\cong\SL(\C^2)/\I(2)$ 
acting on the spacetime translations $\R^4$.
The  irreducible faithful  re\-pre\-sen\-ta\-tions
which are induced by irreducible
$\SU(2)\x\R^4$ re\-pre\-sen\-ta\-tions
are labeled by the integer invariant spin $2J$ 
for rotations with fixgroup (Wigner's `little group') 
$\SU(2)$ in a rest system and 
the continuous  positive mass squared $m^2>0$ as translation invariant 
\begin{eq}{l}
(2J,m^2)\in\irrep[\SL(\C^2)\sx\R^4]\cong\N\x\R_+
\hbox{ (for fixgroup $\SU(2)$)}
\end{eq}With the indefinite Lorentz metric there
are additional  re\-pre\-sen\-ta\-tion types for
trivial mass square, e.g. for photons, and for negative mass square,
not realized with particles. They are not considered here.

The Lorentz scalar matrix elements (no states)
 characteristic for irreducible re\-pre\-sen\-ta\-tions,   
integrate  spacetime translation re\-pre\-sen\-ta\-tions
\begin{eq}{rl}
\SO_0(1,3)\sx\R^4/\SO_0(1,3)\cong\R^4\ni 
 x\mape
  d^{(1,3)}(x)&=\int {d^4q\over  2\pi m^2}\de(q^2-m^2)e^{iqx}\cr
=\int{d^3q\over 2\pi m^2 q_0}\cos q_0 x_0
 \cos \rvec q\rvec x
&=\int{d^3\ty y\over 2\pi}\cos m \ty y x\cr
\hbox{with }{q_0=\sqrt{m^2+\rvec q^2}}

\cr
\end{eq}The energy-momentum `directions' 
on the special relativity forward   3-hy\-per\-bo\-loid
$\cl Y^3\cong\SO_0(1,3)/\SO(3)$ for 
mass $m^2$  and positive energy can be pa\-ra\-me\-tri\-zed by hyperbolic coordinates or, 
more familiar, by  momenta
\begin{eq}{l}
\begin{array}{r}\ty y={\scriptsize\pmatrix{
\cosh\psi\cr{\rvec q\over|\rvec q|}\sinh\psi\cr}}= 
\vth(q^2)\vth(q_0){q\over |q|}\in\cl Y^3\cr
|q|=\sqrt{|q^2|}\end{array}~
\left\{\begin{array}{rl}
\int d^3\ty y&=\int_{0}^\infty \sinh^2\psi~ d\psi \int d^2\om\cr
&= \int {d^3q\over m^2q_0}|_{q_0=\sqrt{m^2+\rvec q^2}}\cr
\de(\ty y)&={1\over\sinh^2\psi}\de(\psi)\de(\rvec \om)\cr
&=m^2q_0\de(\rvec q)\cr
\end{array}\right.\cr

\vth(q^2)e^{iqx}=e^{\ep(q_0)i|q|\ty y x},~~
\vth(q^2)\cos qx=\cos |q|\ty y x\cr
\end{eq}The Hilbert space\cite{S051} is given by 
 the free particle  relevant wave packets $f\in L^2(\cl Y^3)$ 
for energy-momentum `directions' $\ty y$  or for momenta $\rvec q $,
e.g. for trivial rotation invariant $J=0$
where the elements of a distributive basis 
 $\rstate{m_\pi^2;\rvec q}$ can describe a pion with momentum $\rvec q$
and positive energy $q_0 =\sqrt{m_\pi^2+\rvec q^2}$
\begin{eq}{rl}
\hbox{distributive basis:}&\{\rstate{m^2;\ty y}=\rstate{m^2;\rvec q}
\mid\ty y\in\cl Y^3,~\rvec q\in\R^3\}\cr
\hbox{scalar product distribution:}
&\sprod{m^2;\ty y'}{m^2;\ty y}=2\pi\de(\ty y-\ty y')\cr
&\sprod{m^2;\rvec q'}{m^2;\rvec q}=2\pi m^2q_0\de(\rvec q-\rvec q')\cr
\hbox{translation action:}
& \rstate{m^2;\ty y}\mape e ^{im \ty y x}
\rstate{m^2;\ty y}\cr
& \rstate{m^2;\rvec q}\mape e ^{iq x}
\rstate{m^2;\rvec q}\cr
\hbox{Hilbert vectors:}&
\rstate {m^2;f}
=\int{d^3\ty y\over 2\pi}
f(\ty y)\rstate{m^2;\ty y}\cr
&\hskip14mm=\int{d^3q\over 2\pi m^2q_0}f(\rvec q)\rstate{m^2;\rvec q}\cr
&\sprod {m^2;f}{m^2;f'}=\int {d^3\ty y\over 2\pi}~\ol {f(\ty y)} f'(\ty y) \cr
&\hskip26mm=\int {d^3q\over 2\pi m^2q_0}~\ol{ f(\rvec q)}f'(\rvec q)
\end{eq} 

In contrast to the scattering states with compact homogenous group
and finite sphere area $|\Om^2|=4\pi$,
the integral of the distributive basis over the energy-momentum hyperboloid is
no cyclic vector. Because of  the infinite volume
$|\cl Y^3|$, it is  a cyclic vector distribution
\begin{eq}{rl}
\hbox{cyclic vector distribution: }
\rstate{m^2;1}&=\int{d^3\ty y\over 2\pi}\rstate{m^2;\ty y}\cr
\int{d^3\ty y~d^3\ty y'\over (2\pi)^2}
\lstate{m^2;\ty y'}\cos  q x \rstate{m^2;\ty y}
&=\int {d^4q\over  2\pi m^2}\de(q^2-m^2)e^{iqx}
\end{eq}

The fixgroup $\SO(s)$-induced re\-pre\-sen\-ta\-tions  
for general Poincar\'e groups
$\SO_0(1,s)\sx\R^{1+s}$ with  $s=1,2,\dots$
position dimensions integrate translation re\-pre\-sen\-ta\-tions over the hyperboloid
$\cl Y^{s}\cong\SO_0(1,s)/\SO(s)$.
They involve Neumann functions 
for timelike and Macdonald functions for 
spacelike translations, both for integer and halfinteger
index.
The translation invariant $m^2>0$ is used as intrinsic mass unit
\begin{eq}{rl}
\R^{1+s}\ni x\mape d^{(1,s)}(x)&= \int {2d^{1+s}q\over |\Om^{s-1}| }\de(q^2-1)e^{iqx}
=\int{2d^s\ty y\over|\Om^{s-1}| }\cos \ty y x
\cr
&=
{2\over |\Om^{s-1}|}
{-\vth( x^2)\pi\cl N_{-{s-1\over2}}(|x|)+\vth(-x^2)2\cl K_{{s-1\over2}}(|x|)
\over |{x\over 2\pi}|^{{s-1\over2}}}\cr

\hbox{ with }\ty y&=\vth(q^2)\vth(q_0){q\over |q|}\in\cl Y^s,~~|x|=\sqrt{|x^2|}
\end{eq}$L^2(\cl Y^{s})$ is the Hilbert space with the Poincar\'e group action.

The embedded re\-pre\-sen\-ta\-tions of the  time translations 
and  the Euclidean group
\begin{eq}{l}
\SO_0(1,s)\sx\R^{1+s}\supnoteq \R\x [\SO(s)\sx\R^s]
\end{eq}are seen in the partial integration decomposition,
e.g. for $1+s=4$
\begin{eq}{l}
d^{(1,3)}(x)=
\int{d^4q\over 2\pi m^2}\de(q^2-m^2)e^{iqx}=
\int dq_0\vth(q_0^2-m^2)2\cos q_0x_0 {\sin\sqrt{q_0^2-m^2}r\over m^2 r}

\end{eq}

\section{Measures and spacetime coefficients}

(Energy-)momentum measures are used 
in the definition of free particle re\-pre\-sen\-ta\-tions.
The Lebesgue measure ${d^nq\over (2\pi)^n}$
is the Plancherel measure for the irreducible translation 
re\-pre\-sen\-ta\-tions $\R^n\ni x\mape e^{iqx}\in\U(1)$
and Haar measure $d^nx$.
For irreducible re\-pre\-sen\-ta\-tions of
affine groups $G\sx\R^n$ it is modified by     
Dirac  distributions of (energy-)momenta on homogeneous spaces $G/H$.
They describe interaction free structures
with cyclic translation re\-pre\-sen\-ta\-tions.

\subsection{Spherical, hyperbolic, Feynman  and causal measures}

For the circle  one has
different  pa\-ra\-me\-tri\-zations, e.g.
\begin{eq}{l}
\Om^1\ni{\scriptsize\pmatrix{q_0\cr iq\cr}},\hbox{ for semi-circle: }
{\scriptsize\pmatrix{\cos\chi\cr i\sin\chi\cr}}_{-{\pi\over2}}^{{\pi\over2}}
={1\over \sqrt{1+p^2}}{\scriptsize\pmatrix{1\cr ip\cr}}_{-\infty}^\infty
={1\over 1+v^2}{\scriptsize\pmatrix{1-v^2\cr 2iv\cr}}_{-1}^1\cr
\end{eq}Therewith, the  Euclidean group relevant measure for  
the momentum direction sphere, i.e. 
for the compact classes of orthogonal groups $\SO(1+s)/\SO(s)\cong\Om^s$,
has the pa\-ra\-me\-tri\-zations - also for $s=0$ where applicable
\begin{eq}{l}
|\Om^s|=\int d^s\om=\int 2d^{1+s}q~\de(q_0^2+\rvec q^2-1)=
\int_0^\pi(\sin\chi)^{s-1}d\chi\int d^{s-1}\om
=\int {2 d^sp\over (\rvec p^2+1)^{s+1\over2}}\cr
\hbox{polar decomposition: }
q=|q|\rvec \om \hbox{ with }|q|^2=q_0^2+\rvec q^2,~\rvec \om\in\Om^s\cr
\end{eq}For noncompact classes of orthogonal groups 
there is the Poincar\'e group relevant measure 
of the
one shell positive energylike hyperboloid
$\SO_0(1,s)/\SO(s)\cong\cl Y^s$
whose pa\-ra\-me\-tri\-zations can be  obtained with the spherical-hyperbolic
transition $(i\rvec q,i\chi,i\rvec p,i\rvec v)\to(\rvec q,\psi,\rvec p,\rvec v)$ 
 \begin{eq}{l}
\int d^s\ty y=\int 2d^{1+s}q~\vth(q_0)\de(q_0^2-\rvec q^2-1)=
\int_0^\infty(\sinh\psi)^{s-1}d\psi\int d^{s-1}\om
=\int{ d^sq\over \sqrt{\rvec q^2+1}}\cr
\hbox{`polar'decomposition: }
q=|q|\ty y\hbox{ with }|q|^2=q_0^2-\rvec q^2,~\ty y\in\cl Y^s\cr
\end{eq}

The Dirac `on-shell' and the principal  value (with $q^2_\ro P$) 
`off-shell'  distributions are  imaginary 
and real part of the
 (anti-) Feynman distributions
 \begin{eq}{rl}
 \log(q^2\mp io-\mu^2)
&=\log|q^2-\mu^2|\mp i\pi\vth(\mu^2-q^2)
\cr
 {\Ga(1+N)\over (q^2\mp io-\mu^2)^{1+N}}
&=-(-{\p\over\p q^2})^{1+N}\log(q^2\mp io-\mu^2)\cr
= (-{\p\over\p q^2})^N{1\over q^2\mp io-\mu^2}
&={\Ga(1+N)\over (q_{\ro P}^2-\mu^2)^{1+N}}\pm i\pi\de^{(N)}(\mu^2-q^2)\cr
\hbox{ for }\mu^2\in\R&\hbox{and }N=0,1,\dots\cr
\end{eq}Feynman distributions are possible for any signature $\O(t,s)$
with  positive or negative invariant $\mu^2$.

Characteristic for and compatible only with the
orthochronous  Lorentz group
$\SO_0(1,s)$ are the  advanced (future) and retarded (past) causal    
ener\-gy-mo\-men\-tum distributions with
positive invariant $m^2$ only. They are distinguished by their energy $q_0$ behavior
 \begin{eq}{rl}
 \log((q\mp io)^2-m^2)
&=\log|q^2-m^2|\mp i\pi\ep(q_0)\vth(m^2-q^2)
\cr
 {\Ga(1+N)\over ((q\mp io)^2-m^2)^{1+N}}
&=-(-{\p\over\p q^2})^{1+N}\log((q\mp io)^2-m^2)\cr
= (-{\p\over\p q^2})^N{1\over (q\mp io)^2-m^2}
&={\Ga(1+N)\over (q_{\ro P}^2-m^2)^{1+N}}
\pm i\pi\ep(q_0)\de^{(N)}(m^2-q^2)\cr
\hbox{for }m^2\ge 0&\hbox{and }(q\mp io)^2=(q_0\mp io)^2- \rvec q^2
\end{eq}

\subsection{Re\-pre\-sen\-ta\-tion coefficients } 

In this subsection all
re\-pre\-sen\-ta\-tion matrix elements for noncompact operations 
are given\cite{VIL} which will be
relevant for the spacetime theory in the following.
They  can be obtained as Fourier transformed measures
and involve Bessel, Neumann and Macdonald functions.

The compact and noncompact selfdual projections of the exponential $\C\ni z\mape e^z$   
come with real and imaginary complex poles $q^2=\ep$  
\begin{eq}{rl}

\int{dq\over\pi}~{1\over q^2-io-\ep}e^{iqx}&=\left\{\begin{array}{ll}
ie^{i|x|},&\ep=+1\cr
e^{-|x|},&\ep=-1\cr\end{array}\right.
\end{eq}They define the basic  selfdual spherical state $\R\ni x\mape \cos x$
and the basic selfdual hyperbolic  state $\R\ni  x\mape e^{-| x|}$
 (more of that below).

The scalar   distributions for the definite orthogonal groups
in  general dimension
with real and imaginary singularities at $\rvec q^2=\pm 1$ 
give Bessel with  Neumann and Macdonald functions respectively -
wherever the $\Ga$-functions are defined
\begin{eq}{rl}
~~m\in\R:&\hskip10mm \int{dq\over2i\pi}~{\Ga(1-\nu)\over (q-io-m)^{1-\nu}}e^{iqx}=
\vth(x){e^{imx}\over (ix)^\nu}\cr\cr

\begin{array}{c}
\O(s),~s=1,2,3,\dots\cr
r=\sqrt{\rvec x^2},~~\nu\in\R\cr\end{array}
&
\left\{\begin{array}{rl}

\int {d^sq\over \pi^{{s\over2}-\nu}}
 {\Ga({s\over 2}-\nu)\over (\rvec q^2)^{{s\over2}-\nu}}e^{i\rvec q\rvec x}
&={\Ga(\nu)\over({r^2\over 2\pi})^\nu}\cr\cr
\int {d^sq\over \pi^{{s\over2}-\nu}}
 {\Ga({s\over 2}-\nu)\over (\rvec q^2-io-1)^{{s\over2}-\nu}}e^{i\rvec q\rvec x}
&={\pi(i\cl J_\nu-\cl N_\nu)(r)\over({r\over 2\pi})^\nu}
\cr
\cr
\int {d^sq\over \pi^{{s\over2}-\nu}}
 {\Ga({s\over 2}-\nu)\over (\rvec q^2+1)^{{s\over2}-\nu}}
 e^{i\rvec q\rvec x}
&=
{2\cl K_\nu(r)\over({r\over 2\pi})^\nu}=

e^{i\pi\nu}{\pi(i\cl J_\nu-\cl N_\nu)(ir)\over({ir\over 2\pi})^\nu}

\end{array}\right.
\cr

\end{eq}

All  (half)integer index functions arise by derivation
${d\over d{r^2\over 4\pi}}={2\pi\over r}{d\over dr} $,
called 2-sphere spread
\begin{eq}{rl}
\R_+\ni r&\mape {(\pi\cl J_\nu,~\pi\cl N_\nu,~
2\cl K_\nu)(r)\over({r\over 2\pi})^\nu}
=\left\{\begin{array}{l}
\(-{d\over d{r^2\over 4\pi}}\)^N
(\cos r,~\sin r,~e^{-r})\cr
\hskip15mm \nu+{1\over2}=N=0,1,2,\dots\cr
\(-{d\over d{r^2\over 4\pi }}\)^N
\(\pi\cl J_0(r),~\pi\cl N_0(r),~2\cl K_0(r)\)\cr
\hskip15mm \nu=N=0,1,2,\dots\cr\end{array}\right.
\end{eq}The half-integer index start from  the exponentials. 
The integer index Bessel functions begin with $\cl J_0$
which is used for 
scattering in the position plane $\SO(2)\sx\R^2$.
$\cl J_0$ integrates  position  translation states $x\mape \cos px$  
with the invariants on a circle $p=\cos\chi\in\Om^1$
\begin{eq}{l}
\pi\cl J_0(r)=\int d^2q~\de(\rvec q^2-1)e^{-i\rvec q\rvec x}
=\int_0^\pi d\chi~\cos (r\cos\chi)=
\pi{\SUM_{k=0}^\infty}{ (-{r^2\over4})^k\over(k!)^2}\cr
\end{eq}The integer index Neumann and Macdonald functions
start from free particles in rotation 
 free 2-dimensional spacetime $\SO_0(1,1)\sx\R^2$
and involve integrals of  
time translation states $t\mape \cos\om t$
and position  translation states $z\mape e^{-|Qz|}$
with the invariants on a hyperbola $(\om,Q)=\cosh\psi\in\cl Y^1$
\begin{eq}{rl}
\int d^2 q~\de(q^2-1)e^{iqx}&=
\int d\psi~[\vth( x^2)\cos (|x|\cosh\psi)
+\vth(- x^2)e^{-|x|\cosh\psi}]
\cr
&=-\vth( x^2)\pi\cl N_0(|x|)+\vth(- x^2)2\cl K_0(|x|)
\cr
{\scriptsize\pmatrix{
-\pi\cl N_0\cr
2\cl K_0\cr
}}(r)
&=-2{\SUM_{k=0}^\infty}{ (\mp{r^2\over4})^k\over (k!)^2}
[\log{r\over2}-\Ga'(1)-\phi(k)]\cr
\phi(0)&=0,~~\phi(k)=1+{1\over2}+\ldots +{1\over k},~~k=1,2,\dots \cr
-\Ga'(1)&=\lim_{k\to\infty}\brack{\phi(k)-\log k}=0.5772\ldots
\hbox{ (Euler's constant)}\cr
\end{eq}

The (half)integer index 
Bessel, Neumann and Macdonald functions
are relevant for (odd) even  dimensions and rank $R$
\begin{eq}{l}
\SO_0(t,s) \hbox{ with }
t+s=\left\{\begin{array}{llll}
\nu+{3\over2}&=1+N&=1+2R&=1,3,5,\dots\cr
\nu+2&=2+N&=2R&=2,4,6,\dots\cr
\end{array}\right.
\end{eq}For $R=1,2,\dots$ only the Bessel functions are regular at $r=0$. 
The characteristic  difference for even and odd dimension is seen in the
use  of the  rotation free case
with $\R$-re\-pre\-sen\-ta\-tions for $\nu=-{1\over2}$:
The halfinteger 
index   functions arise by derivation, i.e. 2-sphere spread,
with respect to the group parameter, starting with $\nu={1\over2}$, whereas the
integer  index  functions
start from $\nu=0$
and  involve a finite integration  $\ze\in[0,1]$ over $\R$-re\-pre\-sen\-ta\-tions
\begin{eq}{l}
\hbox{for }\nu=-{1\over2}:~\cos r
\begin{array}{rrll}
\nearrow&
{\sin r\over r}&={d\over d{r^2\over 2}}\cos r
&\hbox{for }\nu={1\over2}\cr
\searrow&\pi\cl J_0(r)&=\int_0^1{2d\ze\over\sqrt{1-\ze^2}}~\cos \ze r&\hbox{for }\nu=0\cr
\end{array}

\end{eq}

\section{Bound states of hyperbolic position}

Selfdual spherical $\SO(2)$-coefficients
of translation re\-pre\-sen\-ta\-tions  are
states in $L^\infty(\R)_+$
with  Dirac measures in $\cl M(\R)$
\begin{eq}{l}
\R\ni
t\mape\cos\om t=\int d q~|\om| \de(q^2-\om^2)e^{iqt}

\end{eq}Bound states and interactions 
are characterizable 
by selfdual 
hy\-per\-bo\-lic $\SO_0(1,1)$-coefficients which are square integrable states in 
$L^\infty(\R)_+\cap L^2(\R)$ and have 
a rational function as positive Radon measure
\begin{eq}{l}
\R\ni z
\mape e^{-|Qz|}=\int{dq\over \pi}{|Q|\over q^2+Q^2}e^{-iqz}
\end{eq}The spherical and hyperbolic invariants  
come from  a real and  `imaginary' momentum pair as poles in the
complex momentum plane, i.e. from $q=\pm\om$ and  $q=\pm i|Q|$ respectively. 
In contrast to the hyperbolic state, the spherical state is decomposable into a
direct sum.

The 1-dimensional quantum mechanical 
example is given by  the Schr\"odinger functions of the harmonic  oscillator.
They are  position re\-pre\-sen\-ta\-tion matrix  
elements  with the re\-pre\-sen\-ta\-tion invariant 
the inverse intrinsic length $Q^2={1\over \ell^4}=kM$.
Here the hyperbolic   state $z\mape e^{-|Q|r}$ with positive
definite  coordinate
shows up in  a repa\-ra\-me\-tri\-zation with the square of the usual position
parameter $r={x^2\over 2}$ 
\begin{eq}{l}
[\ell^2{p^2\over 2}+{x^2\over 2\ell^2}]\psi(x)={E\over \om}\psi(x)\then
\psi_0(x)=e^{-{x^2\over2\ell^2}}=e^{-|Q|r}
\hbox{ with } {2E_0\over \om}
=1

\end{eq}

In contrast to free states, 
bound states  for nonabelian groups 
use higher order
momentum poles, where the order depends on the 
position space dimension.
This will be exemplified by the nonrelativistic hydrogen atom
bound states which represent the noncompact nonabelian group 
$\SO_0(1,3)$ and start with  momentum dipoles.

\subsection{The Kepler factor and the Coulomb potential}

The bound state  solutions of the nonrelativistic quantum mechanical
hydrogen atom with the Coulomb potential
in the Hamiltonian $H={\rvec p^2\over 2}-{1\over r}$
(intrinsic units)
are characterized by integers for a rank 2 invariance group:
A principal quantum number $k=1,2,\dots$
and quantum numbers for 
angular momentum $L=0,1,\dots,k-1$ and its direction $L_3\le |L|$.
The compact group invariants determine the integer degree both  of the
spherical harmonics $\ro Y^L$ and of the Laguerre polynomials
$\ro L^N$ with the sum of the degrees
 $L+N=2J=k-1=0,1,\dots $ in the Schr\"odinger wave functions
\begin{eq}{l}
[{\rvec p^2\over 2}-{1\over r}]\psi(\rvec x)=E\psi(\rvec x),~~
\left\{\begin{array}{rl}
\rstate{k;L,L_3}\sim\psi(\rvec x)
&\sim r^L \ro Y_{L_3}^L(\phi,\th)
~ \ro L_{1+2L}^N({2r\over k})e^{-{r\over k}}\cr
&\sim(\rvec x)_{L_3}^L~ \ro L_{1+2L}^N({2r\over k})e^{-{r\over k}}\cr
\hbox{with }2E&=-{1\over k^2},~~k=L+N+1\end{array}\right.
\end{eq}It is more appropriate to combine the
spherical harmonics for the 2-sphere 
re\-pre\-sen\-ta\-tions 
with the matching radial power to 
the harmonic polynomials $(\rvec x)^L_{L_3}\sim r^L \ro Y_{L_3}^L(\phi,\th)$
with $\rvec\p^2(\rvec x)^L_{L_3}=0$,
 acted upon with irreducible rotation group $\SO(3)$-re\-pre\-sen\-ta\-tions.
The remaining factor 
represents the position 
 radial variable $\R_+\ni r\mape \ro L_{1+2L}^N({2r\over k})e^{-{r\over k}}$.

The separation of the harmonic polynomials in 
the general Schr\"odinger equation with rotation symmetry 
leaves equations for  the radial re\-pre\-sen\-ta\-tion coefficients 
$r\mape d_L(r)$
\begin{eq}{rl}
&[{\rvec p^2\over 2}+V(r)]\psi(\rvec x)=E \psi(\rvec x)\cr
 \psi(\rvec x)
 ={\SUM_{L=0}^\infty}~{\SUM_{m=-L}^L}
 (\rvec x)^L_m d_L(r)\then&
[{d^2\over dr^2}+{2(1+L)\over r}{d\over dr}
-2V(r)+2E_L]d_L(r)=0\cr
\end{eq}The condition to obtain 
for the position re\-pre\-sen\-ta\-tion 
the basic  selfdual  hyperbolic
state $r\mape e^{-|Q|r}$ as $L=0$ solution of the Schr\"odinger equation
determines the Kepler factor as  potential  
\begin{eq}{rl}
d_0(r)=e^{-|Q|r}
\hbox{ with}& [{d^2\over dr^2}+{2\over r}{d\over dr}
-2V(r)+2E_0]d_0(r)=0\cr
\then& V(r)=-{|Q|\over r}\hbox{ and } 2E_0=-Q^2
\end{eq}The free hyperbolic `wave' is the hydrogen ground state.
$Q^2$ is the position re\-pre\-sen\-ta\-tion characterizing invariant.

\subsection{The multipoles of the hydrogen atom}

The hyberbolic structure of a nonrelativistic dynamics with  
the Coulomb-Kepler potential 
${1\over r}$
 and the invariance of the Lenz-Runge 
`perihelion' vector has been exploited  quantum mechanically by Fock\cite{FOCK}.
With the additional rotation invariance the bound state vectors
come in irreducible 
$k^2$-dimensional re\-pre\-sen\-ta\-tions 
of the group $\SO(4)={\SU(2)\x\SU(2)\over \I(2)}$,
centrally correlating two $\SU(2)$'s, with the 
integer invariant $k=1+2J=1,2,\dots$.

The  measure of the 3-sphere as the manifold of 
the orientations of the rotation group 
$\SO(3)$ in the invariance group $\SO(4)$ has a momentum pa\-ra\-me\-tri\-zation  
by a dipole
\begin{eq}{l}
\Om^3\cong\SO(4)/\SO(3),~~
{1\over\sqrt{\rvec q^2+1}}{\scriptsize\pmatrix{1\cr i\rvec q\cr}}\in\Om^3\subnoteq
\R^4\then
|\Om^3|=\int d^3\om
=\int {2 d^3q\over (\rvec q^2+1)^2}=2\pi^2
\end{eq}$\Om^3$-integration of the 
pure translation states
$\R^3\ni\rvec x\mape e^{-i\rvec q\rvec x}\in\U(1)$,
i.e. the Fourier transformed $\Om^3$-measure, gives the
hydrogen ground state function as a scalar re\-pre\-sen\-ta\-tion coefficient
of 3-position space
\begin{eq}{l}
\cl Y^3\cong\SO_0(1,3)/\SO(3)\cong\SL(\C^2)/\SU(2)\ni\rvec x\mape
\int { d^3q\over \pi^2 }~{|Q|\over (\rvec q^2+Q^2)^2}e^{-i\rvec q\rvec x}=e^{-|Q|r}
\end{eq}In the bound states,  
3-position space is  
represented  as 3-hy\-per\-bo\-loid with a continuous invariant 
$Q^2$ for the imaginary `momenta' $\rvec q^2=-Q^2$ 
on a 2-sphere $\Om^2$ and a discrete rotation invariant $2J\in\N$.

The bound states are 
matrix elements of 
infinite dimensional cyclic principal $\SL(\C^2)$-re\-pre\-sen\-ta\-tions
where - with the Cartan subgroups $\SO(2)\x\SO_0(1,1)$ -
the irreducible ones are characterized
by one integer and one continous invariant   
\begin{eq}{l}
(2J,Q_\pm^2)\in\irrep\SL(\C^2)\cong \N\x\R_+\hbox{ (principal series)}
\end{eq}In the language of induced re\-pre\-sen\-ta\-tions, 
the bound states of the hydrogen atom  are  
rotation $\SO(3)$-intertwiners on the group $\SO_0(1,3)$
($\cl Y^3$-functions) with
values in Hilbert spaces with 
$\SO(3)$-re\-pre\-sen\-ta\-tions in
$(1+2J)^2$-di\-men\-sio\-nal $\SO(4)$-re\-pre\-sen\-ta\-tions.  

For the nonrelativistic hydrogen atom
bound states, the rotation dependence $\rvec x$ is 
effected by momentum derivation 
of the $\Om ^3$-measure
\begin{eq}{l}
\rvec x e^{-r}=
\int {d^3q\over \pi^2}
{4i\rvec q\over (1+\rvec q^2)^3}e^{-i\rvec q\rvec x}
\hbox{ with }{4\rvec q\over (1+\rvec q^2)^3}=-{\p\over \p\rvec q}
{1\over (1+\rvec q^2)^2}
\end{eq}The  3-vector  factor 
${2\rvec q\over1+\rvec q^2}$
is uniquely supplemented to a normalized 4-vector on the 3-sphere
- a pa\-ra\-me\-tri\-zation of the sphere
\begin{eq}{rl}
{1\over1+\rvec q^2}
{\scriptsize\pmatrix{
\rvec q^2-1\cr
2i\rvec q\cr}}=
{\scriptsize\pmatrix{\cos\chi \cr {\rvec q\over |\rvec q|}i\sin\chi
\cr}}={\scriptsize\pmatrix{p_0\cr i\rvec p\cr}}
\in\Om^3\subnoteq\R^4,~~p_0^2+\rvec p^2=1
\cr
\end{eq}The  normalized 4-vector 
$\ro Y^{(1,1)}(p)\sim
{\scriptsize\pmatrix{p_0\cr i\rvec p\cr}}\in\Om^3$
is the analogue  to the normalized  3-vector 
$\ro Y^1({\rvec q\over |\rvec q|})\sim {\rvec q\over |\rvec q|}\in\Om^2$ 
used for the  build-up of the  2-sphere harmonics 
$\ro Y^L({\rvec q\over |\rvec q|})\sim
({\rvec q\over |\rvec q|})^L$. 
Analoguously, the higher order $\Om^3$-harmonics
arise from the totally symmetric traceless
products 
$\ro Y^{(2J,2J)}(p)\sim(p)^{2J}$, e.g.
 the nine independent components  in the 
 $(4\x4)$-matrix 
\begin{eq}{rl}
\ro Y^{(2,2)}(p)
\sim (p)^2_{jk}&=p_jp_k-{\de_{jk}\over 4}
\cong {\scriptsize\left(\begin{array}{c|c}
{3p_0^2-\rvec p^2\over 4}&ip_0p_a\cr\hline
ip_0p_b&
p_ap_b-{\de_{ab}\over 4}\cr
\end{array}\right)}\cr
\hbox{with }p_ap_b-{\de_{ab}\over 4}
&=p_ap_b-{\de_{ab}\over 3}\rvec p^2-{\de_{ab}\over 3}{3p_0^2-\rvec p^2\over4}
\hbox{ for }p^2=1\cr
\end{eq}

The Kepler bound states in $(1+2J)^2$-multiplets
come with $2J$-dependent $\Om^3$-harmonics and multipoles 
\begin{eq}{l}
\cl Y^3\ni \rvec x\mape
\int {d^3q\over\pi^2}
{1\over (1+\rvec q^2)^2}
(p)^{2J}e^{-i\rvec qQ\rvec x}\hbox{ with }\left\{\begin{array}{rl}
 p&=
{1\over1+\rvec q^2}
{\scriptsize\pmatrix{
\rvec q^2-1\cr
2i\rvec q\cr}}\cr

Q&={1\over 1+2J}\cr
\end{array}\right.
\end{eq}As illustration 
the $k=2$ bound state quartet
with tripole vector 
\begin{eq}{rl}
Q={1\over 2}:~~
\int {d^3q\over\pi^2}
{1\over(1+\rvec q^2)^2}
{\scriptsize\pmatrix{
p_0\cr
i\rvec p\cr}}e^{-i\rvec qQ\rvec x}
&=\int {d^3q\over\pi^2}
{1\over(1+\rvec q^2)^3}
{\scriptsize\pmatrix{
\rvec q^2-1\cr
2i\rvec q\cr}}e^{-i\rvec qQ\rvec x}\cr
={\scriptsize\pmatrix{
{1-Qr\over2}\cr
{Q\rvec x\over2}\cr}}e^{-Qr}&= {\scriptsize\pmatrix{
\hfill{1\over 4}\ro L^1_1(2Qr)\cr
\hfill {Q\rvec x\over2}~\ro L^0_2(2Qr)\cr}}
\end{eq}and the $k=3$ bound state nonet
 with quadrupole tensor measure
\begin{eq}{rl}
Q={1\over 3}:~~
\int {d^3q\over \pi^2}
{1 \over (1+\rvec q^2)^2}
{\scriptsize\pmatrix{
3p_0^2-\rvec p^2\cr
 ip_0\rvec p\cr
3\rvec  p\ox\rvec p-\bl 1_3\rvec p^2 \cr}}e^{-i\rvec qQ\rvec x}
&=\int {d^3q\over \pi^2}
{4 \over (1+\rvec q^2)^4} 
{\scriptsize\pmatrix{
3({\rvec q^2-1\over 2 })^2-\rvec q^2\cr
i\rvec q~{\rvec q^2-1\over 2 }\cr
3\rvec q\ox\rvec q-\bl 1_3\rvec q^2
\cr}}e^{-i\rvec qQ\rvec x}

\cr
=
{\scriptsize\pmatrix{
1-2Qr+{2Q^2r^2\over 3}\cr
{2-Qr\over 3}{Q\rvec x\over2}\cr
{Q^2\over2}(\bl 1_3{r^2\over 3}-\rvec x\ox\rvec x)
}}e^{-Qr}
&={\scriptsize\pmatrix{
\hfill{1\over 3}\ro L^2_1(2Qr)\cr
\hfill {Q\rvec x\over2}~{1\over 6}\ro L^1_3(2Qr)\cr
{Q^2\over2}(\bl 1_3{r^2\over 3}-\rvec x\ox\rvec x)\ro L^0_5(2Qr)\cr}}

\end{eq}

\subsection{Yukawa potentials as tangent coefficients}

States for free translation re\-pre\-sen\-ta\-tions 
obey homogeneous equations $(\p^2+P^2)d=0$ with 1st order selfdual
derivatives. Bound states for a nonabelian action group where
the dimension is strictly larger than the rank have higher order 
poles - e.g. 2nd order
for hyperbolic position space $\cl Y^3$.
The related lower  order poles do not characterize
group re\-pre\-sen\-ta\-tion
coefficients, but re\-pre\-sen\-ta\-tions of tangent translations  for the
nonabelian structure.
For  hyperbolic position $\cl Y^3$ the
tangent translation re\-pre\-sen\-ta\-tions are  Yukawa potentials
with 1st order poles.
For nontrivial invariant, they are 
related to Macdonald functions with halfinteger index
which arise by 2-sphere spread from the $\SO_0(1,3)$-state 
\begin{eq}{rl}
(\rvec\p^2-Q^2)e^{-|Q|r}&=-2|Q|{e^{-|Q|r}\over r}\cr
\R^3\ni\rvec x\mape 2|Q|{e^{-|Q|r}\over r}&=-{\p\over \p {r^2\over
4}}e^{-|Q|r}=\int {d^3q\over \pi^2}~{|Q|\over
\rvec q^2+Q^2} e^{-i\rvec q\rvec x}\cr
&\hbox{ interaction coefficients for $\cl Y^3$}

\end{eq}

More on the general  structure of interactions 
as Lie algebra re\-pre\-sen\-ta\-tion coefficients will be given below.

\section{Residual re\-pre\-sen\-ta\-tions}

The method of 
residual re\-pre\-sen\-ta\-tions
with (energy-)momentum distributions is intended to 
generalize, especially to nonabelian noncompact operations,  
the cyclic Hilbert re\-pre\-sen\-ta\-tions of translations  
via positive (energy-)momentum measures. It uses 
the injection of the Fourier transformed Radon measures
into the essentially bounded function classes
\begin{eq}{l}
\int d^nq~e^{iqx}\cl M(\R^n)\subnoteq L^\infty(\R^n),~~
\int d^n q~ \tilde d(q)e^{iqx}=d(x)\cr
\end{eq}The form of residual re\-pre\-sen\-ta\-tions 
shows a generalization and modification
of the Feynman propagators as used in canonical quantum field theory.

The goal of the residual re\-pre\-sen\-ta\-tion
method  is  to translate
the relevant re\-pre\-sen\-ta\-tion structures 
of homogeneous spaces (real Lie groups) and its tangent translations
(Lie algebras, more on that below) - 
invariants, normalizations, product
re\-pre\-sen\-ta\-tions etc. - into the language of rational complex
(energy-)momentum  functions
with its poles, its residues
and convolution products.

Semisimple and reductive
Lie groups 
have  factorizations $G=KP$ into maximal compact group
$K$ and parabolic subgroups\cite{KNAPP} $P=M A N$
with noncompact Cartan subgroup $A$, its centralizer 
$MA$ and nilpotent $\log N$. The subgroup $MA$ is
similar to the fixgroup defined direct product subgroups 
in affine groups $H_0\x\R^n\sub H\sx\R^n$ where induced re\-pre\-sen\-ta\-tions are used for
free scattering and particle states. 
It remains to be seen if parabolically induced
re\-pre\-sen\-ta\-tions of $G$ can be connected with
the residual re\-pre\-sen\-ta\-tions considered
in the following.

\subsection{Residual re\-pre\-sen\-ta\-tions of symmetric spaces}

Harmonic analysis of a sym\-met\-ric space $G/H$ 
with real Lie groups $G\sup H$ 
 analyzes complex
$G/H$-mappings
with respect to  irreducible $G$-re\-pre\-sen\-ta\-tions 
with the related invariants.
The eigenvalues (weights) of the group $G$-re\-pre\-sen\-ta\-tions are
a subset of the
linear Lie algebra forms  $(\log G)^T$.
For translations all linear forms are weights, the (energy-)momenta. 
For simple groups, the weights 
constitute a subset of the weight space $W^T$
(linear forms of a Cartan Lie algebra $W$)
with the dimension
the Lie algebra rank, $\dim_\R W^T=\rank_\R \log G$. 
The weights are discrete for a compact group.
The Lie algebra  
is acted upon with  the adjoint  re\-pre\-sen\-ta\-tion of the
group in the affine group $G\sx\log G$, its forms with the  coadjoint (dual) one. 
Invariants are multilinear
Lie algebra  forms, e.g. linear for abelian groups or the bilinear 
Killing form for semisimple groups.

The tangent spaces of $G/H$
are isomorphic to the corresponding Lie algebra classes,
denoted by
$\log G/H=\log G/\log H$
with $\dim_\R\log G/H=\dim_\R G-\dim_\R H$. It inherits
the adjoint action
of the group $G$, the linear forms the coadjoint one.

Now, the definition of residual re\-pre\-sen\-ta\-tions:
Functions on (re\-pre\-sen\-ta\-tion coefficients of)
a  symmetric space $G/H$, especially $d\in L^\infty(G)$
\begin{eq}{l}
d:(G/H)_{\rm repr}\map \C,~~x\mape d( x)\cr
\end{eq}are assumed to be pa\-ra\-me\-tri\-zable by
vectors $x\in V$ 
(translations) of an orbit in a  real vector space
with fixgroup $H$  
\begin{eq}{l}
 x\in G\m x_0\cong G/H, ~~G\m x_0\sub V\cong \R^n 
\end{eq}e.g., a group $G$ by
 its Lie algebra $\log G$ (canonical coordinates) 
like $\SU(2)\cong\{e^{i\rvec\si\rvec x}\mid
 \rvec x\in\R^3\}$ 
 or the   hyperboloid 
 $\SO_0(1,3)/\SO(3)\cong\{x\in\R^4\mid  x^2=\ell^2>0,~x_0>0\}$ 
by the vectors of a timelike orbit.
With the dual  space  
$q\in V^T\cong\R^n$ 
(by abuse of language called (energy-)momenta, also in the general case),
e.g. the dual Lie algebra,
the  re\-pre\-sen\-ta\-tions of $G/H$  
are characterizable by $G$-invariants $\{I_1,\dots ,I_R\}$,
with  rational values    for a compact and 
rational or continuous values  for
a noncompact group.
The invariants are given 
by $q$-polynomials
and can be built by multilinear invariants - $q=m$ 
for an abelian group,  quadratic 
 invariants $q^2=\pm m^2$, e.g. Killing form invariants, etc.

If there exists a distribution of the (energy-)momenta,
especially $\tilde d\in\cl M(\R^n)$, e.g. for positive type functions,
whose Fourier transformation 
gives the functions $d$  on the symmetric space
and if the generalized function
$\tilde d$ comes as  quotient of two polynomials 
where the invariant zeros of the denominator polynomial $Q(q)$ 
characterize a $G$-re\-pre\-sen\-ta\-tion 
 \begin{eq}{l}
\tilde d(q)\cong {P(q)\over Q(q)}\hbox{ with $Q(q)$-factors }\{(q-m)^n,
(q^2\pm m^2)^n,(q^k\pm m^k)^n\},~m\in\R
\end{eq}then $d$  is called a 
{residual re\-pre\-sen\-ta\-tion of 
$G/H$} and the
complex rational function $q\mape \tilde d(q)$
a { residual re\-pre\-sen\-ta\-tion  function (distribution)}.

A re\-pre\-sen\-ta\-tion of a symmetric space $G/H$
contains 
re\-pre\-sen\-ta\-tions of subspaces $K$, e.g. of abelian subgroups
$\SO(2)\subnoteq\SU(2)$
or  $\SO_0(1,1)\subnoteq\SL(\C^2)/\SU(2)$.
A residual $G/H$-re\-pre\-sen\-ta\-tion 
with canonical 
tangent space  parameters $x=(x_K,x_\perp)$ 
has a projection  to a
residual $K$-re\-pre\-sen\-ta\-tion 
by integration $\int d^{n-s} x_\perp$ 
over the complementary  space
 ${\log G/H\over\log K}\cong\R^{n-s}$ 
 
 \begin{eq}{l}
K\map\C,~~ x_K\mape  d(x_K,0)

=\int {d^{n-s} x_\perp\over(2\pi)^{n-s}} d(x)
=\int d^s q_K ~\tilde d(q_K,0) e^{i  q_Kx_K }
\end{eq}The integration picks up the Fourier components
for trivial tangent space forms (energy-momenta) $q_\perp=0$ of 
${\log G/H\over\log K}$.

A Fourier integral involves irreducible re\-pre\-sen\-ta\-tions
$x\mape e^{iqx}$ of the underlying translations $x\in V$. 
With that, residual re\-pre\-sen\-ta\-tions with positive
distributions of the (energy-)momenta (characters)
 $q\in V^T$ are cyclic translation
re\-pre\-sen\-ta\-tion coefficients.

With velocities and actions measured in units $(c,\hbar)$
all energy and momentum invariants can be measured in mass units.
A mass unit does not imply a translation invariant. 
Nontrivial invariants $m\ne0$ can be used as intrinsic units by 
a rescaling of translations  
 $x\mape {x\over |m|}$ and  (energy-)momenta $q\mape |m|q$ to obtain 
dimensionless Lie parameters and
eigenvalues. To include the trivial case $m=0$, 
invariants will be kept in most cases -
and somewhat inconsequentially - in the
di\-men\-sio\-nal form. 
 
\subsection
{Rational complex re\-pre\-sen\-ta\-tion functions}

The simplest case of residual re\-pre\-sen\-ta\-tions
is realized  for  time  
and 1-di\-men\-sio\-nal position
with  energy  and
  momentum distributions respectively.
  The re\-pre\-sen\-ta\-tions yield - for linear invariant -
 matrix elements of the real 1-di\-men\-sio\-nal
compact and  noncompact group
 $\U(1)=\exp i\R$ and  $\D(1)=\exp \R$ respectively  and  - for dual invariants -
 of their selfdual spherical und hyperbolic doublings $\SO(2)$ and $\SO_0(1,1)$ 
 respectively.

An irreducible $\R$-re\-pre\-sen\-ta\-tion
is the  residue of a rational complex energy function
or, equivalently, a Fourier transformed  Dirac  distribution supported by 
the linear invariant energy $m\in\R$  
\begin{eq}{rl}
\R\ni t&\mape  e^{ imt}
=\oint ~{dq\over2 i\pi}{1\over q-m}e^{ iqt}=\int dq~\de(q-m)e^{ iqt}
\in\U(1)\cr
\end{eq}This 
gives the prototype of a   residual re\-pre\-sen\-ta\-tion.
The integral $\oint$ circles the singularity
in the  mathematically positive direction.

For the abelian group $\D(1)\cong \R$, where the dimension coincides with the rank and
where the eigenvalues $q$ are the group invariants $m$, the transition
to the residual form is a trivial transcription to the singularity $q=m$.
This will be different for 
nonabelian groups with dimension strictly larger than rank, e.g. for 
the rotations $\SO(3)$, with dimension 3 and rank 1, 
with the invariant  a square $\rvec q^2=m^2$ of the three
$\R^3$-eigenvalues $\rvec q$.

In the  Fourier transformations
of the  future and past distributions  
the real-imaginary  decomposition 
into Dirac  and  principal value  distributions
goes with  the order function
decomposition $\vth(\pm t)={1\pm\ep(t)\over2}$ in the 
functions on future $\R_+$ and past $\R_-$ 
\begin{eq}{l}
\hbox{causal: }\R_\pm\ni \vth(\pm t)t\mape
\pm 
\int {dq\over 2i\pi} {1 \over  q \mp io -m}e^{iqt }
=\vth(\pm t)   e^{imt}\cr
\end{eq}

All those distributions originate  from
the  same re\-pre\-sen\-ta\-tion functions with one  pole
in the compactified complex plane
\begin{eq}{l}
\C\ni q\mape  {1  
\over  q-m}\in\ol\C,~~m\in\R
\end{eq}The position $q=m$
of the singularity is related to the
continuous { invariant}.
The Fourier transforms with 
different contours  around the pole
represent via $\vth(\pm t)$ the causal structure of the reals. 

A re\-pre\-sen\-ta\-tion distribution with nontrivial residue can be normalized
\begin{eq}{l}
\R\ni 0\mape  1
=\oint ~{dq\over2 i\pi}{1\over q-m}
=\res_ {m} 
{1\over  q-m}=\sprod mm
\end{eq}The residual normalization gives, simultaneously, 
both the normalization of the unit $t=0$ re\-pre\-sen\-ta\-tion 
$t\mape e^{imt}$ (pure state)  and
the scalar product  
of the normalized eigenvector (pure cyclic vector) 
$\rstate m$.

\subsection
{Compact and noncompact dual invariants}

Poles at  dual compact  re\-pre\-sen\-ta\-tion invariants
$q^2=m^2$  can be combined from 
linear poles at  $q=\pm |m|$, the invariants for the dual irreducible
subre\-pre\-sen\-ta\-tions. 

The Fourier transforms
of the  causal 
 and
 (anti-)Feynman  energy-distributions
are functions on the cones, the bicone 
and the group with $\SO(2)$ matrix elements
\begin{eq}{rrll}
\hbox{causal:}&\R_\pm\ni\vth(\pm t)t \mape&
 \pm \int {d q\over i\pi}~{{\scriptsize\pmatrix{  
 q\cr |m| \cr}}\over (q\mp io)^2-m^2}
 e^{iqt}&=
 \vth(\pm t)2{\scriptsize\pmatrix{  \cos mt\cr i\sin |m|t \cr}}\cr
\hbox{bicone:}&\R_+\uplus \R_-\ni t\mape&
 \pm\int {d q\over i\pi} ~{{\scriptsize\pmatrix{ |m|\cr q\cr }}
\over q^2\mp io-m^2} e^{iqt}&={\scriptsize\pmatrix{1\cr\pm\ep(t)\cr }}
 e^{\pm i |mt|}\cr
\hbox{group:}&\R\ni t\mape& 
 \int d q~
{\scriptsize\pmatrix{|m|\cr q\cr }}\de(q^2-m^2)e^{iqt}
&={\scriptsize\pmatrix{\cos mt\cr i\sin |m|t \cr  }}\cr
\end{eq}

The normalization  for $t=0$
uses different  matrix elements for the 
  causal residues with two poles with equal imaginary part  
and for the  Feynman 
residues with two poles 
with opposite imaginary part

\begin{eq}{rlll}

\hbox{causal:}&\C\ni q\mape {q\over q^2-m^2}\in\ol\C,&
\sum\oint_{\pm |m|}
{dq\over i\pi}{q\over q^2-m^2}&=\sum\res_ {\pm |m|}{2q\over q^2-m^2}=2\cr
\hbox{Feynman:}
&\C\ni q\mape {|m|\over q^2-m^2}\in\ol\C,&
\pm\oint_{\pm |m|}
{dq\over i\pi}{|m|\over q^2-m^2}&=\res_ {\pm |m|}{\pm 2|m|\over q^2-m^2}=1\cr
\end{eq}

The  functions with  noncompact dual  re\-pre\-sen\-ta\-tion invariants
$q^2=-m^2$ 
give, as Fourier transformed $\Om^1$-measure, 
 noncompact  
matrix elements of faithful cyclic $\D(1)$-re\-pre\-sen\-ta\-tions, not irreducible 
\begin{eq}{rl}
\SO_0(1,1)\cong\R\ni x&\mape
\int {d q\over\pi}~{|m|\over q^2+m^2} e^{iqx}\cr
&=
\oint {d q\over2i\pi}[{\vth(-x)\over q-i|m|} 
-{\vth(x)\over q+i|m|}]e^{iqx}
=e^{-|  mx|}\cr

\end{eq}The re\-pre\-sen\-ta\-tion relevant
residues are taken at  imaginary `momenta' $q=\pm i|m|$ in the complex
momentum plane
\begin{eq}{l}
\C\ni q\mape {|m|\over q^2+m^2}
\in\ol\C,~~
\oint_{\pm i|m|}{dq\over \pi}{|m|\over q^2+m^2}=
\res_{\pm i|m|}{ \pm 2i|m|\over q^2+m^2} =1

\end{eq}

\subsection{Residual re\-pre\-sen\-ta\-tions of hyperbolic positions}

Distributions of $s$-di\-men\-sio\-nal momenta $\rvec q\in\R^s$ 
with the action of the rotation group $\SO(s)$ 
are used for re\-pre\-sen\-ta\-tions\cite{STRICH,SHER}
  of the
hyperboloids $\cl Y^s$ and spheres $\Om^s$.
For $s=1$ `flat' and `hyberbolic' are isomorphic.
The  residual re\-pre\-sen\-ta\-tions of nonabelian noncompact 
hyperboloids and compact spheres  
with $s\ge2$ 
have to  embed
the  nontrivial re\-pre\-sen\-ta\-tions of the abelian groups 
with continuous  and  integer  dual  invariants respectively
\begin{eq}{rlll}
\SO_0(1,1)\cong\cl Y^1\ni  x&\mape \int{dq\over\pi}~
{|m|\over q^2+m^2}e^{iqx}&=e^{-|mx|},&m^2>0\cr
\SO(2)\cong \Om^1\ni e^{ ix}&\mape \pm\int{dq\over i\pi}~{|m|\over q^2\mp
io-m^2}e^{iqx}
&=e^{\pm i |mx|},&|m|=1,2,\dots\cr
\end{eq}The  pole invariants
 $\{\pm i|m|\}$ and $\{\pm |m|\}$ on the discrete sphere $\Om^0=\{\pm 1\}$
are embedded, for the nonabelian case,
into singularity spheres $\Om^{s-1}$ which arise in the Cartan factorization
\begin{eq}{rrlrr}
\SO_0(1,s)/\SO(s)&\cong\cl Y^{s}&\cong&\SO_0(1,1)\o\Om^{s-1}\cr
\SO(1+s)/\SO(s)&\cong\Om^{s}&\cong&\SO(2)\o\Om^{s-1}\cr
\end{eq}

The rank of the orthogonal groups
gives the real (noncompact) rank 1 for the 
odd dimensional hyperboloids, i.e. one continuous noncompact invariant 
\begin{eq}{rl}
\rank_\R\SO_0(t,s)=R\hbox{ for }t+s=2R\hbox{ and }t+s&=2R+1\cr
\rank_\R\SO_0(1,2R-1)-\rank_\R\SO(2R-1)&=1\cr
\rank_\R\SO_0(1,2R)-\rank_\R\SO(2R)&=0\cr
\end{eq}Odd dimensional hyperboloids and spheres, $\cl Y^s$
and $\Om^s$ with  $s={2R-1}$,
will be considered as generalization of the minimal and characteristic 
nonabelian case
$s=3$ with nontrivial rotations for the nonrelativistic
hydrogen atom above.

The coefficients of residual re\-pre\-sen\-ta\-tions of   
hyperboloids $\cl Y^{2R-1}$
 use
the Fourier transformed  measure of the momentum sphere $\Om^{2R-1}$ 
with singularity   sphere $\Om^{2R-2}$ 
for imaginary `momenta' with continuous noncompact
invariant  $\rvec q^2=-m^2<0$.
$\SO(2R)$-multiplets  arise via the sphere pa\-ra\-me\-tri\-zation  
${1\over \rvec q^2+m^2}
{\scriptsize\left(\begin{array}{c}
\rvec q^2-m^2\cr
2i|m|\rvec q\cr
\end{array}\right)}\in\Om^{2R-1}\subnoteq \R^{2R}$
\begin{eq}{l}
\hbox{for }\cl Y^{2R-1},~
R=1,2,\dots\hbox{ with }{2\over |\Om^{2R-1}|}={\Ga(R)\over\pi^R}\cr
\rvec x\mape
\left\{\begin{array}{ll}
 \int
{2 d^{2R-1}q\over |\Om^{2R-1}|}
{|m|\over  (\rvec q^2+m^2)^R}
e^{-i\rvec q\rvec x}&=
e^{-|m|r } \cr
 \int
{2 d^{2R-1}q\over |\Om^{2R-1}|}
{R|m|\over  (\rvec q^2+m^2)^{1+R}}
{\scriptsize\left(\begin{array}{c}
\rvec q^2-m^2\cr
2i|m|\rvec q\cr
\end{array}\right)}
e^{-i\rvec q\rvec x}&=
{\scriptsize\left(\begin{array}{c}
R-1-|m|r\cr \rvec x\cr
\end{array}\right)}
e^{-|m|r } \cr

\end{array}\right.\cr

\end{eq}

Each state  $\{\rvec x\mape e^{-|m|r}\}\in L^\infty(\SO_0(1,2R-1))_+$ 
with $m^2>0$
characterizes an infinite dimensional Hilbert space 
with a faithful cyclic re\-pre\-sen\-ta\-tion of $\SO_0(1,2R-1)$
as familiar for $R=2$ from the 
principal series re\-pre\-sen\-ta\-tions of the Lorentz group $\SO_0(1,3)$.
The positive type function defines the  Hilbert product
\begin{eq}{rl}
\hbox{distributive basis:}&
\{\rstate{m^2;\rvec q}\mid \rvec q\in\R^{2R-1}\}\cr
\hbox{scalar product distribution:}&
\sprod{m^2;\rvec q'}{m^2;\rvec q}
={|m|\over (\rvec q^2+m^2)^R} {|\Om^{2R-1}|\over2}\de(\rvec q-\rvec q')\cr
\hbox{Hilbert vectors:}&
\rstate{m^2;f}=\int {2d^{2R-1}q\over |\Om^{2R-1}|} ~f(\rvec q)\rstate{m^2;\rvec q}\cr 
&\sprod{m^2;f}{m^2;f'}
=\int {2d^{2R-1}q\over |\Om^{2R-1}|} ~\ol{f(\rvec q)}~
{|m|\over (\rvec q^2+m^2)^R} f'(\rvec q)\cr
\end{eq}There is a  re\-pre\-sen\-ta\-tion of   
each abelian noncompact subgroup 
in the Cartan decomposition $ \cl Y^{2R-1}\cong \SO_0(1,1)\o\Om^{2R-2}$
with the action on a distributive basis and, therewith, on the Hilbert vectors
\begin{eq}{l}
\begin{array}{rl}
\hbox{$\SO_0(1,1)$-re\-pre\-sen\-ta\-tions for all }\rvec\om\in\Om^{2R-2}:
e^{-\rvec \om\rvec x}&\mape 
e^{-i|\rvec q|\rvec\om\rvec x}=e^{-i\rvec q\rvec x}\in\U(1)\cr
\hbox{action of all }\SO_0(1,1):~
\rstate{m^2;\rvec q}&\mape e^{-i\rvec q\rvec x}
\rstate{m^2;\rvec  q}\end{array}\cr
\begin{array}{rl}
\hbox{cyclic vector:}&
\rstate{m^2;1}=\int{2 d^{2R-1}q\over |\Om^{2R-1}|}~ \rstate{m^2;\rvec q}\cr
\hbox{with}&\int{4 d^{2R-1}q~ d^{2R-1}q'\over |\Om^{2R-1}|^2}~ 
\lstate{m^2;\rvec q'}e^{-i\rvec q\rvec x}
\rstate {m^2;\rvec q}=e^{-|m|r}\end{array} \cr
\end{eq}

The scalar product is written with the positive type function, e.g. for 
3-dimensional position $R=2$  with intrinsic unit
\begin{eq}{rl}
\sprod{f}{f'}
=\int {d^3q\over \pi^2} ~\ol{f(\rvec q)}~
{1\over (\rvec q^2+1)^2} f'(\rvec q)
&=\int d^3x_1d^2x_2 ~\ol{\tilde f(\rvec x_2)}~
e^{-|\rvec x_1-\rvec x_2|} \tilde f'(\rvec x_1)\cr
\hbox{with }
f(\rvec q)&=\int d^3x~\tilde f(\rvec x) e^{i\rvec q\rvec x}
\end{eq}It can be brought in the form of square integrability $L^2(\R^3)$
by absorption of the square root 
\begin{eq}{l}
\psi(\rvec q)={\sqrt{8\pi}\over \rvec q^2+1} f(\rvec q)
\then
\sprod{f}{f'}
=\int {d^3q\over (2\pi)^3} ~\ol{\psi(\rvec q)}~
\psi'(\rvec q)
=\int d^3x~\ol{\tilde\psi(\rvec x)}~
\tilde \psi'(\rvec x)\cr
\end{eq}

Therefore, all infinite dimensional Hilbert spaces for 
different  continuous invariants $m^2>0$
are subspaces of  one Hilbert space
$L^2(\cl Y^{2R-1})$ with $\cl Y^{2R-1}\cong\R^{2R-1}$. 
States with different invariants are not orthogonal, i.e.,
they have a nontrivial  Schur scalar product\cite{SCHUR}
\begin{eq}{l}
\rsprod{d^{m^2_1}}{d^{m^2_2}}=
\int d^{2R-1}x~ 
e^{-|m_1|r}e^{-|m_2|r}={ |\Om^{2R-1}|\over 2\pi}
\({2\over
|m_1|+|m_2|}\)^{2R-1}
\end{eq}The orthogonality of the $\cl Y^3$-re\-pre\-sen\-ta\-tion coefficients
with different invariant $m^2={1\over (1+2J)^2}$ in the hydrogen atom
is a consequence of the different rotation invariants $J$.

The corresponding {matrix elements of re\-pre\-sen\-ta\-tions of odd dimensional
spheres} are obtained by real-imaginary transition
from noncompact to compact operations $\SO_0(1,1)\to \SO(2)$. They involve  
Feynman distributions with supporting 
singularity  sphere $\Om^{2R-2}$  
for real momenta with integer compact invariant $\rvec q^2=m^2$,
$|m|=1,2,\dots$
\begin{eq}{l}
\begin{array}{l}
\hbox{for }\Om^{2R-1}:\cr
R=1,2,\dots\end{array}~~\rvec x\mape
\left\{\begin{array}{rl}
\pm \int {2d^{2R-1}q\over i|\Om^{2R-1}|}{|m|\over (\rvec q^2\mp io-m^2)^{R}}
e^{-i\rvec q\rvec x}&=e^{\pm i|m|r}\cr
 \int {d^{2R-1}q\over \pi^{R-1} }|m|\de^{(R-1)}(m^2-\rvec q^2)
e^{-i\rvec q\rvec x}&=\cos |m| r\cr
\end{array}\right.
\end{eq}The irreducible re\-pre\-sen\-ta\-tion spaces are finite
dimensional, e.g. for $R=2$ isomorphic to $\C^{1+2L}$. The irreducible spaces
for different discrete invariants, e.g.  
$L=0,1,\dots$, are Schur-orthogonal subspaces\cite{PWEYL}
of the infinite dimensional Hilbert space $L^2(\Om^{2R-1})$.

The residual normalization for   complex re\-pre\-sen\-ta\-tion
functions
 \begin{eq}{l}
\R\x\Om^{2R-1}\inmap \C\x\Om^{2R-1}\ni 
\rvec q=|\rvec q|~{\rvec q\over |\rvec q|}\mape 
{\mu\over(\rvec q^2+\mu^2)^{R}}\in\ol \C ,~~\mu\in\C
\end{eq}has to take into account the sphere degrees of freedom
in $\C\x\Om^{2R-1}$, e.g. for $\cl Y^{2R-1}$
\begin{eq}{l}
\res_{\pm i|m|}{2|m|\over(\rvec q^2+m^2)^R}=
\int {2d^{2R-1}q\over |\Om^{2R-1}|}~{|m|\over (\rvec q^2+m^2)^R}
=\oint_{\pm i|m|}{d q\over \pi}{ |m| (q^2)^{R-1}\over (q^2+m^2)^R}=1
\end{eq}The higher order $q^2$-power is compensated with
the $q^2$-power of the measure.
Nonscalar functions  have trivial  residue.

The tangent translations for the nonabelian 
Lie algebras $\log\SO(1,2R-1)$ for the hyperboloids 
and $\log\SO(2R)$ for the spheres  are represented by
Yukawa potentials and spherical waves
(halfinteger index Macdonald and Hankel functions respectively)
which arise by 2-sphere spread of the states
\begin{eq}{rlrl}
R=2,3,\dots,\hbox{ for }\cl Y^{2R-1}:&\rvec x\mape&
\int {2d^{2R-1}q\over |\Om^{2R-1}|}~~{1\over (\rvec q^2+m^2)^{R-1}}
e^{-i\rvec q\rvec x}
&=2{e^{-|m|r} \over r}\cr
\hbox{ for }\Om^{2R-1}:&\rvec x\mape&
\pm\int {2d^{2R-1}q\over i|\Om^{2R-1}|}~{1\over (\rvec q^2\mp io -m^2)^{R-1}}
e^{-i\rvec q\rvec x}&=2{e^{\pm i|m|r}\over r}
\cr
\end{eq}

\section
{Residual re\-pre\-sen\-ta\-tions of  spacetime}

The re\-pre\-sen\-ta\-tions of the time translations $\R$
and of the hyperboloid $\cl Y^3$ as  model of position space
as seen in the nonrelativistic hydrogen bound states
can be brought together in re\-pre\-sen\-ta\-tions of
homogeneous  spacetime models
whose matrix elements will be given in a residual formulation.

\subsection
{Homogeneous causal spacetimes}

The spacetime translations in the Poincar\'e group
$\SO_0(1,s)\sx\R^{1+s}$, $s\ge1$, 
can be decomposed into  Lorentz group orbits, i.e. symmetric spaces with
characteristic fixgroups
\begin{eq}{lll}
\hbox{trivial}:&\SO_0(1,s)/\SO_0(1,s)&\cong\{0\}
\cr
\hbox{timelike}:&\SO_0(1,s)/\SO(s)&\cong\cl Y^s\cr
\hbox{spacelike}:&\SO_0(1,s)/\SO_0(1,s-1)&\cong\cl Y^{(1,s-1)}
\cr
\hbox{lightlike}:&
\SO_0(1,1)/\{1\}&\cong\R\cr
&\SO_0(1,s)/\SO(s-1)\sx\R^{s-1}&\cong\ro V^s,~s\ge2\cr
\end{eq}$\ro V^s\cong\R\x\Om^{s-1}$  is the tipless foward lightcone.
Symmetric spaces 
with the same dimension $(1+s)$ as the translations
and  with isomorphic fixgroup for all elements
are all future or all past timelike and all spacelike translations.
For $s\ge2$, only the  timelike ones are causally ordered.
Open future $\R^{1+s}_+$ will be taken as the causal homogeneous model for 
 spacetime  with the dilation Poincar\'e group as tangent 
structure
\begin{eq}{rl}
\cl D^{1+s}=&\R^{1+s}_+=\{x\in\R^{1+s}\mid x^2>0,~x_0>0\}\cr
\cong&\D(1)\x\cl Y^s\cr
\hbox{tangent}&\log\cl D^{1+s}\cong\R^{1+s},~~[\D(1)\x\SO_0(1,s)]\sx\R^{1+s}\cr
\end{eq}The fixgroup $\SO(s)$ is  maximal compact in
the reductive Lie group  $\D(1)\x \SO_0(1,s)$.

The Fourier transformations of the advanced 
and retarded causal measures
 are supported by future and past respectively. 
 They involve the off-shell principal
 value  part    
\begin{eq}{rl}
\SO_0(1,s):~~\int {d^{1+s}q\over \pi} {1\over q_\ro P^2-m^2}e^{iqx}&
=i\ep(x_0)
\int d^{1+s}q~ \ep(q_0)\de(m^2-q^2)e^{iqx}\cr
\int{d^{1+s}q\over \pi}{1\over (q\mp io)^2-m^2}e^{iqx}&=
\pm 2i\vth(\pm x_0) \int d^{1+s}q~\ep(q_0)\de(m^2-q^2)e^{iqx}\cr
&=\vth(\pm x_0)2\int {d^{1+s}q\over \pi} {1\over q_\ro P^2-m^2}e^{iqx}\cr
\end{eq}Therewith, the characteristic function for
the future cone  can be written as
Fourier transformed advanced causal measure with trivial invariant,
characteristically different for odd and even dimensions 
\begin{eq}{lrl}
R=0,1,\dots,&\SO_0(1,2R):&
\left\{\begin{array}{rl} 
\vth(x_0)\vth(x^2)2|x|&=
\int {2d^{1+2R}q\over|\Om^{1+2R}|}{1\over 
 [-(q-io)^2]^{1+R} }e^{iqx}\cr
\vth(x_0)\vth(x^2)2{x\over |x|}&=
\int {2d^{1+2R}q\over|\Om^{1+2R}|}{iq\over 
 [-(q-io)^2]^{1+R} }e^{iqx}\cr
 \end{array}\right.\cr\cr
R=1,2,\dots,&  \SO_0(1,2R-1):& 
\left\{\begin{array}{rl} 
\vth(x_0)\vth(x^2)2\pi &=
\int {2d^{2R}q\over|\Om^{2R-1}|}
 {1\over [-(q-io)^2]^{R} }e^{iqx}\cr
\vth(x_0)\vth(x^2)\pi x&=
\int {2d^{2R}q\over|\Om^{2R-1}|}
 {iqR\over [-(q-io)^2]^{1+R} }e^{iqx}\cr
 \end{array}\right.\cr

\end{eq}The linear order function $\vth(\pm x_0)$ for time  
future and past  $\R_\pm$ with $R=0$
is embedded in  order functions which can have Lorentz properties
for $R=1,2,\dots$, e.g. 
scalar and  vector.
In the following, a shorthand notation for the characteristic
future function is used
\begin{eq}{l}
\vth(x)= \vth(x_0)\vth(x^2)\in\{0,1\}
\end{eq}

The minimal even dimensional case $\cl D^2$ is called  abelian
or Cartan spacetime. 4-dimensional spacetime
$\cl D^4\supnoteq\cl D^2$  with nontrivial  rotation degrees of freedom
is the minimal and characteristic nonabelian case. 
$\cl D^4$ can be looked at\cite{S97,S991} also to be the modulus set 
in the polar decomposition of the
general linear group $\GL(\C^2)\ni g=u\o|g|$, i.e. it pa\-ra\-me\-tri\-zes 
the orientation  of the unitary operations $\U(2)$  in all complex linear operations.
In addition to the future translation pa\-ra\-me\-tri\-zation, 
it has also an exponential pa\-ra\-me\-tri\-zation  with 
four Lie algebra parameters  
\begin{eq}{rl}
\cl D^4&=\{x=x_0\bl 1_2+\rvec x

=e^{\psi}\mid \psi=\psi_0\bl 1_2+\rvec \psi
\in\R^4\}
\cong\GL(\C^2)/\U(2)\cr
x&={\scriptsize\pmatrix{x_0+x_3&x_1-ix_2\cr x_1+ix_2&x_0-x_3\cr}} 
,~~\psi={\scriptsize\pmatrix{\psi_0+\psi_3&\psi_1-i\psi_2\cr 
\psi_1+i\psi_2&\psi_0- \psi_3\cr}}\cr
&e^{2\psi_0}=x_0^2-r^2,~~\tanh^2|\rvec \psi|={r^2\over x_0^2},~~
{\rvec\psi\over|\rvec\psi|}={\rvec x\over r}\in\Om^2 \cr
\end{eq}$\psi_0$ pa\-ra\-me\-tri\-zes eigentime $e^{\psi_0}$.

Even dimensional spacetime $\cl D^{2R}$, $R=1,2,\dots$, 
has real rank 2, i.e. two
characterizing continuous invariants 
for the two embedded maximal noncompact abelian operations 
 - for causal eigentime $e^{\psi_0}\in\D(1)$ (`hyperbolic hopping') 
and  for position $e^{\si^3\psi_3}\in \SO_0(1,1)$ (`hyperbolic stretching')
\begin{eq}{rl}
\hbox{Iwazawa decomposition ($G=KAN$): }
\D(1)\x\SO_0(1,s)&=\SO(s)\o\cl D^2\o e^{\R^{s-1}}\cr
\hfill\hbox{with } \cl D^2
=\D(1)\x\SO_0(1,1)\ni e^{\bl 1_2\psi_0+\si^3\psi_3}
&=e^{{\bl1_2+\si^3\over 2}\psi_+}e^{{\bl1_2-\si^3\over 2}\psi_-}
\end{eq}The two re\-pre\-sen\-ta\-tion invariants will be introduced as
masses $(m_0^2,m_\ka ^2)$ in a residual re\-pre\-sen\-ta\-tion.
With the
re\-pre\-sen\-ta\-tion in a unitary group, e.g.
\begin{eq}{l}
\D(1)\x\SO_0(1,1)\map \U(\bl1_{2R})\o \SO(2R)\subnoteq\U(2R)
\end{eq}there arises a real rank $R$-dependent correlation for both
continuous invariants - in analogy to the  central correlation
$\U(1)\cap\SU(R)\cong\I(R)$
of the rational hypercharge-isospin invariants in the standard model of
electroweak interactions  given
above.  $m_0^2$
will be used as intrinsic unit. The  mass ratio
$\ka^2={m_\ka ^2\over m_0^2}$ characterizes the re\-pre\-sen\-ta\-tion 
and   determines the gauge coupling constants (more below).

$\cl D^4$ is the
nonabelian starting point also for another  
chain of causal symmetric spaces $\D(R)$ with real rank $R$,
characterizing unitary relativity
as the manifold of unitary groups $\U(R)$
 in the genral linear group $\GL(\C^R)$
\begin{eq}{l}
\D(R)\cong\GL(\C^R)/\U(R),\hskip20mm
{\begin{array}{rl}
\cr\cr
\cl D^2&\cr
\cap~&\cr
\D(1)\subnoteq\D(2)=\cl D^4&\hskip-2mm\subnoteq\D(3)\subnoteq\dots\cr
\cap~&\cr
\cl D^6&\cr
\cap~&\cr
\dots&\cr
\end{array}}

\end{eq}The causal spaces $\D(R)$ are  real $R^2$-dimensional 
positive cones, pa\-ra\-me\-tri\-zable 
in the  C*-algebras of complex $(R\x R)$-matrices. With the determinants
 as
 $\SL(\C^R)$-invariant 
 multilinear forms (volume forms) - 
 for $\D(2)$ identical with the Lorentz metric -
the  future measure with invariant $m_\ka ^R$ is given by
\begin{eq}{l}
\hbox{for }\D(R):~~
d^{R^2}\ka_+(q)={d^{R^2}q\over [\det(q-io)-m^R_\ka ]^R}={dq\over q-io-m_\ka }
,~~{d^{4}q\over [(q-io)^2-m^2_\ka ]^2},\dots
\end{eq}

\subsection
{Residual re\-pre\-sen\-ta\-tions of  Cartan spacetime}

The Lorentz compatible  embedding of  1-di\-men\-sio\-nal future  into 
 2-di\-men\-sio\-nal Cartan future (even dimensional abelian  spacetime)
is given by  the Fourier transformed pole and dipole distribution
in the scalar and vector future functions 
\begin{eq}{l}
\cl D^2=\R^2_+\ni \vth(x) x\mape \left\{\begin{array}{ll}
\int {d^2q\over \pi}
 {1\over -(q-io)^2 }e^{iqx}&=\vth(x)2\pi\cr
\int {d^2q\over \pi}
 {iq\over [-(q-io)^2]^{2} }e^{iqx}&=\vth(x)\pi x\end{array}\right.
\end{eq}States $z\mape e^{-|Qz|}$   of  position  $\cl Y^1\cong \SO_0(1,1)$
with momentum measures  ${|Q|\over q_3^2+Q^2}$
 are embedded with an energy dependent  invariant for the $\Om^1$-momentum 
 measure
\begin{eq}{l}
{d^2q\over -q_\ro P^2+m^2}=
\vth(m^2-q_0^2)~{dq_0\over|Q|} d\om(Q)
+\vth(q_0^2-m^2)~{d^2q\over -q_\ro P^2+m^2}
\cr
d\om(Q)=
dq_3{|Q|\over q_3^2+Q^2}
\hbox{ with }Q^2={m^2-q_0^2}
\end{eq}They lead to the Lorentz scalar future measures with invariant $m^2$.
The $\cl D^2$-re\-pre\-sen\-ta\-tion coefficients are Bessel functions
\begin{eq}{l}
\R^2_+\ni\vth(x)x\mape \int {d^{2}q  \over \pi[-(q- io)^2+m^2]}
e^{iqx}=
\vth(x)2\pi\cl J_0(|mx|)\cr
\end{eq}The projection 
$x=\bl 1_2 t+\si_3 z$ on  re\-pre\-sen\-ta\-tion 
coefficients of  position $\SO_0(1,1)$
and  of time $\D(1)$  is obtained by time and position 
integration respectively
\begin{eq}{rl}
\SO_0(1,1)\ni z\mape\int{d t\over 2\pi}
\int {d^{2}q \over    \pi[-(q-io)^2+m^2]}
e^{iqx}&=
\int {dq  \over \pi}~{1\over  q^2+m^2}
e^{-iqz}\cr
&={e^{-|mz|}\over|m|}\cr
\R_+\ni \vth(t)t\mape
\int{d z\over 2\pi}\int {d^{2}q  \over \pi[-(q- io)^2+m^2]}
e^{iqx}&=\int {dq  \over \pi}~{1\over - (q- io)^2+m^2}
e^{iqt}\cr
&=\vth(t)2{\sin mt\over m}\cr

\end{eq}

The selfdual re\-pre\-sen\-ta\-tions of causal time $\D(1)$  with invariant $m^2$
are embedded in a Lorentz vector. It is a 
tangent translation distribution of spacetime $\cl D^2$
\begin{eq}{rl}
\R^2\ni x\mape \int {d^{2}q  \over i\pi}~{q\over  (q- io)^2-m^2}
e^{iqx}
&={\p\over\p x}\vth(x)2\pi\cl J_0(|mx|)\cr
\hbox{with }{\p\over\p x}={x\over2}{\p\over\p {x^2\over 4}}
\hfill &=\vth(x_0)\pi x
\left[\de({x^2\over 4})-m^2\vth(x^2){2\cl J_1(|mx|)\over |mx|}\right]
\cr

\end{eq}with the projections to re\-pre\-sen\-ta\-tions of the time group 
and of the position tangent translations
\begin{eq}{rl}
\R_+\ni \vth(t)t\mape& \int{d z\over 2\pi}
\int {d^{2}q  \over i\pi}~{q\over  (q- io)^2-m^2}
e^{iqx}
=\int {dq  \over i\pi}~{q\over  (q- io)^2-m^2}
e^{iqt}\cr
&=\vth(t)2\cos mt\cr
\R\ni z\mape& \int{d t\over 2\pi}
\int {d^{2}q  \over i\pi}~{q\over  (q- io)^2-m^2}
e^{iqx}
=-\int {dq  \over i\pi}~{q\over  q^2+m^2}
e^{-iqz}\cr
&
=-{1\over |m|}{\p\over\p z}e^{-|mz|}
=\ep(z)e^{-|mz|}
\end{eq}

The residual Lorentz vector re\-pre\-sen\-ta\-tions 
of 2-di\-men\-sio\-nal spacetime are characterized by 
{two  Lorentz invariants} $(m_0^2,m_\ka ^2)$
which can be written with an invariant singularity  integrated  
over a finite interval
\begin{eq}{rl}
{2\over q^2-m_\ka ^2}~{q\over q^2-m_0^2}
&=\int_{m_\ka^2}^{m_0^2}{dm^2\over m_0^2-m_\ka^2}
~{2q\over (q^2-m^2)^2}\cr
=-{\p\over \p q}
\int_{m_\ka^2}^{m_0^2}{dm^2\over m_0^2-m_\ka^2}
~{1\over q^2-m^2}
&=-{\p\over \p q}\int_0^1 d\zeta
{1\over q^2-\zeta m_0^2-(1-\ze)m_\ka^2}
=-{\p\over \p q}{\log{q^2-m_0^2\over q^2-m_\ka^2}\over
m_0^2-m_\ka^2}

\end{eq}By the Lorentz compatible embedding
both invariants contribute to the re\-pre\-sen\-ta\-tions 
of both the causal group $\D(1)$ and the position
hyperboloid $\SO_0(1,1)$.
The Lorentz vector  Cartan  spacetime energy-momentum distribution 
\begin{eq}{rl}
\hbox{for }\cl D^2=\D(1)\x\SO_0(1,1):&
{d^2q\over i\pi[-(q-io)^2+m_\ka ^2]}~{q\over (q-io)^2-m_0^2}\cr
\end{eq}leads to the residual re\-pre\-sen\-ta\-tion coefficients
\begin{eq}{rl}
\R_+^2\ni \vth(x)x\mape& 
 
\int 
{d^2q\over i\pi[-(q-io)^2+m_\ka ^2]}~{q|m_0|\over (q-io)^2-m_0^2}
e^{iqx}\cr

&=-\vth(x)
{\pi |m_0|x\over2}
 {\p\over\p {x^2\over4}}{\cl J_0(|m_0x|)
-\cl J_0(|m_\ka x|)\over m_0^2-m_\ka ^2}\cr
\end{eq}On the lightcone $x^2=0$, where
time and position translations coincide $x^3=\pm x^0$,
the contributions  from both invariants  cancel each other.

The  projections
on re\-pre\-sen\-ta\-tions of the  causal group $\D(1)$
and of  position  $\cl Y^1$ are
\begin{eq}{rrl}
\hbox{time future:}&\R_+\ni\vth( t)  t\mape& 
\int {dz\over2\pi}
\int 
{d^2q\over i\pi[-(q-io)^2+m_\ka ^2]}~{q|m_0|\over (q-io)^2-m_0^2}
~e^{iqx}\cr

& &= \vth( t)2|m_0| {\cos m_0 t-\cos m_\ka  t\over m_0^2-m_\ka ^2}\cr

\hbox{position:}&\SO_0(1,1)\cong \R\ni z\mape&
 \int {dt\over2\pi}
 \int 
{d^2q\over i\pi[-(q-io)^2+m_\ka ^2]}~{q|m_0|\over (q-io)^2-m_0^2} ~e^{iqx}\cr
&&=-{2|m_0|\over m_0^2-m_\ka ^2}\ep(z){\p\over\p |z|}V(|z|)\cr
 \end{eq}The position projection
displays   exponential  interactions  
\begin{eq}{l}
V(|z|)=
{e^{-|m_\ka z|}\over|m_\ka |}-{e^{-|m_0z|}\over|m_0|},~~
{\p\over \p |z|}V(|z|)=e^{-|m_\ka z|}-e^{-|m_0z|}
\end{eq}

\subsection
{Residual re\-pre\-sen\-ta\-tions of nonabelian spacetime}

Cartan spacetime is the abelian noncompact substructure of 
even dimensional spacetimes 
\begin{eq}{l}
R=1,2,\dots:~~\cl D^{2R}=\D(1)\x\cl Y^{2R-1}\cong{\D(1)\x\SO_0(1,2R-1)\over
\SO(2R-1)}
\end{eq}Higher order  poles
have to be used for  the states $\rvec x\mape e^{-|Q|r}$ of   
position hyperboloids $R\ge2$ with nontrivial rotation degrees of freedom,
e.g. dipoles for 4-dimensional spacetime
\begin{eq}{rl}{d^{2R}q\over (m^2-q_\ro P^2)^R}
&=\vth(m^2-q_0^2){dq_0\over |Q|}d^{2R-1}\om(Q)
+\vth(q_0^2-m^2){d^{2R}q\over (m^2-q_\ro P^2)^R}\cr
d^{2R-1}\om(Q)&= d^{2R-1}q {|Q|\over (\rvec q^2+Q^2)^R}
\hbox{ with }Q^2={m^2-q_0^2}\cr
\end{eq}The Lorentz scalar causal measures of spacetime
 \begin{eq}{l}
\cl D^{2R}\ni\vth(x)x\mape
\int {2d^{2R}q \over  |\Om^{2R-1}| ~[-(q- io)^2+m^2]^R}e^{iqx}=
\vth(x)2\pi\cl J_0(|mx|)\cr

\end{eq}embed the re\-pre\-sen\-ta\-tion of hyperbolic position 
\begin{eq}{rl}
\cl Y^{2R-1}\ni\rvec x\mape& \int{dt\over 2\pi}
\int {2d^{2R}q  \over |\Om^{2R-1}|~  [-(q- io)^2+m^2]^R}
e^{iqx}\cr
&=
\int {2d^{2R-1} q  \over |\Om^{2R-1}|}~{1 \over  (\rvec q^2+m^2)^R}
e^{-i\rvec q\rvec x}={e^{-|m|r}\over|m|} \cr
\end{eq}with the projection on  
time re\-pre\-sen\-ta\-tion coefficients\cite{S89}
\begin{eq}{rl}

\R_+\ni\vth(t)t\mape& 
\int{|\Om^{2R-1}|~d^{2R-1}x\over(2\pi)^{2R}}
\int {2d^{2R}q  \over |\Om^{2R-1}|~[-(q- io)^2+m^2]^R}
e^{iqx}\cr
&=\int {dq\over \pi}  ~{1\over  [-(q- io)^2+m^2]^R}
e^{iqt}
=\vth(t){1\over\Ga(R)}(-{\p\over\p m^2})^{R-1}{2\sin mt\over m}\cr
\hbox{ for }R=2\hfill&=\vth(t)2{\sin mt-mt\cos mt\over m^3}\cr

\end{eq}

In general, the time and
position projections of even dimensional spacetime 
re\-pre\-sen\-ta\-tion coefficients, given by integer index Bessel functions,
are halfinteger Bessel, Neumann and Macdonald functions.

The Lorentz vector embedded selfdual $\D(1)$-re\-pre\-sen\-ta\-tions 
are re\-pre\-sen\-ta\-tions of tangent spacetime translations
\begin{eq}{rl}
\R^{2R}\ni x\mape 
\int {2d^{2R}q  \over i|\Om^{2R-1}|}~{q\over  (q- io)^2-m^2}
e^{iqx}
&=\pi x\Ga(R)({\p\over\p {x^2\over 4}})^R
\vth(x)\cl J_0(|mx|)\cr
\end{eq}They are familiar from the interaction inducing 
off-shell contribution of  Feynman
propagators which are proportional to Bessel functions and 
 supported by the causal bicone
 \begin{eq}{rl}
\int {2d^{2R}q\over i|\Om^{2R-1}|}{1\over q^2- io-1}e^{iqx}
&=\Ga(R)\({\p\over \p{x^2\over 4}}\)^{R-1}
\left[ \pi\vth(x^2)(i\cl J_0-\cl N_0)(|x|)+\vth(-x^2)2\cl K_0(|x|)\right]\cr
&=\Ga(R)\({\p\over \p{x^2\over 4}}\)^{R-1}
\int d\psi~[ \vth(x^2)e^{i|x|\cosh\psi}+\vth(-x^2)e^{-|x|\cosh\psi}]\cr
\end{eq}e.g. for 4-dimensional spacetime
\begin{eq}{l}
\R^4\ni x\mape \int {d^{4}q  \over i\pi^2}~{q\over  q_\ro P^2-m^2}
e^{iqx}
={\pi x\over 2}
\left[
\de'({x^2\over 4})
-m^2\de({x^2\over 4})
+m^2\vth(x^2){4\cl J_2(|mx|)\over x^2}\right]\cr
\end{eq}The off-shell contributions of  Feynman
propagators are no coefficients of Poincar\'e group re\-pre\-sen\-ta\-tions.

The $\D(1)$-re\-pre\-sen\-ta\-tion shows up  in  the time  projection
\begin{eq}{rl}
\R_+\ni\vth(t)t\mape &\int{|\Om^{2R-1}|~d^{2R-1}x\over
(2\pi)^{2R}}
\int {2d^{2R}q  \over i|\Om^{2R-1}|}~{q\over  (q- io)^2-m^2}
e^{iqx}\cr
&=\int {dq  \over i\pi}~{q\over  (q- io)^2-m^2}
e^{iqt}=\vth(t)2\cos mt\cr
\end{eq}The projection on position tangent coefficients
\begin{eq}{rl}
\R^{2R-1}\ni\rvec x\mape&
\int{dt\over2\pi}\int {2d^{2R}q  \over i|\Om^{2R-1}|}~{q\over  (q- io)^2-m^2}
e^{iqx}
=-\int {2d^{2R-1}q  \over i|\Om^{2R-1}|}~{\rvec q\over  \rvec q^2+m^2}
e^{-i\rvec q\rvec x}\cr
&=-{\p\over\p\rvec x }
\int {2d^{2R-1}q  \over |\Om^{2R-1}|}~{1\over  \rvec q^2+m^2}
e^{-i\rvec q\rvec x}
={\rvec x\over 2|m|}
\Ga(R)(-{\p\over \p{r^2\over 4}})^R e^{-|m|r}\cr
\end{eq}involves Yukawa forces, e.g.
for 3-dimensional position translations
in 4-dimensional spacetime  
\begin{eq}{l}
\R^3\ni\rvec x\mape
{\rvec x\over 2|m|}
({\p\over \p{r^2\over 4}})^2 e^{-|m|r}=
-\rvec x
{\p\over \p{r^2\over 4}} {e^{-|m|r}\over r}=
{\rvec x\over r}~{1+|m|r\over r^2}e^{-|m|r}\cr
\end{eq}

In an $\SO_0(1,2R-1)$-compatible framework
with two  Lorentz invariants $(m_0^2,m_\ka ^2)$,
both the invariant $m_0^2$ 
in the pole distribution for the Lorentz vector 
 and the invariant $m_\ka ^2$
in the Lorentz scalar causal measure  with pole of order $R$  
are used in re\-pre\-sen\-ta\-tions 
of $\D(1)$ and $\cl Y^{2R-1}$
\begin{eq}{rl}
{2\over (q^2-m_\ka ^2)^R}~{q\over q^2-m_0^2}
&=\int_{m_\ka^2}^{m_0^2} {dm^2\over m_0^2-m^2}
~({ m_0^2-m^2\over m_0^2-m_\ka^2})^R
~{ 2qR\over (q^2-m^2)^{1+R}}\cr
=-{\p\over \p q}
\int_{m_\ka^2}^{m_0^2}
{dm^2\over m_0^2-m^2}
~({ m_0^2-m^2\over m_0^2-m_\ka^2})^R
~{ 1\over (q^2-m^2)^{R}}
&=-{\p\over \p q}
\int_0^1 d\zeta
{(1-\zeta)^{R-1}\over 
[q^2-\zeta m_0^2-(1-\ze)m_\ka^2]^R}

\end{eq}This defines
the Lorentz vector advanced causal distribution
as product of the Lorentz scalar measure 
 \begin{eq}{rl}
\hbox{for }\cl D^{2R}\cong\D(1)\x\cl Y^{2R-1}:~~
d^{2R}\ka_+(q)&
{q|m_0|\over  (q-io)^2-m_0^2}\cr
\hbox{with } d^{2R}\ka_+(q)&=
{2d^{2R}q\over i|\Om^{2R-1}|[-(q-io)^2+m_\ka ^2]^R}
\cr
\hbox{e.g. }R=2:~~d^4\ka_+(q)&={d^{4}q\over i\pi^2 [(q-io)^2-m_\ka ^2]^2}\cr
\end{eq}with the $\D(1)$-related simple pole factor.
The Fourier transform
gives the  coefficients  of residual  re\-pre\-sen\-ta\-tions 
of even dimensional spacetime.  
E.g., for 4-dimensional spacetime with dipole\cite{HEI} 
for the Lorentz scalar future measure 
\begin{eq}{l}
\D(2)=\cl D^4=\R_+^{4}\ni \vth(x)x\mape 
\int d^4\ka_+(q) ~{q|m_0|\over (q-io)^2-m_0^2}
~e^{iqx}
\cr
\hskip20mm=\vth(x){\pi|m_0| x\over2}{1\over m_0^2-m_\ka ^2}{\p\over\p {x^2\over
4}}\left[
{\p\over\p{x^2\over 4}}
{\cl J_0(|m_0x|)
-\cl J_0(|m_\ka x|)\over m_0^2-m_\ka ^2}
-\cl J_0(|m_\ka x|)\right]\cr
\end{eq}the projections $x= t\bl1_2+\rvec x$ on 
re\-pre\-sen\-ta\-tion coefficients of time future
and on 3-di\-men\-sio\-nal hyperbolic position 
\begin{eq}{rrl}
\hbox{time:}&\R_+\ni \vth( t) t\mape&
\int { d^3x\over 8\pi^2}
\int d^4\ka_+(q)
 ~{q|m_0|\over (q-io)^2-m_0^2}
~e^{iqx}
\cr
&&=\vth( t){2|m_0|\over  m_0^2-m_\ka ^2}\(
{\cos m_0 t-\cos m_\ka  t\over m_0^2-m_\ka ^2}+{m_\ka  t\sin m_\ka  t\over 2m_\ka ^2}\) 
\cr
\hbox{position:}&\cl Y^3\ni \rvec x\mape&
\int {dt\over2\pi}
\int d^4\ka_+(q)  ~{q|m_0|\over (q-io)^2-m_0^2}~e^{iqx}
\cr
&&={4|m_0|\rvec x\over ( m_0^2-m_\ka ^2)^2}({\p\over \p r^2})^2V_1(r)\cr
 \end{eq}involve  Yukawa and  exponential interactions
\begin{eq}{rl}
 V_1(r) &=
 {e^{-  |m_\ka |r}\over |m_\ka |}-{e^{-  |m_0|r}\over |m_0|}
-{m_0^2-m_\ka ^2\over 2|m_\ka |^3}(1+|m_\ka |r)e^{-  |m_\ka |r}
\cr

V_3(r)={\p\over \p r^2}V_1(r)
&=
{e^{-  |m_0|r}-e^{-  |m_\ka |r}\over 
r}+{m_0^2-m_\ka ^2\over 2|m_\ka |}e^{-  |m_\ka |r}
\cr

\end{eq}An exponential interaction 
is the 2-sphere spread
of a  1-di\-men\-sio\-nal position re\-pre\-sen\-ta\-tion 
${1\over 2}e^{-r}=-{\p\over\p{r^2\over 4}}(1+r)e^{-  r}$
with an  $r$-proportional contribution\cite{BOE,S89} from a dipole.

\section{Product re\-pre\-sen\-ta\-tions of spacetime}

Product re\-pre\-sen\-ta\-tions come with 
the product of re\-pre\-sen\-ta\-tion coefficients, i.e. in a residual formulation
with the convolution $*$
of  (ener\-gy-)mo\-men\-tum distributions. The convolution 
 picks up a residue itself
 \begin{eq}{l}
 *\cong \de(q_1+q_2-q)\cong\res_{q_1+q_2=q}
 \end{eq}It defines the residual product which leads to  product re\-pre\-sen\-ta\-tions.
The convolution adds (energy-)momenta of singularity
manifolds as imaginary and real eigenvalues  for
compact and  noncompact re\-pre\-sen\-ta\-tion invariants.

The Radon (energy-)momentum measures are a convolution algebra
which reflects the pointwise multiplication $\m$ property of
the essentially bounded function classes
\begin{eq}{l}
\int d^nq~e^{iqx}\cl M(\R^n)\subnoteq L^\infty(\R^n),~~\left\{\begin{array}{cl}
\cl M(\R^n)*\cl M(\R^n)&\sub \cl M(\R^n)\cr
L^\infty(\R^n)\m L^\infty(\R^n)&\sub L^\infty(\R^n)\cr\end{array}\right.
\end{eq}

In the Feynman integrals of special relativistic quantum field theory
 as convolutions  of 
ener\-gy-mo\-men\-tum distributions,
the on-shell parts for 
 translation  re\-pre\-sen\-ta\-tions 
give product re\-pre\-sen\-ta\-tion coefficients
of the Poincar\'e group, i.e. 
energy-momentum distributions  for free  states
(multiparticle  measures, below). 
The off-shell interaction contributions  are not convolutable.
 This is the origin of the
`divergence' problem in quantum field theories with interactions.
With respect to Poincar\'e group re\-pre\-sen\-ta\-tions,
the convolution of Feynman propagators
makes no sense.

\subsection
{Convolutions with linear invariants}

The convolution products for  energy 
distributions can be read off directly from the pointwise products of 
re\-pre\-sen\-ta\-tion matrix elements of time
with the sum of the $\D(1)$-invariants as invariant of the product 
\begin{eq}{l}
\begin{array}{rl}
e^{im_1t}\m e^{im_2t}&=e^{im_+t}\cr
\hbox{ with }m_+&=m_1+m_2\cr\end{array}\then\left\{\begin{array}{rl}
{1\over q\mp io -m_1} 
(\pm {*\over2i\pi}){1  \over q\mp io -m_2}&=
{1\over q\mp io -m_+}\cr
{1\over q- io -m_1} 
*{1 \over q+ io -m_2}&=0\cr\end{array}\right.
\end{eq}The {normalization of the residual product}
 is the 1-sphere measure as used in the residue 
 \begin{eq}{l}
\oint {dq\over 2i\pi}=\res_{},~~
{*\over 2\pi}
\cong {1\over |\Om^1|}\de(q_1+q_2-q)\cr
\end{eq}

The residual product for the two  causal function algebras,
conjugated and orthogonal to each other,
and the Dirac convolution algebra is summarized 
with  the {residually normalized re\-pre\-sen\-ta\-tion functions}
and the integration contours 
\begin{eq}{c}
\vth(\pm t)e^{imt}=\pm\int{dq\over 2i\pi}~{1\over q\mp io-m}e^{iqt}\hskip10mm
{\scriptsize\begin{array}{|c|}\hline
\hbox{causal time $\D(1)$}\hbox{ and  energies }\R\cr\hline\hline
({\stackrel1*},q)=(\pm{*\over 2i\pi},q\mp io)
\hbox{ causal, orthogonal}\cr\hline
 {1\over q-m_1}  
~{\stackrel1*}~{1 \over q-m_2}={1\over q-m_+ }\cr
\hline
\de(q-m_1)*\de(q-m_2)=\de(q-m_+)\cr\hline
\end{array}}
\end{eq}

\subsection
{Convolutions with selfdual invariants}

The causal  distributions with compact dual invariants
\begin{eq}{rl}
\pm {1\over i\pi}{q\over (q\mp io)^2 -m^2}
&=|m|\de(q^2 -m^2)\pm {1\over i\pi}{q\over q^2_\ro P -m^2}
=\pm{1\over 2i\pi}({1\over q\mp io-|m|}+{1\over q\mp io+|m|})\cr

\end{eq}keep the property to constitute 
orthogonal convolution algebras, conjugated to each
other
\begin{eq}{c}
2\cos m_1t\m 2\cos m_2t= 2\cos m_+t+
2\cos m_-t \hbox{ with }m_\pm=|m_1|\pm |m_2|\cr\cr
\vth(\pm t)2\cos mt=\pm\int {dq\over i\pi}{q\over (q\mp io)^2 -m^2}e^{iqt}~~~
\hfill{\scriptsize

\begin{array}{|c|}\hline
\hbox{causal time $\D(1)$}\hbox{ and  energies }\R\cr\hline\hline
({\stackrel1*},q^2)=(\pm{*\over i\pi},(q\mp io)^2)
\hbox{ causal, orthogonal}\cr\hline
{q\over q^2-m_1^2}~{\stackrel1*}~{q\over q^2-m_2^2}=
{q\over q^2-m_+^2}+{q\over q^2-m_-^2}\cr\hline
\end{array}}\cr
\end{eq}The residual normalization for selfdual invariants
uses half the 1-sphere measure
\begin{eq}{l}
{*\over \pi}\cong{2\over |\Om^1|}\de(q_1+q_2-q)
\end{eq}

Since the Feynman energy  distributions 
combine future  and past 
distributions 
\begin{eq}{rl}
\pm{1\over i\pi}{|m|\over q^2\mp io -m^2}&=
|m|\de(q^2 -m^2)\pm{1\over i\pi}{|m|\over q^2_\ro P -m^2}
=\pm{1\over 2i\pi}({1\over q\mp io-|m|}-{1\over q\pm io+|m|})\cr
\end{eq}they
constitute  convolution algebras,
conjugated to each other, however not orthogonal
$e^{+i|m_1t|}\m e^{- i|m_2t|}\ne0$
\begin{eq}{c}

e^{\pm i|mt|}=\pm\int{dq\over i\pi}
{|m|\over q^2\mp io -m^2}e^{iqt}~~
{\scriptsize
\begin{array}{|c|}\hline

\hbox{bicone time }\R_+\uplus \R_-\hbox{ and energies } \R\cr\hline\hline
(\stackrel 1*,q^2)
=(\pm{*\over i\pi}, q^2\mp io)\hbox{ Feynman, not orthogonal}\cr\hline
 {|m_1|\over q^2-m^2_1}  ~{\stackrel 1*}~ {|m_2|\over q^2-m^2_2}  
={|m_+|\over q^2-m^2_+}\cr  
\hline
\end{array}}

\end{eq}

The faithful Hilbert re\-pre\-sen\-ta\-tions
of $\cl Y^1$ (1-di\-men\-sio\-nal abelian position) 
with Fourier transformed $\Om^1$-measures and noncompact 
dual invariants constitute a  convolution algebra
\begin{eq}{c} 
e^{-|mz|}=
\int{dq\over \pi}
{|m|\over q^2+m^2}e^{-iqz}\hskip10mm
 {\scriptsize
 \begin{array}{|c|}\hline

\hbox{position }\cl Y^1 \hbox{ and `momenta' } \R\cr\hline\hline
|\Om^1|=2\pi,~~
\stackrel 1*~={*\over\pi}\cr\hline
{|m_1|\over q^2+m^2_1}  ~\stackrel 1*~{|m_2|\over q^2+m^2_2} 
={|m_+|\over q^2+m^2_+}\cr
\hline
\end{array}}
\end{eq}

\subsection{Product re\-pre\-sen\-ta\-tions of free particles}

Interaction free product structures have to convolute Dirac distributions for 
cyclic translation re\-pre\-sen\-ta\-tions.

In contrast 
to the convolution of Dirac distributions for 
selfdual invariants $m^2_{1,2}>0$ with
basic  selfdual  spherical 2-dimensional re\-pre\-sen\-ta\-tions 
\begin{eq}{rl}
\hbox{abelian }\R:&2|m_1|\de(q^2-m_1^2)*~2|m_2|\de(q^2-m_2^2)\cr
&\hskip20mm=2|m_-|\de(q^2-m_-^2)+2|m_+|\de(q^2-m_+^2)
\end{eq}the convolution of Dirac distributions for 
 the infinite dimensional re\-pre\-sen\-ta\-tions of the  Euclidean groups, $s\ge2$, 
with the  sphere radii as momentum invariants $\rvec q^2=m^2>0$
is nontrivial for all momentum invariants  
between the  `sum and difference sphere' of the factors
$m_+^2\ge\rvec q^2\ge m_-^2$ 
\begin{eq}{l}
\SO(s)\sx\R^s:~~\de(\rvec q^2-m_1^2)
{*\over |\Om^{s-2}|}\de(\rvec q^2-m_2^2)= {2|Q|^{s-3}\over |\rvec q|}
\vth(m_+^2-\rvec q^2)\vth(\rvec q^2-m_-^2)\cr
s=2,3,\dots
\end{eq}There arises
a momentum dependent normalization factor $|Q|$ 
 which contains the characteristic two particle convolution function
 \begin{eq}{l}
Q^2={-\De(\rvec q^2)\over 4\rvec q^2},~~
\De(\rvec q^2)=\De(\rvec q^2,m^2_1,m^2_2)=
(\rvec q^2- m_+ ^2)(\rvec q^2- m_-^2)\cr
\end{eq}It is symmetric in the three invariants involved
\begin{eq}{l}
\De(a,b,c)=a^2+b^2+c^2-2(ab+ac+bc)=(a+b-c)^2-4ab
\end{eq}The minimal cases $s=2,3$ are characteristic
for even and odd dimensional position
\begin{eq}{l}
\SO(3)\sx\R^3:~\de(\rvec q^2-m_1^2)
~{*\over 2\pi}~ \de(\rvec q^2-m_2^2)= {2\over |\rvec q|}
\vth(m_+^2-\rvec q^2)\vth(\rvec q^2-m_-^2)
\end{eq}

Product re\-pre\-sen\-ta\-tions of  Poincar\'e groups 
with the hyberboloid `radii' as energy-momentum invariants $q^2=m^2\ge0$
\begin{eq}{rl}
\SO_0(1,s)\sx\R^{1+s}:&
\vth(\pm q_0)\de(q^2-m_1^2)~{*\over |\Om^{s-1}|}~
\vth(\pm q_0)\de(q^2-m_2^2)\cr
s=1,2,3,\dots
&\hfill ={2|Q|^{s-2}\over|q|}\vth(\pm q_0)\vth(q^2-m_+^2)
\end{eq}involve the two particle threshold factor
\begin{eq}{l}
Q^2={\De(q^2)\over 4|q|^2},~~
\De(q^2)=\De(q^2,m_1^2,m_2^2)=(q^2- m_+ ^2)(q^2- m_-^2)
\end{eq}The minimal cases $s=1, 2$ are
characteristic for even and odd dimensional spacetime.
$1+s=4$ is the minimal case with nonabelian rotations
\begin{eq}{rl}
\SO_0(1,3)\sx\R^{4}:&
\vth(\pm q_0)\de(q^2-m_1^2)~{*\over4\pi}~
\vth(\pm q_0)\de(q^2-m_2^2)\cr
&\hfill ={2|Q|\over  |q|}\vth(\pm q_0)\vth(q^2-m_+^2)
\end{eq}For nontrivial position, the convolution  of  
 $s$-di\-men\-sio\-nal on-shell  hyperboloids (particle measures)
does not lead to $s$-di\-men\-sio\-nal on-shell  hyperboloids
 $\de(q^2-m_+^2)$.
 The squared  sum of the invariants as product invariant
 gives
  the {threshold} for  energy-mo\-men\-ta
 $q^2=(q_1+q_2)^2\ge  m_+ ^2$  in the  2-particle product measure.
Here, the energy is enough to produce 
 two free particles with masses $m_{1,2}$ and momentum 
 $(\rvec q_1+\rvec q_2)^2\ge0$.

\subsection{Product re\-pre\-sen\-ta\-tions of hyperbolic positions}

The residual re\-pre\-sen\-ta\-tions
of  $3$-dimensional hyperbolic  position $\cl Y^3$
use the Fourier transformed  $\Om^{3}$-measure,
familiar from
the nonrelativistic hydrogen Schr\"odinger functions.
The radii of the `momentum' spheres as invariants
are added up in the  convolution

\begin{eq}{c}
 
e^{-|m|r } 
=\int {d^3q\over \pi^2}{|m|\over (\rvec q^2+m^2)^2}
e^{-i\rvec q\rvec x}\hskip10mm
{\scriptsize\begin{array}{|c|}\hline
\hbox{position }\cl Y^3\cong\SO_0(1,3)/\SO(3)
\cr
 \hbox{and `momenta' }\R^3
 \hbox{ with }\SO(3)\cr\hline
 
|\Om^3|=2\pi^2,~~
\stackrel3*~={*\over  \pi^2}\cr\hline
{|m_1|\over (\rvec q^2+m^2_1)^2} \stackrel3*
{|m_2|\over (\rvec q^2+m^2_2)^2}=
{|m_+|\over (\rvec q^2+m^2_+)^2}\cr\hline
\end{array}}\cr

\end{eq}

In general, the re\-pre\-sen\-ta\-tions
of odd dimensional hyperboloids $\cl Y^{2R-1}$ come with Fourier transformed
$\Om^{2R-1}$-measures and imaginary singularity sphere $\Om^{2R-2}$
for the `momentum' eigenvalues.
The sphere measures can be obtained by invariant
momentum derivatives
\begin{eq}{l}
 {1\over \Ga(R)}(-{\p\over\p\rvec q^2})^{R-1}{|m|\over \rvec q^2+m^2}
={|m|\over (\rvec q^2+m^2)^{R}},~~R=1,2,\dots\cr
\end{eq}Product re\-pre\-sen\-ta\-tions arise by the convolution
with  the sphere volume as residual normalization
\begin{eq}{c}
e^{-|m|r } 
=\int {2 d^{2R-1}q\over |\Om^{2R-1}|}{|m|\over (\rvec q^2+m^2)^{R}}
e^{-i\rvec q\rvec x}\cr\cr

{\scriptsize\begin{array}{|c|}\hline
\hbox{position }\cl Y^{2R-1}\cong\SO_0(1,2R-1)/\SO(2R-1),~R=1,2,\dots\cr
 \hbox{and `momenta' }\R^{2R-1}
 \hbox{ with }\SO(2R-1)\cr\hline\hline

|\Om^{2R-1}|=
{2\pi^{R}\over\Ga(R)}
,~~\stackrel{2R-1}*~={*~2\over |\Om^{2R-1}|}\cr\hline
({\p\over  \p \rvec q})^{L_1}
{|m_1|\over (\rvec q^2+m^2_1)^{R}} \stackrel{2R-1}*
({\p\over\p \rvec q})^{L_2}
{|m_2|\over (\rvec q^2+m^2_2)^{R}}=({\p\over \p \rvec q})^{L_1+L_2}
{|m_+|\over (\rvec q^2+m^2_+)^{R}}\cr
\hfill\hbox{for } L=0,1,\dots\cr\hline
\end{array}}\cr
\end{eq}

\noindent 
The convolution may involve tensor products
for $\SO(2R-1)$-re\-pre\-sen\-ta\-tions.
In general, nontrivial $\SO_0(t,s)$-properties
are effected by  the convolution compatible 
(energy-)momentum derivatives
\begin{eq}{rl}
{\p\over \p  q}= 2q {\p\over\p q^2},~~
{\p\over \p  q}\ox  q
&=\bl1_{t+s}+ q\ox q~2{\p\over\p q^2}\cr
{\p\over \p  q}\ox {\p\over \p  q}&=(
\bl1_{t+s}+ q\ox q~2{\p\over\p q^2})
2{\p\over\p q^2},~~\dots\cr
\end{eq}which - acting on multipoles - 
raise the pole order 
\begin{eq}{l}
-{\p\over \p  q}
{\Ga(R)\over( q^2+\mu^2)^{R}}
={ 2q~\Ga(1+R) \over(q^2+\mu^2)^{1+R}}
\end{eq}

The re\-pre\-sen\-ta\-tions of odd dimensional spheres 
use a singularity sphere $\Om^{2R-2}$ with real momentum eigenvalues
in the convolutions
\begin{eq}{c}
e^{\pm i|m|r } 
=\pm\int {2d^{2R-1}q\over i |\Om^{2R-1}|}{|m|\over
 (\rvec q^2\mp io-m^2)^{R}}e^{-i\rvec q\rvec x}\cr\cr

{\scriptsize\begin{array}{|c|}\hline
\hbox{sphere }\Om^{2R-1} \cong \SO(2R)/\SO(2R-1),~R=1,2,\dots\cr
 \hbox{and momenta }\R^{2R-1}
 \hbox{ with }\SO(2R-1)\cr\hline\hline
|\Om^{2R-1}|=
{2\pi^{R}\over\Ga(R)},~~
(\stackrel{2R-1}*,\rvec q^2)
=(\pm  {*~2\over i|\Om^{2R-1}|},\rvec q^2\mp io)
\hbox{ not orthogonal}\cr\hline

({\p\over  \p \rvec q})^{L_1}
{|m_1|\over (\rvec q^2-m^2_1)^{R}} \stackrel{2R-1}*
({\p\over\p \rvec q})^{L_2}
{|m_2|\over (\rvec q^2-m^2_2)^{R}}=({\p\over \p \rvec q})^{L_1+L_2}
{|m_+|\over (\rvec q^2-m^2_+)^{R}}\cr
\hfill\hbox{for } L=0,1,\dots\cr\hline
\end{array}}\cr
\end{eq}

\noindent 
The abelian case $R=1$ with the 1-sphere $\Om^1\cong\SO(2)$
has been used above
for compact time $\R$-re\-pre\-sen\-ta\-tions by Feynman distributions. 
For $R=2$ there arise re\-pre\-sen\-ta\-tions of
the spin group $\Om^3\cong\SU(2)$.

\subsection
{Residual products for spacetime}

The convolution product of
causal and  Feynman  distributions on even dimensional spacetimes,
can be computed with the familiar methods.
For the causal distributions, the product residue is defined as causal too.

The convolutions of Cartan
ener\-gy-mo\-men\-tum pole distributions are 

{\scriptsize\begin{eq}{|c|}\hline
\hbox{spacetime  $\cl D^{2}=\D(1)\x\cl Y^1$}  \cr
\hbox{with }\SO_0(1,1),~~|\Om^1|=2\pi\cr\hline\hline
({\stackrel 2*},q^2)=
\left\{\begin{array}{ll}
(\pm{*\over 2i\pi},(q\mp io)^2),&
\hbox{causal, orthogonal}\cr
(\pm{*\over i\pi},q^2\mp io),
&\hbox{Feynman, not orthogonal}\cr       
\end{array}\right.\cr\hline
{1\over q^2-m_1^2}
\stackrel 2*
{1 \over q^2-m_2^2}=
\int_0^1 d\ze  {1\over \ze(1-\ze)q^2-\ze m_1^2 -(1-\ze)m_2^2}\cr\hline
\end{eq}}

\noindent The residual products 
display pole  distributions  only before the 
 integration
over finite $\ze\in[0,1]$, characteristic for even-dimensional spaces 
\begin{eq}{l}
\int_0^1  d\ze~{1\over
\ze(1-\ze)(q^2\mp io)-\ze m_1^2 -(1-\ze)m_2^2}\cr

\hfill={2\over\sqrt {|\De(q^2)|}}\Biggl[\pm i\pi
 \vth(q^2-m_+^2)-
\vth(\De(q^2))\log\Big|{ \Si(q^2)-\sqrt{4\De(q^2)}
 \over  m_+^2-m_-^2}  \Big|~\cr
\hfill -\vth(-\De(q^2))\arctan{\sqrt{-4\De(q^2)}\over  \Si(q^2)}
\Biggr]\cr
 \hbox{with }
 \De(q^2)=(q^2 -m_+ ^2)(q^2- m_-^2),~~ 
 \Si(q^2)=(q^2 -m_+ ^2)+(q^2- m_-^2)
\cr
\end{eq}

The pole  distributions  can be written with spectral functions, 
e.g. 
\begin{eq}{cl}
\int_0^1 {d\ze\over  q^2-\ze m_0^2-(1-\ze)m_\ka^2  }&
=\int_{m_\ka^2}^{m_0^2}
 {dM^2\over m_0^2-m_\ka^2}~{1 \over q^2-M^2  }\cr
\int_0^1 {d\ze \over \ze q^2-m^2  }&=\int_{m^2}^\infty
{ dM^2\over M^2}~{1 \over q^2-M^2  }\cr
\int_0^1 {d\ze(1-\ze)\over \ze q^2-m^2  }&=\int_{m^2}^\infty
 {dM^2\over M^2}~{M^2-m^2\over M^2}{1 \over q^2-M^2  }\cr
\end{eq}The $\ze$-dependent $q^2$-singularities 
disappear after $\ze$-integration, there arise
logarithms,  no energy-momentum poles.
The logarithm  is typical for 
a finite integration\cite{BESO}, e.g. for a function
holomorphic on the integration curve 
\begin{eq}{l}
\int_a^bdz~f(z)={\sum\res_{}}[f(z)\log{z-b\over z-a}],~~
\left\{
\begin{array}{rl}
\int_a^\infty dz~f(z)&=-{\sum\res_{}}[f(z)\log( z-a)]\cr
\int dz~f(z)&={\sum\res_{}}f(z)\cr\end{array}\right.
\end{eq}with the {sum of all residues} in the closed complex plane,
cut along the integration curve, here
\begin{eq}{rll}
\int_0^1 {d\ze\over  q^2-\ze m_0^2-(1-\ze)m_\ka^2  }&
=\sum\res[{1\over  q^2-\ze m_0^2-(1-\ze)m_\ka^2  }
\log{\ze-1\over \ze}]&={1\over m_0^2-m_\ka^2}\log{m_0^2-q^2\over m_\ka^2-q^2}\cr
\int_{0}^{1}
{d\ze\over \ze q^2-m^2}&=\sum\res_{}[{1\over \ze q^2-m^2}
\log{\ze-1\over \ze}]&={\log (1- {q^2\over m^2})\over q^2}\cr 
\int_{0}^{1}
{d\ze~(1-\ze)\over \ze q^2-m^2}&=\sum\res_{}{1-\ze\over \ze q^2-m^2}
\log{\ze-1 \over \ze}
&={(1-{m^2\over q^2})\log(1- {q^2\over m^2})-1\over q^2}\cr
\end{eq}In the 3rd case there is a nontrivial residue at the holomorphic point
$\ze=\infty$.

The corresponding structures for Minkowski spacetime
as minimal case with nontrivial rotation degrees of freedom
 are  
 
{\scriptsize\begin{eq}{|c|}\hline
\hbox{spacetime  $\cl D^{4}=\D(2)=\D(1)\x\cl Y^3$}\cr
\hbox{with }\SO_0(1,3),~~|\Om^3|=2\pi^2\cr\hline\hline
({\stackrel 4*},q^2)=
\left\{\begin{array}{ll}
(\mp{*\over 2i\pi^2},(q\mp io)^2)
,&\hbox{causal, orthogonal}\cr
(\mp{*\over i\pi^2},q^2\mp io),&
\hbox{Feynman, not orthogonal}\cr       
\end{array}\right.\cr\hline
{1\over(q^2-m_1^2)^2}
\stackrel 4*
{1 \over(q^2-m_2^2)^2} =
\int_0^1 d\ze  {\ze(1-\ze)\over [
\ze(1-\ze)q^2-\ze m_1^2 -(1-\ze)m_2^2]^2}\cr\hline

\end{eq}}

In general,  one obtains the even dimensional spacetime
$\cl D^{2R}$ distributions  of energy-momenta 
by derivations
\begin{eq}{l}
{1\over \Ga(R)}(-{\p\over\p q^2})^{R-1}{1\over q^2-m^2}
={1\over (q^2-m^2)^{R}},~~R=1,2,\dots
\end{eq}For the Feynman distributions,
the residual normalization is 
half  the volume ${|\Om ^{2R-1}|\over 2}$ of the position related sphere
with the sign $(-1)^R$.
For the  causal distributions,  
 the full  volume is taken 
\begin{eq}{c}
\int {2(-1)^Rd^{2R}q \over  |\Om^{2R-1}| ~[(q- io)^2-m^2]^R}e^{iqx}=
\vth(x)2\pi\cl J_0(|mx|)\cr\cr
{\scriptsize\begin{array}{|c|}\hline
\hbox{spacetime  $\cl D^{2R}=\D(1)\x\cl Y^{2R-1}$},~R=1,2,\dots\cr
\hbox{with }\SO_0(1,2R-1)
,~~|\Om^{2R-1}|={2\pi^{R}\over\Ga(R)}\cr\hline\hline
({\stackrel{2R}*},q^2)=
\left\{\begin{array}{rl}
(\mp{*(-1)^R\over i|\Om^{2R-1}|},(q\mp io)^2),&\hbox{causal, orthogonal}\cr
(\mp{* ~2(-1)^R\over i|\Om^{2R-1}|}, q^2\mp io),&\hbox{Feynman, not orthogonal}\cr
\end{array}\right.\cr\hline
({\p\over\p q})^{L_1}
{1\over(q^2-m_1^2)^R}
\stackrel {2R}*
({\p\over\p q})^{L_2}
{1 \over(q^2-m_2^2)^R} =
({\p\over\p q})^{L_1+L_2}{1\over\Ga(R)}
(-{\p\over\p q^2})^{R-1}\int_0^1 d\ze  {1\over 
\ze(1-\ze)q^2-\ze m_1^2 -(1-\ze)m_2^2}\cr\hline
\end{array}}
\end{eq}with the
examples - wherever the $\Ga$-functions are defined for $\nu\in\R$
\begin{eq}{rcll}
{\Ga(R+\nu_1)\over (q^2-m_1^2)^{R+\nu_1}}
&{\stackrel {2R}*\over\Ga(R)}&
{\Ga(R+\nu_2)\over (q^2-m_2^2)^{R+\nu_2}}
&= (-{\p\over\p q^2})^{R-1}[\nu_1,\nu_2](q^2)\cr
&&&=
\int_0^1 d\ze  
{\ze^{R-1+\nu_1}(1-\ze)^{R-1+\nu_2}\Ga(R+\nu_1+\nu_2)\over 
[\ze(1-\ze)q^2-\ze m_1^2 -(1-\ze)m_2^2]^{R+\nu_1+\nu_2}}\cr
{2q~\Ga(1+R+\nu_1)\over (q^2-m_1^2)^{1+R+\nu_1}}
&{\stackrel {2R}*\over\Ga(R)}&
{\Ga(R+\nu_2)\over (q^2-m_2^2)^{R+\nu_2}}
&=2q(-{\p\over\p q^2})^{R}[\nu_1,\nu_2](q^2)\cr
&&&=2q \int_0^1 d\ze  
{\ze^{R+\nu_1}(1-\ze)^{R+\nu_2}\Ga(1+R+\nu_1+\nu_2)\over 
[\ze(1-\ze)q^2-\ze m_1^2 -(1-\ze)m_2^2]^{1+R+\nu_1+\nu_2}}\cr

{2q~\Ga(1+R+\nu_1)\over (q^2-m_1^2)^{1+R+\nu_1}}
&{\stackrel {2R}*\over\Ga(R)}&
{2q~\Ga(1+R+\nu_2)\over (q^2-m_2^2)^{1+R+\nu_2}}
&= -2{\p\over\p q}\ox q~(-{\p\over\p q^2})^{R}[\nu_1,\nu_2](q^2)\cr

\end{eq}

Nontrivial $\SO_0(1,2R-1)$-properties are effected by 
energy-momentum derivatives
$({\p\over \p q})^L$.


\section{Interactions and tangent structure of spacetime}

From a re\-pre\-sen\-ta\-tion of a real Lie group and its
 functions, 
one can obtain the re\-pre\-sen\-ta\-tion of its Lie algebra
with the corresponding 
functions.

Complex functions  of a symmetric space $G/H$ (section 2), 
in short $\cl G\sub L^\infty (\R^n)$,
are re\-pre\-sen\-ta\-tion  coefficients
 of the group $G$, 
e.g states of $\SO_0(1,3)$ in the case of hyperbolic positions $\cl Y^3\ni\rvec
x\mape d(\rvec x)=e^{-|m|r}$.
Their pointwise multiplication property
$\cl G\m\cl G\sub\cl G$ allows  - with Fourier 
transformation to (energy-)momentum Radon measures $\cl M(\R^n)$, 
e.g. $\R^3\ni\rvec q\mape\tilde d(\rvec q)=  {|m|\over(\rvec q^2+m^2)^2}$
  - the definition of a convolution  algebra
$\tilde{\cl G}* \tilde{\cl G}\sub \tilde{\cl G}$.

The transition from  Lie group re\-pre\-sen\-ta\-tion  coefficients
to those of the  Lie algebra  is effected
by derivations $(\p d)\o \ol d$ with respect to Lie algebra parameters,
e.g. for time $\D(1)$ by ${de^{imt}\over dt}e^{-imt}=im$
or for rotations $\SU(2)$
by $(\p^ae^{i|m|\rvec x\rvec\si})\o e^{-i|m|\rvec x\rvec\si}=
i|m|\si^bu(\rvec x)^a_b$.
The tangent   Lie algebra functions from orthogonally symmetric space  functions  
involve the corresponding  derivatives with the 2-sphere spread.
An important and for hyperbolic position characteristic 
example are the  Yukawa potentials. They arise as the orbit 
of the Coulomb potential $\om^{-1}(\rvec x)={2\over r}$
acted upon with  the  $\SO_0(1,3)$-states
\begin{eq}{rl}
\R^3\ni\rvec x\mape \rvec \p e^{-|m|r}&={\rvec x\over2}
{\p\over \p{r^2\over4}}e^{-|m|r}\cr 
-{1\over |m|}{\p\over \p{r^2\over4}}e^{-|m|r}&={2 e^{-|m|r}\over r}={2\over r}\m e^{-|m|r}\cr
\end{eq}{Tangent functions}
for a symmetric space 
are  obtained with   inverse derivatives $\om^{-1}$,
familiar as Green functions of differential equations.
They give distributions $\tilde\om^{-1}$ of the
translation characters,
e.g.  energy-momentum distributions for spacetime translations.
The action of { inverse derivative distributions}
upon the re\-pre\-sen\-ta\-tion coefficients defines the
 associated {tangent module}
 \begin{eq}{rl}
\om^{-1} :{\cl G}\map \log{\cl G},~~ d\mape
 {{\om}}=
\om^{-1}\m  d&\cr 
\tilde\om^{-1}: \tilde{\cl G}\map \log\tilde{\cl G},~~\tilde d\mape
\tilde {{\om}}=
 \tilde\om^{-1}*\tilde d&\cr 
\end{eq}Tangent functions,
 $q\mape \tilde {{\om}}(q)$, are defined with the same integration contour
as the re\-pre\-sen\-ta\-tion functions
$q\mape\tilde d(q)$.

The general situation with respect to group functions - always on
 space(-time) translations  $x\in\R^n$
or on (energy-)momenta $q\in\R^n$: The Fourier transformed Radon 
(energy-)momentum distributions $\cl M$ 
are $L^\infty$-func\-tions of space(-time) translations  - they are used for group 
re\-pre\-sen\-ta\-tion coefficients. 
The Fourier transformed space(-time) function classes $L^1$
 are a dense subspace of the  continuous  functions $\cl C_\infty$ of (energy-)momenta
which vanish for infinity\cite{TREV,FOL} 
 - they are used for Lie algebra 
re\-pre\-sen\-ta\-tion coefficients. 

$L^\infty$ is the dual of $L^1$.
$\cl C_\infty$ contains the compactly supported
functions $\cl C_c $ whereof the Radon  
distributions   are the dual. 

All those spaces 
with hydrogen ground state and
Yukawa potential as an example for 
re\-pre\-sen\-ta\-tion coefficients are summarized
 with their  Fourier transformation and
multiplicative properties - pointwise or convolutive -
as follows

\begin{eq}{c}
{\scriptsize \begin{array}{|c||c|c|c|}\hline
&\hbox{Fourier
transformation}&\hbox{(energy-)momentum}&\hbox{space(-time)}\cr
&&\hbox{functions}&\hbox{functions}\cr
\hline\hline
\hbox{for Lie group }G&
\cl M \map L^\infty,~\tilde d\rightarrow d
&\cl M *\cl M \sub \cl M
&L^\infty \m L^\infty \sub L^\infty \cr
\hbox{re\-pre\-sen\-ta\-tion coefficients}
&\int{d^3q\over\pi^2}~e^{i\rvec q\rvec x}{|m|\over(\rvec q^2+m^2)^2}=e^{-|m|r}
& \tilde d_1*\tilde d_2(q)&d_1\m d_2(x)\cr
\hline
\hbox{for Lie algebra }\log G&
\cl C_\infty\lmap L^1
,~\tilde\om \leftarrow\om
&\cl C_\infty \m\cl C_\infty \sub\cl C_\infty
&L^1 *L^1 \sub L^1\cr
\hbox{re\-pre\-sen\-ta\-tion coefficients}
&{1\over \rvec q^2+m^2}
=\int{d^3x\over 4\pi}~ e^{-i\rvec q\rvec x}{e^{-|m|r}\over r}

& \tilde \om_1\m\tilde \om_2(q)&
 \om_1*\om_2(x)\cr
\hline
\end{array}} \cr\cr

\begin{array}{lrcl}
\hbox{for Lie group:}&\cl M&\rightarrow&L^\infty\cr
&\cup\hskip1mm&&\hskip1mm\cup\cr
\hbox{for Lie algebra:}&L^1&\rightarrow&\cl C_\infty \cr
\end{array}\hbox{with}

\begin{array}{cl}
\cl M&\hskip-8mm=(\cl C_c )'\cr
\tilde \om^{-1}*\downarrow\hskip9mm&
\cr
\cl C_\infty&\hskip-8mm\supnoteq \cl C_c \cr\end{array}
\hbox{and}\begin{array}{cl}
 L^\infty&\hskip-8mm=(L^1)'\cr
 \om^{-1}\m \downarrow\hskip9mm&\hskip-8mm\cr
L^1&\hskip-8mm\cr\end{array}

\end{eq}

For a nonabelian noncompact group, 
the re\-pre\-sen\-ta\-tion coefficients $\log\cl G$ of the
Lie algebra $\log G$ describe interactions. 
Their Fourier transforms $\log\tilde{\cl G}$ constitute
 a subspace of the (energy-)momentum functions
$\cl C_\infty$
\begin{eq}{l}
\tilde{\cl G}\sub \cl M,~~\log \tilde{\cl G}=\tilde \om^{-1}*\tilde {\cl G}\sub\cl C_\infty\cr
\end{eq}$\log\tilde{\cl G}$ is
a  module for the  convolutive action with the group re\-pre\-sen\-ta\-tion distributions
$\tilde{\cl G}$
\begin{eq}{l}
\log\tilde{\cl G}*
\tilde{\cl G}\sub\log\tilde{\cl G}
\hbox{ with }
\tilde \om_1 *\tilde d_2= 
\tilde\om^{-1}*\tilde d_1 *\tilde d_2=\tilde \om_{1*2} 
\end{eq}A requirement of convolutive stability for 
(energy-)momentum tangent functions themselves or pointwise multiplicative
stability for their space(-time) dependent Fourier transforms  
does not make sense
as seen, e.g., in the pointwise 
multiplication of the off-shell contributions of
Feynman propagators (`divergencies') or in
the pointwise square ${e^{-2|m|r}\over r^2}$ of a 
Yukawa potential as tangent re\-pre\-sen\-ta\-tion coefficient which  is not used as basic 
potential. Its convolution square, however, 
makes sense as group re\-pre\-sen\-ta\-tion coefficient  
\begin{eq}{rcl}
\cl C_\infty(\R^3)\ni 
{|m|\over \rvec q^2+m^2}
&\lrmap& {e^{-|m|r}\over r}\in L^1(\R^3)\cr
{1\over \rvec q^2+m^2}\m {1\over \rvec q^2+m^2}
={1\over (\rvec q^2+m^2)^2}&\lrmap&
{e^{-|m|r}\over r}~*~{e^{-|m|r}\over r}= { 2\pi\over |m|}e^{-|m|r}
\end{eq}

\subsection{Tangent modules for abelian groups}

There is not much interaction for abelian groups:
With the isomorphy of the abelian group $\D(1)$
(1-di\-men\-sio\-nal future $\R_+$)
 and
 its Lie algebra $\R$ also the re\-pre\-sen\-ta\-tion 
 algebra $\tilde\cl D^1$ is isomorphic to its tangent  module
 $\log\tilde\cl D^1$. It is generated with the inverse derivative 
 $({d\over  dt})^{-1}\sim {1\over q}=\tilde \om^{-1}(q)$ 
\begin{eq}{c}
{1\over im}{d\over dt} e^{imt}=e^{imt}
=\oint {dq\over 2i\pi}{ 1\over q-m}e^{iqt}\cr\cr
{\scriptsize\begin{array}{|c|}\hline
\hbox{time }{1\over q}\stackrel1*\tilde {\cl D}^1\map\log\tilde {\cl D}^1\cr
\hbox{with }(\stackrel1*,q)=({*\over 2i\pi},q-io) \cr\hline 
\begin{array}{rcll}
{1\over q }&\stackrel1*&{ 1\over q-m}&={1\over q-m}\cr
{1\over q-m_1 }&\stackrel1*&{ 1\over q-m_2}&={1\over q-(m_1+m_2)}
\end{array}
\cr
\hline\end{array}}\cr
\end{eq}

The  hyperboloid  
$\cl Y^1$ (1-dimensional abelian position) with poles for dual invariants 
has the inverse  derivative distribution 
$({d\over d z})^{-1}\sim {q\over  q^2}=\tilde \om^{-1}(q)$
with the sign distribution as Fourier transform
\begin{eq}{c}
-{1\over |m|}{\p\over \p z}e^{-|mz|}
=\ep(z)e^{-|mz|}=
\int {dq\over\pi}~{iq\over q^2+m^2}e^{-iqz}
\cr\cr
{\scriptsize\begin{array}{|c|}\hline

\hbox{1-position }
{q\over q^2}\stackrel1*  \tilde {\cl Y}^1\map\log\tilde {\cl Y}^1\cr
\hbox{with }(\stackrel1*,q^2)=({*\over \pi},q^2+o^2)\cr\hline 
\begin{array}{cccl}
 {q\over q^2}
   &{\stackrel 1*}&{|m|\over q^2+m^2}
 &={q\over q^2+m^2}\cr  
    {q\over q^2+m_1^2}
 &{\stackrel 1*}&{|m_2|\over q^2+m_2^2}&={q\over q^2+m_+^2}\cr  
\end{array}\cr\hline
\end{array}}\cr

\end{eq}

The distributions of the abelian sphere $\Om^1\cong \SO(2)$ 
have a different integration contour
\begin{eq}{c}
\pm{1\over i|m|}{\p\over \p z}e^{\pm i|mz|}
=\ep(z)e^{\pm i|mz|}=
\pm\int {dq\over \pi}~{q\over q^2\mp io -m^2}e^{-iqz}
\cr\cr

{\scriptsize\begin{array}{|c|}\hline

\hbox{circle }
{q\over q^2}\stackrel1*\tilde {\Om }^1\map\log\tilde {\Om }^1\cr
\hbox{with }(\stackrel1*,q^2)=(\pm{*\over i\pi},q^2\mp io)\cr\hline 
\begin{array}{cccl}
 {q\over q^2}
   &{\stackrel 1*}&{|m|\over q^2-m^2}
 &={q\over q^2-m^2}\cr  
    {q\over q^2-m_1^2}
 &{\stackrel 1*}&{|m_2|\over q^2-m_2^2}&={q\over q^2-m_+^2}\cr  
\end{array}\cr\hline
\end{array}}\cr

\end{eq}

\subsection{Interactions of hyperbolic position}

The tangent functions (interactions) for 
 hyperbolic position $\cl Y^3$ 
 with $\rvec \p={\rvec x\over 2}{\p\over \p{r^2\over 4}}$
are the orbit of  
the inverse scalar derivative distribution 
$(\rvec \p^2)^{-1}\sim {1\over \rvec q^2}=\tilde \om^{-1}(q)$
with the Kepler factor (Coulomb and Newton potential) 
as Fourier transform. They are the Yukawa potentials
\begin{eq}{c}
-{1\over |m|}{\p\over \p{r^2\over 4}}e^{-|m|r}=
2{e^{-|m|r} \over r}=\int {d^3q\over\pi^2}
{1\over \rvec q^2+m^2}e^{-i\rvec q\rvec x}
\cr\cr
{\scriptsize\begin{array}{|c|}
\hline
\hbox{position }
{1\over \rvec q^2}\stackrel3*\tilde {\cl Y}^3
\map\log\tilde {\cl Y}^3\cr
\hbox{ with }\SO(3)\hbox{ and }(\stackrel3*,\rvec q^2)=
({*\over \pi^2},\rvec q^2+o^2)\cr\hline 
\begin{array}{ccll}
{1\over \rvec q^2}&\stackrel 3*&
{|m|\over (\rvec q^2+m^2)^2}&={1\over \rvec q^2+m^2}\cr
{1\over \rvec q^2+m_1^2}
&\stackrel 3*&
{|m_2|\over (\rvec q^2+m_2^2)^2}
&={1\over \rvec q^2+m_+^2}\cr
\end{array}\cr\hline
\end{array}}\cr
\end{eq}

The characteristic minimal nonabelian case is generalized to 
the odd dimensional hyperboloids  
\begin{eq}{c}
-{1\over |m|}{\p\over \p{r^2\over 4}}e^{-|m|r}=
2{e^{-|m|r}\over r}=
\int {2d^{2R-1}q\over |\Om^{2R-1}|}{1\over (\rvec q^2+m^2)^{R-1}}
e^{-i\rvec q\rvec x}\cr\cr

{\scriptsize\begin{array}{|c|}
\hline
\hbox{position }
{1\over (\rvec q^2)^{R-1}}\stackrel{2R-1}*
\tilde {\cl Y}^{2R-1}\map \log\tilde {\cl Y}^{2R-1}
,~R=2,3,\dots\cr
\hbox{with }\SO(2R-1)
\hbox{ and }(\stackrel{2R-1}*,\rvec q^2)=
({*~2\over |\Om^{2R-1}|},\rvec q^2+o^2)\cr\hline 
\begin{array}{ccll}
{1\over (\rvec q^2)^{R-1}}
&\stackrel{2R-1}*&
{|m|\over (\rvec q^2+m^2)^{R}}
&=
{1\over (\rvec q^2+m^2)^{R-1}}
\cr
{1\over (\rvec q^2+m_1^2)^{R-1}}
&\stackrel{2R-1}*&
{|m_2|\over (\rvec q^2+m_2^2)^{R}}
&=
{1\over (\rvec q^2+m_+^2)^{R-1}}
\cr
\end{array}\cr\hline
\end{array}}\cr
 
\end{eq}and with the real-imaginary transition  to  the odd dimensional  spheres
\begin{eq}{c}
\mp{1\over i|m|}{\p\over \p{r^2\over 4}}e^{\pm i|m|r}=
2{e^{\pm i|m|r}\over r}
=\pm\int {2d^{2R-1}q\over i|\Om^{2R-1}|}{1\over (\rvec q^2\mp io -m^2)^{R-1}}
e^{-i\rvec q\rvec x}\cr
\cr\cr
{\scriptsize\begin{array}{|c|}
\hline
\hbox{spheres }
{1\over (\rvec q^2)^{R-1}}\stackrel{2R-1}*
\tilde {\Om}^{2R-1}\map\log\tilde {\Om}^{2R-1}
,~R=2,3,\dots\cr
\hbox{with }\SO(2R-1)
\hbox{ and }(\stackrel{2R-1}*,\rvec q^2)=
(\pm{*~2\over i|\Om^{2R-1}|},\rvec q^2\mp io)\cr\hline 
\begin{array}{ccll}
{1\over (\rvec q^2)^{R-1}}
&\stackrel{2R-1}*&
{|m|\over (\rvec q^2-m^2)^{R}}
&=
{1\over (\rvec q^2-m^2)^{R-1}}
\cr
{1\over (\rvec q^2-m_1^2)^{R-1}}
&\stackrel{2R-1}*&
{|m_2|\over (\rvec q^2-m_2^2)^{R}}
&=
{1\over (\rvec q^2-m_+^2)^{R-1}}
\cr
\end{array}\cr\hline
\end{array}}\cr

\end{eq}


\subsection{Interactions of causal spacetime}

The convolutive action of  re\-pre\-sen\-ta\-tion coefficients
of 2-di\-men\-sio\-nal  spacetime 
on the   inverse derivative $\p^{-1}\sim {q\over q^2}=\tilde \om^{-1}(q)$  
\begin{eq}{l}
\int {d^2 q\over i\pi}{  q\over (q- io)^2}e^{iqx}=
  \vth(x_0)\pi x\de({x^2\over 4}),~~\left\{\begin{array}{rl}
    \int{dz\over2\pi}\int {d^2 q\over i\pi}~{  q\over (q- io)^2}e^{iqx}&=2\vth(t)\cr
\int{dt\over2\pi}\int {d^2q  \over i\pi}~{q\over  (q- io)^2}e^{iqx}&=\ep(z)\cr
\end{array}\right.
\end{eq}produces   pole structures for the
spacetime tangent functions with an additional  $\ze$-integration 
\begin{eq}{c}

{\scriptsize\begin{array}{|c|}\hline
\hbox{spacetime }{q\over q^2}\stackrel{2}*
\tilde {\cl D}^{2}\map\log\tilde {\cl D}^{2}\cr
\hbox{with }\SO_0(1,1)\hbox{ and }(\stackrel{2}*, q^2)=
({*\over 2i\pi},(q-io)^2)
\cr\hline 
\begin{array}{cccl}
{q\over q^2} &\stackrel {2}*& { 1 \over q^2-m^2}
&= 
\int_0^1 d\ze ~ { q \over \ze q^2-m^2  }
\cr{q\over q^2-m_1^2} &\stackrel {2}*& { 1 \over q^2-m_2^2}
&= 
\int_0^1 d\ze ~ {(1-\ze)~q \over \ze(1-\ze) q^2-\ze m_1^2 -(1-\ze) m_2^2  }\cr
\end{array}\cr\hline
\end{array}}\cr
\end{eq}The inverse derivative distribution for general 
even dimensional spacetime
has the generalized Coulomb force as position projection
\begin{eq}{rl}
\int {2d^{2R} q\over i|\Om^{2R-1}|}{  q\over (q- io)^2}e^{iqx}&=
  \vth(x_0)\pi x\Ga(R)\de^{(R-1)}(-{x^2\over 4})\cr
\int{|\Om^{2R-1}|~d^{2R-1}x\over
(2\pi)^{2R}}
\int {2d^{2R}q  \over i|\Om^{2R-1}|}~{q\over  (q- io)^2}
e^{iqx}
&=2\vth(t)\cr
\int{dt\over2\pi}\int {2d^{2R}q  \over i|\Om^{2R-1}|}~{q\over  (q- io)^2}
e^{iqx}
&={\rvec x\over r}{\Ga(2R-1)\over (r^2)^{R-1}}\cr
\end{eq}and the action on the re\-pre\-sen\-ta\-tion coefficients 
\begin{eq}{c}
{\scriptsize\begin{array}{|c|}\hline
\hbox{spacetime }
{q\over q^2}\stackrel{2R}*\tilde {\cl D }^{2R}\map \log\tilde {\cl D }^{2R}
,~R=1,2,3,\dots\cr
\hbox{with }\SO_0(1,2R-1)
\hbox{ and }(\stackrel{2R}*, q^2)=
(-{*(-1)^R\over i|\Om^{2R-1}|},(q-io)^2)
\cr\hline 
\begin{array}{cccl}
{ q \over q^2}
&\stackrel {2R}*&{1\over (q^2-m^2)^{R}}  &=
\int_0^1 d\ze ~ { (1-\ze)^{R-1}q \over \ze q^2-m^2  }
=\int_{m^2}^\infty {dM^2\over M^2}
({M^2-m^2\over M^2})^{R-1}{q\over q^2-M^2}\cr

 { q \over q^2-m_1^2} &\stackrel {2R}*&{1\over (q^2-m_2^2)^{R}}
&= \int_0^1 d\ze ~ { (1-\ze)^Rq \over \ze(1-\ze) q^2-\ze m_1^2  -(1-\ze)m_2^2 }\cr
{q~\over q^2-m_1^2}
&\stackrel {2R}*&
{2q~R\over (q^2-m_2^2)^{1+R}}
&=-{\p\over\p q}\ox q~
\int_0^1 d\ze  
{(1-\ze)^R\over 
\ze(1-\ze)q^2-\ze m_1^2 -(1-\ze)m_2^2}\cr

\end{array}\cr\hline
\end{array}}\cr
\end{eq}The time projection displays the embedded
time translation re\-pre\-sen\-ta\-tions, the
position projection the   Yukawa forces
\begin{eq}{rl}
\int {2d^{2R}q  \over i|\Om^{2R-1}|}~{q\over  (q- io)^2-m^2}
e^{iqx}
&=\pi x\Ga(R)({\p\over\p {x^2\over 4}})^R
\vth(x)\cl J_0(|mx|)\cr
\int{|\Om^{2R-1}|~d^{2R-1}x\over
(2\pi)^{2R}}
\int {2d^{2R}q  \over i|\Om^{2R-1}|}~{q\over  (q- io)^2-m^2}
e^{iqx}
&=\vth(t)2\cos mt\cr
\int{dt\over2\pi}\int {2d^{2R}q  \over i|\Om^{2R-1}|}~{q\over  (q- io)^2-m^2}
e^{iqx}
&={\rvec x\over 2|m|}\Ga(R)(-{\p\over \p{r^2\over 4}})^R e^{-|m|r}\cr
\end{eq}

\section{Projective energy-momenta}

The (energy-)momentum dependent 
tangent functions $\log\tilde{\cl G}\sub\cl C_\infty$
with space(-time) functions $\log{\cl G}\sub L^1$
 can be multiplied 
$\log\tilde{\cl G}\m\log\tilde{\cl G}\sub\log\tilde{\cl G}$
and convoluted  $\log{\cl G}*\log{\cl G}\sub\log{\cl G}$.

A  tangent function $q\mape\tilde\om(q)$
is  {projection valued at $q_0$}
if it has value 1 for this (energy-)momentum, called 
a projective point of $\tilde\om$    
\begin{eq}{l}
 \tilde\om \in
 \log\tilde{\cl G}
\hbox{ projective at $q_0$}
\iff \tilde{\om}(q_0)=\bl 1\cr
\hbox{with }\tilde\om=\plint d^n q~\tilde \om(q)~
\rstate q\lstate q \hbox{ and }\sprod{q'}q=\de(q-q')
 
\end{eq}Projection valued tangent functions will be used 
to define dual pairs 
of re\-pre\-sen\-ta\-tion and inverse derivative  distributions
and for eigenvalue equations of re\-pre\-sen\-ta\-tion invariants.

\subsection
{Duality of group and Lie algebra re\-pre\-sen\-ta\-tions}

The invariants of a group re\-pre\-sen\-ta\-tion coincide with those of 
the corresponding  Lie algebra re\-pre\-sen\-ta\-tion.
A pair with a corresponding group and inverse derivative distribution
defines a projector.

The re\-pre\-sen\-ta\-tion algebra 
$\cl G\sub L^\infty=(L^1)'$ is 
dual to the   tangent  module $\log \cl G\sub L^1$.
 The dual product of  re\-pre\-sen\-ta\-tion coefficients
of the  symmetric space $G/H$ and
its  tangent space is  the convolution  at 
trivial  (energy-)momentum $q=0$
 \begin{eq}{lll}
\log\cl G\x\cl G\map\C,&\dprod {{\om}} d=\int d^n x~{\om}(x)d(x)=\cr
\log\tilde{\cl G}\x\tilde{\cl G}\map\C,\hskip10mm &
\dprod {\tilde{{\om}}}{\tilde d }
=\int d^n q~ \tilde {{\om}}(-q)\tilde d (q)
=\tilde {{\om}} * \tilde d (0)\cr
\end{eq}

A {re\-pre\-sen\-ta\-tion distribution
is   dual to the inverse derivative  distribution} $\tilde\om^{-1}$ 
if it has a  projective point for trivial (energy-)momentum
$q=0$ 
\begin{eq}{l}
\hbox{ projective at $q_0=0$}\iff
\hbox{ dual pair }
(\tilde\om^{-1},\tilde d)\iff
\tilde \om(0)=\tilde\om^{-1}*\tilde d(0)=1\cr
\end{eq}

A dual pair of re\-pre\-sen\-ta\-tions for time and nonabelian 
hyperboloids  and spheres  consists of 
one basic re\-pre\-sen\-ta\-tion distribution 
and an inverse derivative  distribution
with the same  continuous real invariant as singularity
and normalization resp.
\begin{eq}{rl}
\hbox{abelian $\D(1)$:}~~&
-{m\over q}\stackrel1*{1\over q-m}
=-{m\over q-m}\stackrel {q=0}=1\cr\cr

\left.\begin{array}{l}
\hbox{$\cl Y^1$ with $\ep=+1$}\cr
\hbox{$\Om^1$ with $\ep=-1$}\cr
\end{array}\right\}
&
{q\ep|m|\over q^2}\stackrel1*{q\over q^2+\ep m^2}
={\ep m^2\over q^2+\ep m^2}\stackrel {q^2=0}=1\cr\cr
\left.\begin{array}{r}
\hbox{hyperboloids $\cl Y^{2R-1}$, $\ep=+1$}\cr
\hbox{spheres $\Om^{2R-1}$,  $\ep=-1$}\cr
\hbox{for }R=2,3,\dots\cr
\end{array}\right\}
&
{(\ep m^2)^{R-1} \over (\rvec q^2)^{R-1}}
\stackrel{2R-1}*{|m|\over (\rvec q^2+\ep m^2)^{R}}
={(\ep m^2)^{R-1}\over (\rvec q^2+\ep m^2)^{R-1}}\stackrel {\rvec q^2=0}=1\cr

\end{eq}

\subsection{The ratio of the invariant masses for spacetime}

In the dual product for 
 homogeneous   even dimensional spacetimes 
$\cl D^{2R}$,
starting with  Cartan spacetime $\D(1)\x\SO_0(1,1)$
and Minkowski spacetime $\D(2)\cong \GL(\C^2)/\U(2)$,
the residual vector  spacetime  re\-pre\-sen\-ta\-tion distribution 
  involves  {two continuous real invariants}
  for real rank 2
in contrast to
the time and position re\-pre\-sen\-ta\-tions with only one invariant. 
The spacetime inverse derivative  distribution 
${q|m_0|\over q^2}$ comes with the
intrinsic mass unit $m_0^2$ of the residual re\-pre\-sen\-ta\-tion   
of homogeneous spacetime with $R=1,2,\dots$ 
\begin{eq}{rl}
{2d^{2R}q\over (q^2-m_\ka ^2)^R}~{ q|m_0|\over  q^2-m_0^2}
&\mape 
{2d^{2R} 
q\over (q^2-\ka^2)^R}~{q\over  q^2-1}\hbox{ with }\ka^2={m_\ka ^2\over
m_0^2}\cr
{2\over (q^2-\ka^2)^R}~{q\over  q^2-1}
&= \int_{\ka^2}^1 {d\eta^2\over 1-\eta^2} ~
\({1-\eta^2\over 1-\ka^2}\)^{R}~{2q~R\over 
(q^2-\eta^2)^{1+R}}\cr
&=-{\p\over \p  q}\int_0^1 d\xi ~
{(1-\xi)^{R-1}\over 
[q^2-\xi -(1-\xi)\ka^2]^R}\cr

\end{eq}The invariant singularities in $q^2-\xi m_0^2 -(1-\xi)m_\ka^2$
are on a finite line 
with $\xi\in[0,1]$.

Compatible with the finite integration and in contrast to the energy and momentum pole functions 
for time and position, 
the  residual product  of the even dimensional  spacetime
re\-pre\-sen\-ta\-tion and  the inverse derivative
distribution
 \begin{eq}{l}
{q\over q^2}~\stackrel{2R}*~ {2\over (q^2-\ka^2)^R}~~{q\over q^2-1}
={\p\over\p q}\ox q~\tilde\om^0
_{2R}(q^2,\ka^2)\cr
\end{eq}does not produce a rational complex function
with a $q^2$-pole, but a finite integration over the square
 $(\ze,\xi)\in[0,1]^2$ for a pole distribution with 
$\ze q^2-\xi m_0^2-(1-\xi)m_\ka^2$
 \begin{eq}{rl}
\tilde\om^0 _{2R}(q^2,\ka^2)&=
-\int_{\ka^2}^1 {d\eta^2\over 1-\eta^2}\({1-\eta^2\over 1-\ka^2}\)^{R}
\int_0^1d\ze~{(1-\ze)^{R-1}
\over \ze q^2-\eta^2}\cr
&=-\int_0^1 d\xi
\int_0^1d\ze~{(1-\xi)^{R-1}(1-\ze)^{R-1}
\over \ze q^2-\xi-(1-\xi)\ka^2}\cr
\end{eq}

The  inverse derivative
distribution  and the residual spacetime re\-pre\-sen\-ta\-tion function
are  dual to each other by fulfilling the condition for the 
mass ratio $\ka^2$ in the scalar causal measure
 \begin{eq}{l}
\tilde\om^0 _{2R}(0,\ka^2)=-{1\over R}\log_{R}\ka^2=1\cr
\end{eq}The {$R$-tails of the  logarithm}
in the $(m_0^2-m_\ka ^2)\sim (1-\ka^2)$-expansion
\begin{eq}{rl}
\R_+\ni \ka^2\mape~~\log_{R}\ka^2&
=-{1\over (1-\ka^2)^R}\int_{\ka^2}^1 
d\eta^2~ {(1-\eta^2)^{R-1}\over \eta^2}
\hbox{ for }R=1,2,\dots\cr

&={1\over(1-\ka^2)^R}[\log \ka^2+{\SUM_{k=1}^{R-1}}{(1-\ka^2)^k\over k}]\cr
&=-{\SUM_{k=R}^{\infty}}{(1-\ka^2)^{k-R}\over k}
\end{eq}increase monotonically  for $\ka^2<1 $ 
\begin{eq}{l}

\ka^2\in(0,1):~~{d\over d\ka^2}\log _R\ka^2>0,~~\left\{\begin{array}{rlcl}
{\ka^2\ll1}:&\log_{R}\ka^2&\sim &\log \ka^2+ {\SUM_{k=1}^{R-1}}{1\over k}\cr
{\ka^2=1}:&\log_R1&=&-{1\over R}\cr\end{array}\right.
\end{eq}For Cartan spacetime without rotation degrees 
both invariants coincide
\begin{eq}{l}
R=1:~~-1=\log_1\ka^2={\log\ka^2\over 1-\ka^2}\then \ka^2={m_\ka ^2\over m_0^2}=1 
\end{eq}For  spacetimes $R\ge2$ with nontrivial rotations 
the mass ratio goes with the exponential of the spacetime rank
\begin{eq}{rll}

  R=2:&-2= {\log\ka^2+ 1-\ka^2\over (1-\ka^2)^2}

&\then\ka^2={m_\ka ^2\over m_0^2}\sim e^{-3}\sim{1\over 20.1}\cr
R=2,3,\dots:

&-R=\log_R\ka^2&\then
 \ka^2={m_\ka ^2\over m_0^2}
 \sim\exp[-R - {\SUM_{k=1}^{R-1}}{1\over k}] \hbox{ for }\ka^2\ll1\cr
 \end{eq}

\section
{Product invariants and masses}

Starting from one  defining re\-pre\-sen\-ta\-tion
for time, position or spacetime,
its convolution products define  product re\-pre\-sen\-ta\-tions.
The projective points  of 
the tangent (energy-)momenta functions give the translation invariants.
To motivate the general concept of a 
{polynomial  re\-pre\-sen\-ta\-tion algebra}
with its {associated tangent  module} and
the related {eigenvalue equations},  these structures 
are exemplified  first for
the abelian time translations $\D(1)\cong\R$.  

\subsection{The linear spectrum for time translations}

The re\-pre\-sen\-ta\-tion 
energy distribution ${1\over q-m}\in\tilde{\cl D}^1$ for  time 
translations $\R\ni t\mape e^{imt}$ with a frequency $\R\ni m\ne0$ 
generates, by the  convolution powers, a polynomial  re\-pre\-sen\-ta\-tion algebra 
\begin{eq}{l}
\tilde \cl D^1(m):~\{
\({1\over q-m}\)^{*k}=
{\scriptsize\underbrace{{1\over q-m}\stackrel 1*\cdots\stackrel 1*{1\over q-m}}
_{k~\rm{times}}}
={1\over q-km}\mid k=1,2,\dots\}
\end{eq}The poles give the  invariants for the power re\-pre\-sen\-ta\-tions  
$t\mape e^{ikmt}$  - the equidistant oscillator energies.

The inverse time derivative
as derivative distribution 
is dual  to the generating re\-pre\-sen\-ta\-tion
\begin{eq}{l}
-{m\over q}\stackrel 1 *{1\over q-m}=-{m\over q-m}
\stackrel{q=0}=1
\end{eq}Its action on the polynomial re\-pre\-sen\-ta\-tion algebra
defines the associated tangent module
\begin{eq}{l}
\log\tilde \cl D^1(m):~~\{-{m\over q} \stackrel 1*{1\over q-km}
=-{m\over q-km}\mid k=1,2,\dots\}
\end{eq}

The eigenvalues  as  invariants
for the product re\-pre\-sen\-ta\-tions
are the projective energy  points of the tangent functions, i.e. the 
solutions of the eigenvalue equations 
\begin{eq}{l}
k=1,2,\dots:~~-{m\over q-km}=1\then q=m_{k-1}=(k-1)m=0,m,2m,\dots
\end{eq}

\subsection
{Eigenvalue equations for  product invariants}

In general, a  re\-pre\-sen\-ta\-tion (energy-)momentum distribution 
  $\tilde d(m)\in\tilde {\cl G}$
with invariants $m$  generates, by its convolution  powers (involving tensor powers),
the associated {polynomial 
re\-pre\-sen\-ta\-tion algebra}
\begin{eq}{l}
\tilde{\cl G}(m):~~
\{\tilde d^{*k}(m)\mid \tilde d^{*k}(m,q)=
\underbrace{\tilde d(m)*\cdots*\tilde d(m)}
_{k~\rm{times}}(q),~ k=1,2,\dots\}\cr
\end{eq}The inverse derivative distribution, dual to $\tilde d(m)$  
\begin{eq}{l}
\tilde \om^{-1}(m)*\tilde d(m)(q)\stackrel{q=0}=\bl 1
\end{eq}defines the associated { tangent  module}  
 \begin{eq}{l}
\tilde\om^{-1}(m)*\tilde{\cl G}(m)= \log\tilde{\cl G}(m):~~
\{\tilde\om^{k-1}(m)=\tilde\om^{-1}(m)*\tilde d^{*k}(m)\mid
k=1,2,\dots\}
 \end{eq}
  
The invariants for the  power re\-pre\-sen\-ta\-tions
are the projective (energy-)momenta of the tangent functions
given by the solutions of the  eigenvalue equations 
\begin{eq}{l}
k=1,2,\dots:~~
\tilde \om ^{k-1}(m,q)
=\bl 1 \then q=m_{k-1}\cr
\end{eq}


\subsection{The  quadratic spectrum for hyperbolic position}

For  position $\cl Y^3\cong\SO_0(1,3)/\SO(3)$,
the polynomial re\-pre\-sen\-ta\-tion algebra  
is visible in the  nonrelativistic
hydrogen bound states.
The   Coulomb
potential is the dual tangent function 
for the  $\SO_0(1,3)$-state $\rvec x\mape e^{-|m|r}$
\begin{eq}{rl}
 d_0(\rvec x)&=e^{-|m|r}=\int {d^3 q\over \pi^2}~
{|m|\over(\rvec q^2+m^2)^2}e^{-i\rvec q\rvec x}\cr
\om^{-1}(\rvec x)&=-|m|{\p\over \p{r^2\over4}}d_0(\rvec x)=2{m^2\over r}=
\int {d^3 q\over \pi^2}~{m^2\over\rvec q^2}e^{-i\rvec q\rvec x}\cr
\then
\tilde\om^{-1}\stackrel3*\tilde d_0(q)&={m^2\over\rvec q^2}
\stackrel3*
{|m|\over(\rvec q^2+m^2)^2}
={m^2\over \rvec q^2+m^2}\stackrel{\rvec q^2=0}= 1=\int_0^\infty r^2
dr~{m^2\over r}e^{-|m|r}
\end{eq}The Hamiltonian $H={\rvec p^2\over 2}-{|m|\over r}$ relates to each other
 the $\SO_0(1,3)$-invariants 
   for position $\cl Y^3$ and the 
 time translation invariants (binding energies).
The time translation action can be written with  
the convolution product of the  inverse derivative distribution 
and the Fourier transformed  state  
$\tilde d_0$
\begin{eq}{rl}
2H d_0(\rvec x)=2E_0 d_0(\rvec x)\iff& 
(\rvec q^2 -{|m|\over \rvec q^2}
\stackrel3*)
\tilde d_0=2E_0\tilde d_0\cr
\iff& (\rvec q^2
-{|m|\over \rvec q^2}
\stackrel3*){|m|\over (\rvec q^2+m^2)^2}
={\rvec q^2|m|\over (\rvec q^2+m^2)^2}-{|m|\over \rvec q^2+m^2}
=2E_0{|m|\over (\rvec q^2+m^2)^2}\cr
\then&
2E_0=-m^2\cr
\end{eq}

The  polynomial re\-pre\-sen\-ta\-tion algebra
for general odd dimensional nonabelian hyperboloids 
 and spheres with the states $\rvec x\mape (e^{-kr},e^{\pm ikr})$ and 
intrinsic unit
\begin{eq}{l}
\tilde{\cl Y}^{2R-1},~\tilde{\Om}^{2R-1}:~~
\{\({1\over(\rvec q^2+\ep)^R}\)^{*k}
={k\over(\rvec q^2+\ep k^2)^R}\mid k=1,2,\dots\}\cr
\hbox{with } \ep=\pm 1 \hbox{ and }R=2,3,\dots
\end{eq}gives - via the convolutive  action -
the momentum dependent tangent  functions
(from Kepler potential 
to Yukawa potentials 
and spherical waves)
\begin{eq}{rl}
\log\tilde{\cl Y}^{2R-1},~~\log\tilde{\Om}^{2R-1}:&
 \hfill\{{2\over r}\m (e^{-kr},e^{\pm ikr})=
   2{(e^{-kr},e^{\pm ikr})\over r}\mid k=1,2,\dots\}\cr 

&\{{\ep^{R-1}\over (\rvec q^2)^{R-1}}\stackrel{2R-1}*
{k\over(\rvec q^2+\ep k^2)^{2R-1}}
={\ep^{R-1}\over (\rvec q^2+\ep k^2)^{R-1}}
\mid k=1,2,\dots\}\cr
\end{eq}

The  eigenvalue equations for the projective momenta $\rvec q$
 give the invariants
for the re\-pre\-sen\-ta\-tions of 
the Lorentz group $\SO_0(1,2R-1)$ as used for $\cl Y^{2R-1}$
and of the rotation group $\SO(2R)$ as used for $\Om^{2R-1}$  
\begin{eq}{rl}
{1\over (\ep\rvec q^2+k^2)^{R-1}}=1
\then \ep \rvec q^2=
-k^2+1&=-4J(1+J)=0,-3,-8\cr
\hbox{for }k&=1+2J=1,2,3,\dots
\end{eq}

Via the Hamiltonian, the  $\SO_0(1,3)$-invariants
can be transmuted into  energy eigenvalues
\begin{eq}{l}
2E_J=-{m^2\over k^2},~~k=1+2J=1,2,\dots
\end{eq}

\subsection{Invariants of spacetime translations - the mass zero solution}

The residual  re\-pre\-sen\-ta\-tion
of $2R$-dimensional  spacetime $\cl D^{2R}$, $R=1,2,\dots$
with the $\D(1)$-pole generates, by its convolution powers, the 
re\-pre\-sen\-ta\-tion algebra
\begin{eq}{l}
\tilde \cl D^{2R}(\ka^2):~~\{
\left(\ka^{2R}(q)~{q\over q^2-1}\right)^{*k}
\mid k=1,2,~\dots\}\hbox{ with }\ka^{2R}(q)={2\over (q^2-\ka^2)^R}
\end{eq}The mass ratio $\ka^2={m_\ka ^2\over m_0^2}$ is determined 
by $\log_R\ka^2=-R$
from duality
with the spacetime translation distribution ${q\over q^2}$.

The  convolution powers of the residual  spacetime re\-pre\-sen\-ta\-tion
have as associated energy-momentum tangent functions 
\begin{eq}{l}
\log\tilde \cl D^{2R}(\ka^2):~~
\{\tilde\om _{2R}^{k-1}(q^2,\ka^2)=
{q\over q^2}\stackrel {2R}*
\left( \ka^{2R}(q)~{q\over q^2-1}
\right)^{*k}

\mid k=0,1,2,\dots\}\cr

\end{eq}There arises the nonabelian causal convolution
for even dimensional spacetime
where the full energy-momentum dependent scalar causal measure 
is included
\begin{eq}{rl}
\tilde\om _{2R}^{-1}(q^2,\ka^2)&={q\over q^2}\cr
\tilde\om _{2R}^{k-1}(q^2,\ka^2)&=
{q\over q^2}\ccon {q \over q^2-1}\ccon\cdots\ccon {q \over q^2-1}\cr
\hbox{with }
\ccon&\cong {i\over |\Om^{2R-1}|}\de(q_1+q_2-q)~{2\over (-q_2^2+\ka^2)^R}

\end{eq}

The solutions of the eigenvalue equations 
\begin{eq}{l}
k=0,1,2,\dots:~~\tilde\om ^{k-1}_{2R}(q^2,\ka^2)=\bl 1
\end{eq}give invariants
of tangent spacetime translations. For a 
positive residual normalization (below) they describe particle masses
in Poincar\'e group re\-pre\-sen\-ta\-tions. 

The simplest nontrivial eigenvalue equation  for $k=1$
 is decomposable with the
two nondecomposable projectors $\{\bl1_{2R}-{q\ox q\over q^2},{q\ox q\over q^2}\}$
\begin{eq}{rl}
{q\over q^2}\ccon
{q\over q^2-1}
&={q\over q^2}\stackrel {2R}*
{2\over (q^2-\ka^2)^R}~{q\over q^2-1}
=
[\bl1_{2R}+q\ox  q~2 {\p\over\p q^2}]~\tilde\om^0 _{2R}(q^2,\ka^2)\cr
&=(\bl1_{2R}-{q\ox q\over q^2})~\tilde\om^0 _{2R}(q^2,\ka^2)
+{q\ox q\over q^2}(1+2 q^2{\p\over\p q^2})~\tilde\om^0 _{2R}(q^2,\ka^2)\cr

\end{eq}e.g. for the rotation free case
\begin{eq}{l}
R=1:~~\tilde\om^0 _{2}(q^2,\ka^2)
=-\int_{\ka^2}^1 
{d\eta^2\over 1-\ka^2}~
\int_0^1{ d\ze\over \ze q^2-\eta^2}
\stackrel{\ka^2=1}=-
\int_0^1{ d\ze\over \ze q^2-1}=-{\log(1-q^2)\over q^2} 
\cr
\end{eq}

The equation 
\begin{eq}{l}
\hbox{for } q^2=0:~~\tilde\om^0_{2R}(q^2,\ka^2)=\bl 1_{2R}
\end{eq}has been used above as duality 
 condition to determine the ratio 
 $\ka^2$ of the spacetime invariants. It is also interpretable 
 as 
 eigenvalue equation
 having as solution a trivial  invariant $q^2=m^2=0$.
The  eigenvalue $m^2=0$ for the vector field in Minkowski spacetime
will be related to  massless vector fields with their residual normalization
(below) as gauge coupling constant.

\section
{Normalization of translation re\-pre\-sen\-ta\-tions}

Starting from a generating  re\-pre\-sen\-ta\-tion, the
residue of a product re\-pre\-sen\-ta\-tion defines its normalization.
For spacetime, the determination of the residues requires the transition from
inverse derivative  energy-momentum distributions
to the associated distributions for the re\-pre\-sen\-ta\-tions
of the spacetime translations.  

The exponential  from the Lie algebra $\R$ (time translations) to
the group $\exp \R=\D(1)$
can be reformulated in the  language 
of residual re\-pre\-sen\-ta\-tions with energy functions
by a {geometric series}
\begin{eq}{rl}
 e^{ im t}
&=\oint {dq \over 2i\pi }{1\over  q-m } e^{i qt}\cr
={\SUM_{k=0}^\infty}{(im t)^k\over k!}
&=\oint {dq\over 2i\pi }{1\over q} 
{\SUM_{k=0}^\infty}({m\over q})^k e^{i qt}
\end{eq}$-{m\over q}$ is the inverse derivative energy function for
the re\-pre\-sen\-ta\-tion function ${1\over q-m}$.

\subsection 
{Geometric transformation and Mittag-Leffler sum}

Translation re\-pre\-sen\-ta\-tions are characterized by (energy-)momentum distributions
with simple poles.
Meromorphic complex functions have only pole singularities. In
the compactified complex  plane $\ol\C$ they constitute the field of  rational
functions. The  re\-pre\-sen\-ta\-tion distributions
for one dimension (pole functions)  
have negative degree 
\begin{eq}{l}
\ol \C\ni q\mape \rho(q)= 
{P^n(q)\over P^k(q)}={\al_0+\al_1q+\dots+\al_nq^n\over
\ga_0+\ga_1q+\dots+\ga_kq^k}\in\ol \C,~~
\al_j,\ga_j\in\C,~\ga_k\ne 0,~~k\le n
\end{eq}

The geometric  transformations for $\D(1)$ (time)
with $z={q\over m}$
\begin{eq}{l}
z\mape {1\over z}=\tilde\om(z)\mape {\tilde\om(z)\over 1-\tilde\om(z)}
={1\over z-1}
\end{eq}are  elements  of the broken rational (conformal)
 bijective transformations of the closed complex plane 
\begin{eq}{l}
\ol\C\ni z\mape {\al z+\be\over \ga z+\de}\in\ol\C
\end{eq}with real
 coefficients  as group isomorphic to 
\begin{eq}{l}
g={\scriptsize \pmatrix{\al&\be\cr\ga&\de\cr}}\in
 \SL(\R^2)\cong \SU(1,1)\sim \SO_0(1,2)
 \end{eq}For $\det g=1$
upper and lower half plane $x\pm io$ remain stable.
The eigenvalue $\tilde\om(z_0)=1$ becomes a pole 
\begin{eq}{l}
{\scriptsize \pmatrix{\al&\be\cr\ga&\de\cr}}
={\scriptsize \pmatrix{1&0\cr -1&1\cr}} :~~
\tilde\om\mape {\tilde\om\over 1-\tilde\om},~~(1,0)\mape(\infty,0)
\end{eq}With one fixpoint $\tilde\om=0$
 the transformation is parabolic, i.e.
an element of the $\R$-isomorphic subgroup 
${\scriptsize \pmatrix{1&0\cr\ga&1\cr}}$.
 
The geometric transformation will be generalized in order
to associate   functions with pole singularities
to the complex eigenvalue functions $\tilde\om(z)$ for spacetime 
\begin{eq}{l}
z\mape \tilde\om(z)\mape {\tilde\om(z)\over 1-\tilde\om(z)}
\end{eq}An eigenvalue $z_0\in\{z\mid \tilde\om(z)=1\}$
gives a pole. If the zero $z_0$  
 is simple with $\tilde\om$ holomorphic there, 
 it defines, by geometric transformation of its
Taylor series, a {Laurent series}\cite{BESO} and a 
 residue 
 \begin{eq}{rcl}
\tilde\om(z)=&1+(z-z_0)\tilde\om'(z_0)&+{\SUM_{k=2}^\infty}
{(z-z_0)^k\over k!}\tilde\om^{(k)}(z_0)\cr
{\tilde\om(z)\over 1-\tilde\om(z)}=&
{\res_{}(z_0)\over z-z_0}&+{\SUM_{k=0}^\infty}
(z-z_0)^ka_k(z_0)\cr
{\res_{}}(z_0)=&-{1\over \tilde\om'(z_0)}\hfill&
\end{eq}Each eigenvalue $\{z_k\mid \tilde\om(z_k)=1\}$ has its own 
principal part. Their sum, called  {Mittag-Leffler sum},
replaces the simple pole for $\D(1)$
\begin{eq}{rll}
z\mape \tilde\om(z)&\mape {\tilde\om(z)\over 1-\tilde\om(z)}
&= {\SUM_{z_k}}{\res(z_k)\over z-z_k}+\dots\cr
\hbox{generalizing }z\mape {1\over z}&\mape {{1\over z}\over 1-{1\over z}}&= 
{1\over z-1}
\end{eq}

Therewith one obtains
for an eigenvalue function  for  spacetime $\cl D^{2R}$
and its  projectors at the invariant solutions
\begin{eq}{l}
\tilde\om(q^2)=\tilde\om^{-1}*\tilde d(q^2)=\bl 1\then q^2\in\{m^2\}
\end{eq}the transition to   complex  re\-pre\-sen\-ta\-tion functions $\tilde \cl G_0$ 
- assumed with simple poles 
\begin{eq}{l}
\tilde{\cl G}\map\log\tilde{\cl G}\map \tilde{\cl G}_0 ,~~~
 \tilde d\mape \tilde \om(q^2)\mape {\tilde\om(q^2)\over
  \bl 1- \tilde\om(q^2) }
= {\SUM_{m^2}}{{\res_{}}(m^2)\over q^2-m^2}+\dots\cr
\end{eq}The residue is the negative inverse of the derivative of
the energy-momentum tangent function at the invariant
\begin{eq}{l}
\tilde\om(q^2)=\bl 1+(q^2-m^2)
{\p\tilde\om\over \p q^2}(m^2)
+\dots\then {\res_{}}(m^2)=-{1\over {\p\tilde\om\over \p q^2}(m^2)}
\cr
\end{eq}A simple pole with positive normalization  can be used 
for the  re\-pre\-sen\-ta\-tion of the Poincar\'e group  $\SO_0(1,2R-1)\sx\R^{2R}$.
The residue gives the normalization of the
associated  re\-pre\-sen\-ta\-tion
\begin{eq}{l}
\hbox{on-shell part }{i\over\pi}{{\res_{}}(m^2)\over q^2+io -m^2}
={\res_{}}(m^2) \de(q^2 -m^2)
\end{eq}

\subsection{Gauge coupling constants as residues at mass zero}

In the residual product  
of the spacetime re\-pre\-sen\-ta\-tion with the dual inverse derivative
\begin{eq}{rl}
{q\over q^2}\ccon
{q\over q^2-1}
&=
[\bl1_{2R}+q\ox  q ~2{\p\over\p q^2}]~\tilde\om^0 _{2R}(q^2,\ka^2)\cr
{\p\tilde\om^0 _{2R}\over\p q^2}(q^2,\ka^2)&=\int_{\ka^2}^1 
{d\eta^2\over 1-\eta^2}~\({1-\eta^2\over 1-\ka^2}\)^{R}
\int_0^1d\ze~{\ze(1-\ze)^{R-1}
\over (\ze q^2-\eta^2)^2}\cr
\end{eq}the residual normalization ${\res_{}}(0,\ka^2)$
for the massless solution
$\tilde\om^0 _{2R}(0,\ka^2)=1$ is given 
by the inverse of the  negative derivative of the
eigenvalue function there
\begin{eq}{rl}
-{1\over {\res_{}}(0,\ka^2)}=
{\p\tilde\om^0 _{2R}\over\p q^2}(0,\ka^2)
&={1\over R(R+1)(1-\ka^2)^R}\int_{\ka^2}^1 
d\eta^2{(1-\eta^2)^{R-1}\over \eta^4}\cr
&={1\over R(R+1)}[{1\over\ka^2}+(R-1)\log_{R}\ka^2] \cr
&=
{1-R(R-1)\ka^2\over R(R+1)\ka^2}~~
\hbox{ with }-{1\over R}\log_R\ka^2=1\cr
\end{eq}With a small mass ratio
\begin{eq}{l}
\hbox{ for }
 \ka^2\ll1:~~
- {\res_{}}(0,\ka^2)
\sim  R(R+1)\ka^2
\sim
 R(R+1)\exp[-R - {\SUM_{k=1}^{R-1}}{1\over k}]
 \end{eq}one has the numerical values  for Cartan and Minkowski spacetime
\begin{eq}{rl}
-\res(0,\ka^2)&\sim R(R+1){m_\ka ^2\over m_0^2}
=\left\{\begin{array}{rll}
 2{m_\ka ^2\over m_0^2}&=2,&R=1\cr
6{m_\ka ^2\over m_0^2-2m_\ka ^2 }&\sim
{6\over e^3-2}\sim {1\over3},&R=2\cr\end{array}\right.\cr
\end{eq}

With the  geometric  transformation
the Laurent series  gives
 an ener\-gy-mo\-men\-tum 
 distribution for a spacetime translation re\-pre\-sen\-ta\-tion
with invariant   zero 
and residual normalization. 
With appropriate integration contour, it can be  used as
propagator for a mass zero spacetime vector field
 with coupling constant $-{\res_{}}(0,\ka^2)$
 \begin{eq}{l}
\SO_0(1,2R-1)\sx\R^{2R}:
\hbox{ on-shell part }
{i\over\pi} {\eta_{jk}\res(0,\ka^2)\over q^2+io }=\eta_{jk} \res (0,\ka^2) \de(q^2)
\end{eq}This vector field has, in addition to an $\SO_0(1,1)$-related 
  pair with
 neutral signature,  $2R-2$  particle interpretable
  degrees of freedom which are related to the spherical  
  degrees of freedom $\Om^{2R-2}\subnoteq\cl D^{2R}$
  and the compact fixgroup $\SO(2R-2)$ in the massless particle 
  fixgroup $\SO(2R-2)\sx\R^{2R-2}$. 
  Those degrees of freedom have
   a positive scalar product  
 \begin{eq}{l}
 -\eta_{jk}=\left\{\begin{array}{rll}
 {\scriptsize\pmatrix{-1&0\cr0&\bl 1_{2R-1}\cr}}
 &\cong {\scriptsize\pmatrix{0&0&1\cr 
 0&\bl1_{2R-2}&0\cr1&0&0\cr}}
 &\hbox{for }\SO_0(1,2R-1)\sx\R^{2R}\cr  
 {\scriptsize\pmatrix{-1&0\cr0&1\cr}}&\cong {\scriptsize\pmatrix{0&1\cr 1&0\cr}}
 &\hbox{for }\SO_0(1,1)\sx\R^2\cr  
 {\scriptsize\pmatrix{-1&0\cr0&\bl 1_3\cr}}&
 \cong {\scriptsize\pmatrix{0&0&1\cr 
 0&\bl1_2&0\cr1&0&0\cr}}
 &\hbox{for }\SO_0(1,3)\sx\R^4\cr  
\end{array}\right.
\end{eq}The particle interpretable degrees of freedom start
with Minkowski spacetime $2R=4$. There, the  two 
degrees of freedom 
with a positive scalar product 
have left and right polarization
for the axial $\SO(2)$-rotations.

If adjoint re\-pre\-sen\-ta\-tions of  compact 
internal  degrees of freedom, e.g. of $\U(2)$ hypercharge-isospin,
are included,
the accordingly normalized 
residues of the arising  mass zero solutions in 4-di\-men\-sio\-nal spacetime
 may be compared with
the coupling constants
in the propagators of a massless gauge fields
in the standard model of electroweak interactions
\begin{eq}{l}
\SO_0(1,3)\sx\R^4:~{-\eta^{jk}G^2\over q^2+io}
\hbox{ with } G^2\sim ( g^2_1, g^2_2|g^2,\ga^2),~g_1g_2
\sim {1\over 4.6}
\end{eq}Without the introduction of the internal degrees of freedom\cite{S982}
only the order of magnitude of the normalizations $G^2$
can be compared with the residues above for the simple massless poles from 
tangent re\-pre\-sen\-ta\-tions of  spacetime 
$\D(2)\cong \GL(\C^2)/\U(2)$ 
\begin{eq}{l}
\hbox{for }\log\D(2)\cong\R^4:~~
 G^2\lrmap -{\res_{}}(0,\ka^2)\sim {1\over 3}  
\end{eq}

\end{document}